\documentclass[a4paper,11pt]{article}
 \pdfoutput=1
\usepackage{amsmath,amssymb,mathtools,fullpage,slashed,array}
\usepackage{simplewick}
\usepackage{float}
\usepackage{color,xcolor}
\usepackage{graphicx}
\usepackage{cite}
\usepackage{multirow}
 \usepackage{pifont}
 \definecolor{blue}{rgb}{0,0,0.5}
   \restylefloat{table} 
      
\usepackage{verbatim}

\usepackage{jheppub}

\usepackage{siunitx}

\usepackage{dcolumn,booktabs}
\newcolumntype{d}[1]{D{.}{.}{#1}}
\newcommand\mc[1]{\multicolumn{1}{c|}{#1}} 
\newcolumntype{C}{>{$}c<{$}} 
\newcolumntype{L}{>{$}l<{$}} 
\newcolumntype{R}{>{$}r<{$}} 

\allowdisplaybreaks   
   
\definecolor{violet}{rgb}{0.94, 0.2, 0.8}
\definecolor{lightblue}{rgb}{0.2, 0.2, 0.93} 
\definecolor{lightgreen}{rgb}{0.1, 0.73, 0.33}
\definecolor{asparagus}{rgb}{0.53, 0.66, 0.42}
\definecolor{darkgreen}{rgb}{0,0.6,0}

\newcommand{\com}[1]{#1 }

\renewcommand{\arraystretch}{1.5}

\newcommand{\twist}{(\phi_\ga)}
\newcommand{\ksub}{k_0^2}

\newcommand{\ORD}{{\cal O}}

\newcommand{\phimq}{\phi^{(m_q)}}

\newcommand{\VEV}[1]{\left\langle #1 \right\rangle} 
\newcommand{\state}[1]{|#1\rangle}

\newcommand{\al}{\alpha}

\newcommand{\be}{\beta}

\newcommand{\ga}{\gamma}
\newcommand{\Ga}{\Gamma}

\newcommand{\gaf}{{\gamma_5}}
\newcommand{\de}{\delta}
\newcommand{\eps}{\epsilon}

\newcommand{\la}{\lambda}

\newcommand{\sig}{\sigma}

\newcommand{\der}{ d}

\newcommand{\intu}{\int_0^1 du}
\newcommand{\fg}{ f_{3\gamma}}
\newcommand{\qbarq}{\langle \bar{q}q\rangle}
\newcommand{\sbars}{\langle \bar{s}s\rangle}
\newcommand{\intv}{\int_0^1 dv}
\newcommand{\inta}{\int\mathcal{D}\underline{\alpha}}

\newcommand{\au}{\underline{\alpha}}

\newcommand{\wg}[1]{\omega_{\gamma}^{#1}}

\newcommand{\para}{\parallel}
\newcommand{\matel}[3]{\langle #1|#2|#3\rangle}

\newcommand{\vev}[1]{\langle #1 \rangle}
\newcommand{\GeV}{\,\mbox{GeV}}
\newcommand{\GeVs}{\, \mbox{\small GeV}}
\newcommand{\MeV}{\,\mbox{MeV}}
\newcommand{\MeVs}{\,  \mbox{\small MeV}}
\newcommand{\LaQCD}{\Lambda_{\textrm{QCD}}}
\newcommand{\m}[1]{$m_b^{-#1}$}

\newcommand{\SumQ}{\overline{ Q}}
\newcommand{\DelQ}{Q_{\bar{B}}}
\newcommand{\DelQq}{Q_{\bar{B}_q}}
\newcommand{\DelQu}{Q_{\bar{B}_u}}

\newcommand{\Pperp}[1]{P_\perp^{#1}}
\newcommand{\Ppara}[1]{P_\parallel^{#1}}

\newcommand{\pperp}[1]{P^\perp_{#1}}
\newcommand{\ppara}[1]{P^\parallel_{#1}}
\newcommand{\plow}[1]{P^{\textrm{Low}}_{#1}}

\newcommand{\pperpparalow}[1]{P^{\perp,\parallel,\textrm{Low}}_{#1}}
\newcommand{\sD}{}
\newcommand{\FBlow}{ f^{(pt)}_{B_q}}
\newcommand{\FBlowno}{ f^{(pt)}_{B_q}}
\newcommand{\FBlowu}{ f^{(pt)}_{\bar{B}_u}}

\newcommand{\mBz}{m_{B_d}}
\newcommand{\mBu}{m_{B_u}}
\newcommand{\mBs}{m_{B_s}}

\newcommand{\mBsz}{m_{B^{*}_d}}
\newcommand{\mBsu}{m_{B^{*}_u}}
\newcommand{\mBss}{m_{B^*_s}}

\newcommand{\mBoz}{m_{B_{1d}}}
\newcommand{\mBou}{m_{B_{1u}}}
\newcommand{\mBos}{m_{B_{1s}}}

\newcommand{\lscale}[1]{{ |#1\!\GeV}}

\newcommand{\Bor}{{\hat{\mathcal{B}}}}

\newcommand{\Ima}{\textrm{Im}}

\newcommand{\CondQQ}[1]{{ \left\langle \bar{#1} #1 \right \rangle}}

\newcommand{\CT}{CT }
\newcommand{\CTs}{CTs }

\newcommand{\pB}[1]{(p_B)_{#1}}

\newcommand{\DV}{{\cal D}_\perp}
\newcommand{\DA}{{\cal D}_\parallel}
\newcommand{\FV}{V_\perp}
\newcommand{\FA}{V_\parallel}

\newcommand{\FTV}{T_\perp}

\newcommand{\FTA}{T_\parallel}

\newcommand{\gonepl}{g_{BB_1\gamma}}
\newcommand{\gonemi}{g_{BB^*\gamma}}

\newcommand{\fBst}[1]{f#1_{B^*}}
\newcommand{\fBone}[1]{f#1_{B_1}}

\usepackage{accents}

\newcommand{\doublehat}[1]{#1^{(2)}}  
\newcommand{\singlehat}[1]{#1^{(1)}}   

\newcommand{\Cdot}{\!\cdot \!}
\newcommand{\type}{\Gamma}
\newcommand{\calpha}{a}
\newcommand{\cbeta}{b}
\newcommand{\cgamma}{c}

\newcommand{\den}{S}
\newcommand{\denbar}{\overline{S}}
\newcommand{\one}{{\cal U}}

\newcommand{\Corr}{\Pi}

\newcommand{\TAB}{Tab.~}
\newcommand{\TABs}{Tabs.~}
\newcommand{\FIG}{Fig.~}
\newcommand{\SEC}{Sec.~}
\newcommand{\SECs}{Secs.~}
\newcommand{\APP}{App.~}
\newcommand{\APPs}{Apps.~}
\newcommand{\EQ}{Eq.~}

\newcommand{\sSum}{S(\underline{\alpha})\!+\!\tilde{S}(\underline{\alpha})}
\newcommand{\tSum}{\sum\limits_{{i=1}}^{4}T_i^{(1)}(\underline{\alpha})}

\newcommand{\muUV}{\mu_{\text{UV}}}
\newcommand{\mukin}{\mu_{\textrm{kin}}}

\newcommand{\MSbar}{\overline{\text{MS}}}

\newcommand{\Cz}{C_0\left(0,p_B^2,q^2,m_b^2,m_b^2,0\right)}
\newcommand{\Czz}{\widetilde{C}_0\left(0,p_B^2,q^2,0,0,m_b^2\right)}

\begin{document}

\hypersetup{pageanchor=false}

\preprint{CP3-Origins-2020-14 DNRF90}

\title{\boldmath Charged and Neutral  $\bar B_{u,d,s} \to \ga$ Form Factors from \\
Light Cone Sum Rules at NLO
}

\author[1]{Tadeusz Janowski}
\author[1]{Ben Pullin,}
\author[1]{Roman Zwicky}

\affiliation[1]{Higgs Centre for Theoretical Physics, School of Physics and Astronomy, University of Edinburgh, Edinburgh EH9 3JZ, Scotland}

\emailAdd{b.pullin@ed.ac.uk, roman.zwicky@ed.ac.uk}

\bigskip
\pagestyle{empty}

\abstract{
We present the first  analytic ${\cal O}(\alpha_s)$-computation  at twist-$1$,$2$ 
of the $\bar B_{u,d,s} \to \ga$ form factors
within the framework of sum rules on the light-cone. These form factors describe 
the \emph{charged} decay
$\bar B_u \to \gamma \ell^- \bar{\nu}$,  contribute to the flavour changing \emph{neutral} currents 
$\bar B_{d,s} \to \gamma \ell^+ \ell^-$ and serve as inputs to more complicated processes. 
We provide a fit in terms of a $z$-expansion with correlation matrix and extrapolate the form factors 
to the kinematic endpoint by using  the  $g_{BB^*\gamma}$  couplings as a constraint. 
Analytic results are available in terms of multiple polylogarithms as ancillary files.
We give binned predictions for the $\bar B_u \to \gamma \ell^- \bar{\nu}$ branching ratio along with the associated 
correlation matrix. By comparing with three SCET-computations  we extract the 
inverse moment $B$-meson distribution amplitude parameter
$ \la_B = 360 (110)  \textrm{\small MeV}$. The uncertainty thereof  
could be  improved by a more dedicated analysis.
In passing, we extend the photon distribution amplitude to include quark mass corrections 
with a prescription for the magnetic vacuum susceptibility, $\chi_q$, compatible with the twist-expansion. 
The values   
$ \chi_q  =  3.21 (15) \textrm{\small GeV}^{-2}  $ and $ \chi_s  =   3.79  (17) \textrm{\small GeV}^{-2} $ are obtained.} 
\maketitle

\setcounter{page}{1}
\pagestyle{plain}
\hypersetup{pageanchor=true}

\section{Introduction}
\label{sec:intro}

In this work we provide the (photon) \emph{on-shell} form factors  contributing to 
semileptonic $\bar B_u \to \ga \ell^- \bar\nu$ and flavour changing neutral decay (FCNC) 
$\bar B_{d,s}\to \ga \ell^+ \ell^-$ with the method of  light-cone sum rules (LCSR). 
This is the first complete next-to leading order (NLO) computation in $\al_s$ 
including twist-$1$,$2$ and leading order (LO) twist-$3$ and partial twist-$4$. 
The dynamics, contrary to naive expectation, are more involved than for $\bar B\to V(\rho, \omega,K^*,\phi)$\cite{BB98b,BZ04b}.  This comes as the point-like photon, of leading twist-$1$,  
has no counterpart for the mesons and yet the photon is to be described by a photon 
distribution amplitude (DA) (analogous to the vector meson case and cf. \TAB\ref{tab:dict} for a dictionary).
The photon DA can be interpreted as  $\rho_0, \omega_0, \phi$-$\ga$ conversion 
and  is normalised by  
the QCD magnetic vacuum susceptibility $\chi$, for which 
we propose  a prescription to include quark mass effects in \SEC\ref{sec:mqDA}. 
The twist-$1$ part leading in the heavy quark scaling $F^{\bar B \to \ga} \propto m_b^{-1/2}$, differing 
from $F^{\bar B \to V} \propto m_b^{-3/2}$, \com{which we discuss in the hadronic 
picture
in the large-$N_c$ limit  using a dispersion   in \SEC\ref{sec:scaling}.}

In the literature the \emph{charged} $\bar B_u \to \ga $ form factors have been considered in LCSR at  
leading order (LO) in $\al_s$ in  \cite{Khodjamirian:1995uc,Ali:1995uy,Eilam:1995zv} and a partial 
(no twist-$1$)  NLO  contribution in \cite{Ball:2003fq}. 
The decay has received considerable attention in QCD-factorisation and
soft-collinear effective theory (SCET), 
in part, as a 
toy model for factorisation and for its prospect to extract 
parameters of the  $B$-meson's distribution amplitude parameters 
\cite{Korchemsky:1999qb,Lunghi:2002ju,DescotesGenon:2002mw,Bosch:2003fc,Beneke:2011nf,
Braun:2012kp,Wang:2016qii,Beneke:2018wjp,Wang:2018wfj,Galda:2020epp,Shen:2020hsp}.
In \SEC\ref{sec:BDA} we extract the $B$-meson DA parameters  $\la_B$ by comparing to a SCET evaluation in 
three representative models. 
SCET-investigations include  
\cite{Korchemsky:1999qb,Lunghi:2002ju,DescotesGenon:2002mw,Bosch:2003fc} at leading power 
and \cite{Beneke:2011nf} at next-to-leading log and partial sub-leading powers.  
Subleading powers, (partially) corresponding   to the photon DA, were accounted for by dispersion techniques 
\cite{Braun:2012kp,Wang:2016qii,Beneke:2018wjp}. Other approaches include perturbative QCD 
\cite{Shen:2018abs} and a hybrid approach of QCD factorisation and 
LCSR  \cite{Wang:2018wfj} to which we compare in \SEC\ref{sec:CC}. 
A first lattice form factor evaluation has been reported for $D_{(s)},K ,\pi \to \ga$ in \cite{Desiderio:2020oej} and further 
evaluations of $D_s(K)  \to \ga$  are in progress \cite{Kane:2019jtj}.

The \emph{neutral} $\bar B_{d,s}\to \ga $ form factors describe flavour changing neutral currents which 
are generic probes for physics beyond the Standard Model.
The activity in this particular corner has been revived, 
since it has been 
pointed out \cite{Dettori:2016zff}  that, in a certain kinematic region, $\bar B_s \to\ga  \mu^+\mu^- $ may
be extracted from the very same dataset as $\bar B_s \to  \mu^+\mu^- $ at LHC experiments \cite{Aaij:2017vad,Aaboud:2018mst,CMS:2019qnb} (cf. also \cite{Aditya:2012im}). This triggered a number 
of phenomenological investigations \cite{GRZ17,Kozachuk:2017mdk,Beneke:2020fot,Carvunis:2021jga}.
Previously these form factors had been considered, in LCSR at LO \cite{Aliev:1996ud}, 
in \cite{Kruger:2002gf} using aspects of heavy quark symmetry (with input from  \cite{Beyer:2001zn}),  
quark model computations \cite{MN04,Kozachuk:2017mdk} and more recently in SCET \cite{Beneke:2020fot}.
In addition, in this decay, the dipole operator $m_b \bar q_L \sigma \Cdot F b $ necessitates 
photon \emph{off-shell} form factors.  These have been investigated for the first time,
at LO with QCD sum rules \cite{Albrecht:2019zul},  utilising   the on-shell form factors of  this paper as an input.

The paper is organised as follows. 
In \SEC\ref{sec:general}  the form factors are defined and their 
 heavy quark scaling is discussed in a  hadronic picture.  In \SEC\ref{sec:LCSR} the 
 correlation functions entering the computation, Ward identities and contact terms (\CTs) are discussed extensively. 
 The evaluation at LO and NLO for the perturbative and soft parts is documented in \SEC\ref{sec:comp}.  
 The numerics  \SEC\ref{sec:numerics} contains our main practical results in terms of fits, plots, the extraction of the inverse moment of the  $B$-meson DA and the binned prediction 
 for the $\bar B_u \to \gamma \ell^- \bar{\nu}$ rate in \SECs \ref{sec:inputs}, \ref{sec:plots}, \ref{sec:rate} and \ref{sec:BDA} respectively.
 The paper ends with  summary and conclusions in \SEC\ref{sec:conclusion}.
 Appendices include conventions and comments on form factor definitions (\APPs\ref{app:conv}), gauge invariance of the amplitude (\APP\ref{app:GI}),
 coordinate space  equation of motion  (including their non-trivial   renormalisation) (\APP\ref{app:eom}), 
 a compendium  of photon DAs (\APP\ref{app:DAs}), results of correlation functions (\APP \ref{app:results}) 
 and discontinuities of multiple-polylogarithms in (\APP\ref{app:mpl}).

\section{Generalities  on the  \texorpdfstring{$\bar B \to \gamma$}{} Form Factors}
\label{sec:general}
The  section  provides the method-independent information on the $B \to \gamma$ form factors (FFs), such as the definition and the heavy quark scaling.

\subsection{Definition of the Form Factors}
\label{sec:defFF}

Amongst the  dimension six operators of the $b \to q$ effective Hamiltonian there are the standard vector and tensor
operators. Note that the scalar operator vanishes by parity conservation of QCD and the pseudoscalar one only contributes  for an off-shell photon (cf. reference \cite{Albrecht:2019zul} for a complete discussion). 
We consider, in addition, an operator which is redundant by the
equation of motion (EOM) but that will prove useful as a check and is of interest per se \cite{Hambrock:2013zya,BSZ15}. 
In complete generality, with conventions  as in 
\cite{GRZ17,Albrecht:2019zul}, extended to include the charged case,
 we define the  FFs\footnote{It is important to keep in mind that the results quoted throughout are for the $\bar B$-meson.
For example, by  applying a charge  $C$- or $CP$-transformation, one infers that the vector 
 (but not the axial) parts  change sign $V,T_\perp \to - V,T_\perp$ (assuming $C\state{\bar B} = \state{ B}$ 
and or $CP\state{\bar B} = \state{ B}$) when passing from the $\bar B$- to the $B$-FFs.
The adaption to generic phases under $C$ and $CP$ is straightforward.}
\begin{alignat}{3}
\label{eq:ffs}
& \matel{ \gamma(k,\epsilon) }{O^V_\mu }{ \bar{B}_q (p_B) } 
 &\;=\;&   s_e ( \pperp{\mu} \, \FV (q^2)  &\;-\;&  \ppara{\mu} \, (\FA(q^2)+\DelQq \FBlow )  -   \plow{\mu} \DelQq \FBlow
  )   ~, \nonumber \\[0.1cm]
& \matel{ \gamma(k,\eps) }{ O^T_\mu  }{ \bar{B}_q (p_B) }
 &\;=\;& s_e( \pperp{\mu }   \, \FTV(q^2) &\; -\;& \ppara{\mu }  \, \FTA(q^2)   )
  ~,\\[0.2cm]
 & \matel{ \gamma(k,\eps) }{ O^{\cal D}_\mu  }{ \bar{B}_q (p_B) }
 &\;=\;& s_e(  \Pperp{\mu} \, \DV (q^2)  &\; - \;&  \Ppara{\mu} \, (\DA(q^2) + \hat{m}_- \,   \DelQq \FBlow) -   \plow{\mu}  \hat{m}_- \, \DelQq \FBlow )   \;,   \nonumber 
\end{alignat}
where   $ \hat{m}_\pm \equiv m_\pm/m_{B_q}$, $m_\pm \equiv (m_b \pm m_q)$ and $\DelQq=Q_b-Q_q$ is the charge of the $\bar{B}_q$-meson.
Comparison with literature-conventions are deferred to the next section.
Hereafter, 
\begin{equation}
\label{eq:pt-like}
 \FBlow = \frac{m_{B_q} f_{B_q} }{ k \cdot p_B} =  \frac{2  f_{B_q}/m_{B_q}}{1-q^2/m_{B_q}^2} 
\end{equation}
denotes  the point-like (or scalar QED) contribution, proportional to the decay constant,
and consists of 
the  infrared (IR) sensitive  Low-term, which  captures the gauge variance of the matrix element on the LHS (for charged $\bar B_q$) 
and assures that the FFs themselves are gauge invariant (cf.~\SEC\ref{sec:QEDWI} and 
\APP\ref{app:GV}).   The gauge variance of the matrix element is 
cancelled in the full $\bar B_u \to \ga \ell^- \bar \nu$ amplitude when the photon-emission from the lepton is 
added  cf.  \APP\ref{app:GIrestored}. 
The same terms appears in  $ \Ppara{\mu} $-structure 
whose subtraction from the FF was stressed in \cite{Carrasco:2015xwa,Desiderio:2020oej} 
cf. \SEC\ref{sec:QED}. Furthermore, $s_e$ is the sign-convention  of  the covariant derivative (cf. \APP\ref{app:conv}),
$p_B  = q+ k$, on-shell 
momenta $p_B^2 = m_B^2$, $k^2 = 0$, $f_{B_q}$  the decay constant
\begin{equation}
\label{eq:fB}
\matel {0 }{ \bar{q}  \gamma^{\mu} \gamma_{5}  b  }{ \bar{B}_q (p_B) } 
= i p_B^{\mu} f_{B_q} \;,
\end{equation}
local operators  (describing the $b \to q$ transition)
\begin{equation}
\label{eq:weak_op}
 O^V_\mu \equiv -  \frac{1}{e} m_{B_q} \bar{q} \gamma_{\mu} (1- \ga_5) b \;, \quad O^T_\mu  \equiv 
 \frac{1}{e}  \bar{q} i q^\nu \sigma_{\mu \nu} ( 1 + \ga_5) b \;, \quad 
 O^{\cal D}_\mu \equiv 
  \sD   \frac{1}{e} \bar q  ( i \!  \overset{\leftrightarrow}{D})_{\mu}(1 + \gamma_5) b \;,
\end{equation}
and Lorentz structures ($\varepsilon_{0123} = 1$ convention, $e  = \sqrt{ 4 \pi \al} >0 $ and  
 $\pperpparalow{\mu}    \equiv \eps^{*\rho} \pperpparalow{\mu \rho }$)
\begin{equation}
\label{eq:P} 
 \pperp{\mu\rho} \equiv   \varepsilon_{\mu \rho \be \ga } (p_B)^\be k^\ga  \;,  \quad    \ppara{\mu \rho }   \equiv  i  \,  (   p_B \cdot k \, g_{\mu \rho} -   k_{\mu} \,\pB{\rho} ) \;, 
 \quad P^{\textrm{Low}}_{\mu\rho} \equiv i  \pB{\mu} \pB{\rho} \;.
\end{equation}

 The  Lorentz structures are transverse, $q \cdot P^{\perp,\parallel} = 0$, due to
the $\la = \pm 1$-helicity of the photon. As usual the  Lorenz gauge, $k \cdot \eps(\la,k)  = 0$,  is assumed 
and $\eps \to \eps  + k$ is the residual gauge invariance encoded as 
$P^{\perp,\parallel}|_ {\eps \to k} =0$. 
By writing the Lorentz decomposition of 
$\matel{ \gamma(k,\eps) }{ \, \bar{q} i \sigma^{\mu \nu} ( 1 \!+\! \ga_5) b \, }{ \bar{B}_q (p_B) }$
and using the algebraic relation \eqref{eq:5sigma} the FF relation
\begin{equation}
\label{eq:algebraic}
 \FTV^{B \to \ga} (0) =  \FTA^{B \to \ga} (0) \;, 
\end{equation}
 can be shown to hold, in complete analogy to the $B \to V$ FFs, $T_1^{B \to V}(0) = T_2^{B \to V}(0)$.

A short comment on  the $\perp,\para$ subscripts.
 Interpreting the FF
 as a  $\bar B  \to 1^{-(+)} \ga$ transition,   
 $\perp(\para)$ correspond to perpendicular (parallel) $3$-dimensional polarisation vectors of the photon
 and the $1^{-(+)}$ state (as can be seen from \eqref{eq:P}).
In the $B \to V \ell \ell$ literature, e.g.  \cite{Kou:2018nap},  the $\perp,\para$ combinations are known 
as the  transversity amplitudes and are useful because of definite CP-parities.

\subsubsection{Form Factor Conventions in the Literature}
\label{sec:FFconvLit}

We wish to explain why we have adopted a different notation w.r.t. to some of the literature. There are three reasons:
firstly, in the off-shell case there are $7$ FFs \cite{Albrecht:2019zul} for which the $F_{A,V}$ notation becomes 
inefficient. Second, the subtraction of the point-like term in the charged case following 
\cite{Carrasco:2015xwa,Desiderio:2020oej} (cf. \FIG\ref{fig:pt-like} for comparison).
Third, in \cite{GRZ17} the  sign of the FFs was chosen such that the $\bar B_{d,s} \to \ga$ FFs are positive, and since 
the spectator quark is dominant this implies that the $\bar B_u \to \ga$  FFs are negative. 
The latter explains the sign difference
\begin{equation}
\label{eq:above2}
V^{\bar{B}_u \to \ga}_{\perp} = - F^{ \bar{B}_u \to \ga}_{V} \;, \quad 
V^{\bar{B}_u \to \ga}_{\para}  +\DelQu \FBlowu   = - F^{ \bar{B}_u \to \ga}_{A} \;,
\end{equation}
 w.r.t.  to the SCET-literature with focus  on the charged transition.
 Compared with \cite{Desiderio:2020oej} we find
\begin{equation}
V^{D_s \to \ga}_{\perp} = - F^{ D_s \to \ga}_{V} \;, \quad V^{D_s \to \ga}_{\para}     = - F^{ D_s \to \ga}_{A} \;,
\end{equation}
assuming their translation to the SCET-literature.   Essentially these are the same conventions as in \eqref{eq:above2} with the point-like part separated, which hopefully will become 
the new standard.
When comparing with  \cite{MN04,Kozachuk:2017mdk}  we believe the following map should be used
\begin{equation}
V^{\bar B_{d,s} \to \ga}_{\perp,\para} =  F^{ \bar B_{d,s} \to \ga}_{V,A} \;, \quad 
T^{\bar B_{d,s} \to \ga}_{\perp,\para} =  FT^{ \bar B_{d,s} \to \ga}_{V,A}  \;.
\end{equation}
 Note that in comparing analytic results one needs to keep in mind that the signs of
 $\chi$ and $s_{e}$ (and $s_g$) are author-dependent as well.

\subsection{\texorpdfstring{$B \to \ga$}{} Form Factor  in the context of QED-corrections to leptonic decays}
\label{sec:QED}

 Since the $B \to \ga$ FFs are the real emission corrections to the leptonic decay 
$B^+ \to \ell^+ \nu$ or  part of the FCNC $B_{d,s} \to \ell^+ \ell^-$, 
we consider it worthwhile to clarify a few aspects with regards to QED-corrections to leptonic decays: the point-like term indicated in \eqref{eq:ffs}, the absence $\ln m_\ell$-terms, the importance of the kinematic endpoint and extrapolations.
The arguments below go beyond  $B$-physics since infrared effects are rather universal.

In the charged case the photon couples to the charged $B$-meson, there is a point-like or monopole contribution which can be identified by matching the axial current 
 $A_{5\mu} = \bar q \ga_\mu \ga_5 b $ to 
a  meson current $A_{5\mu}  \to - f_B D_\mu \bar B^\dagger$, with
$\matel{0}{B^\dagger(x)}{\bar B} =  e^{-i p_B \cdot x}$  such that \eqref{eq:fB} holds. 
One then easily finds 
\begin{equation}
\label{eq:Jmu5}
\matel{\ga}{A_{5\mu}}{{\bar B}} =  f_B (-i e s_e \DelQ) \eps^{*\rho} \left( g_{\rho\mu} + \frac{(2 p_B -k)_\rho q_\mu  }{2 k  \cdot  p_B -k^2 } \right) \;,
\end{equation}
where the metric term comes from the photon field in the covariant derivative. 
Upon using  $k^2 = 0$ and $\eps^* \cdot k =0 $  one recovers the form in \eqref{eq:ffs}. 

It seems worthwhile to point out that leptonic decays are free from hard-collinear logs of the form
$\ORD(\al \ln m_\ell)$, as these have been shown to vanish from 
 structure-dependent corrections by gauge invariance  (cf. section 3.4 in  \cite{Isidori:2020acz}). 
 Thus  they can only arise from the point-like contribution which is itself free from 
 $\ln m_\ell$-terms. This can be seen as follows. 
 In the real emission the collinear terms would arise from the 
  Low-term which is proportional to the  helicity-suppressed LO term and  therefore the logs are of the form  $m^2_\ell \ln m_\ell$ at worst.  Finally, since real and  virtual point-like contributions cancel each other in the photon-inclusive rate the virtual corrections must be  free from such terms altogether. 
This  completes the argument.

In the actual measurement of $B^+ \to \ell^+ \nu$ the undetected photon is soft 
and $q^2$ is very close to the kinematic endpoint, which is beyond the limit of the direct LCSR computation. This underlines the importance of the extrapolation as 
discussed in \SEC\ref{sec:fit} (based on the  $g_{BB*\ga}$ 
couplings \cite{Pullin:2021ebn}).
A more direct sum rule analysis or the lattice approaches mentioned earlier  are of course an alternative thereto.  We will report on the impact of 
$B_{s} \to  \ga \ell^+ \ell^-$ as a background to $B_d \to \ell^+ \ell^-$ elsewhere.

\subsection{Heavy Quark Scaling of Form Factors}
\label{sec:scaling}

The heavy quark scaling, $m_b \to \infty$, of the FFs has been of wide interest in connection with 
factorisation theorems.  
In some cases, such as for the  $B$-meson decay constant $f_{B_q} \sim m_b^{-1/2}$  \eqref{eq:fB}, 
the scaling   can be traced back to the (relativistic) normalisation fo the $B$-meson state.    
For the $B \to V$ FFs the same reasoning would lead to $m_b^{-1/2}$  but is altered to $m_b^{-3/2}$ by the dynamics  of a fast moving $s$-quark, released by the heavy $b$,  combining with a slow-moving spectator quark $\bar q$. 
However,  the  $\bar B \to \ga$ FFs scale like $m_b^{-1/2}$
 because the perturbative photon does not need to hadronise from two 
asymmetric quarks.\footnote{Note, that the photon DA part, of the   $B \to \ga$ FF,   scales like $m_b^{-3/2}$, 
as the computation is formally analogous to the one of $B \to V$ (cf. \TAB\ref{tab:dict}).} 

The question we would like to pose here is whether this scaling can be understood from a hadronic picture. In the heuristic discussion put forward below we argue that the scaling can be understood from a certain number of low lying vector resonances coupling to  the strange part of the electromagnetic  current.
For this purpose it is useful to consider the photon off-shell FF at first, $k^2 \neq 0$, and write a dispersion relation in the photon momentum
\begin{equation}
T^*_\perp(q^2,k^2)  = \frac{1}{\pi}\int_{\textrm{cut}} \!\!\! du \frac{ \textrm{Im}_u [ T^*_\perp(q^2,u) ]} {u- k^2-i0}  + \textrm{s.t.} \;,
\end{equation}
where ``s.t." stands for subtraction terms and  by $T_\perp(q^2) =  \lim_{k^2 \to 0} T^*_\perp(q^2,k^2) $ the on-shell FF is recovered. 
It is convenient to assume the limit of large colours, $N_c \to \infty $, for which the spectrum is entirely 
described by vector mesons of zero width and the above dispersion representation assumes the
 form\footnote{The dispersion integral needs one subtraction, cf. \APP A.2. \cite{Albrecht:2019zul}, but in what follows we will ignore since it is consistent with our assumption that very large terms in the sum do not affect the leading power behaviour.} 
\begin{equation}
\label{eq:summy}
T^{\bar B_s \to \ga} _\perp(q^2)  =  
( \sum_{V_n = \phi,\phi',\dots }  \frac{  f^{\textrm em}_{V_n}} {m_{V_n}}  \, 
T_1^{\bar B_s \to {V_n}}(q^2)    
+ \textrm{s.t.}  ) +\ORD(Q_b) \;,
\end{equation}
where $f^{\textrm{em}}_{\phi} =  Q_{s} f_{\phi}$ etc, with 
$f_{\phi}$ the  standard decay constant cf. \cite{Albrecht:2019zul} 
and we have separated the $Q_b$-term which is suppressed by one power in $1/m_b$ 
(cf. \TAB\ref{tab:HQ}).
 Up to a certain number of resonances  $\bar n$, 
where $ m_{V_{\bar n}}   \ll m_B$ holds,  every single term in \eqref{eq:summy} scales like $m_b^{-3/2}$.  \com{It is difficult to say anything precise for the state above 
$\bar n$ since by assumption the LC-OPE technique does not apply to this case 
and we cannot infer the behaviour in $n$ of the form factor.  
However, it would be rather surprising if this contribution 
was leading in the heavy quark limit since the finite sum contribution gives a result compatible with no heavy quark suppression other than 
kinematic factors.} 
How does the sum of  terms from $n=1$ to $n = \bar n$ affect the overall scaling? 
In the limit of large colours the Regge ansatz ($n = 1,2,3,\dots$)  $m_{V_n}^2 \propto n$ with $f_{V_n} \propto n^0 $ is phenomenologically successful in that it
reproduces the correct logarithmic behaviour of the vacuum polarisation \cite{Shifman:2000jv}. 
From explicit computation we know that $T_1^{\bar B_s \to V_n}  \propto n^0$ thus every summand 
scales like $n^{-1/2} m_b^{-3/2}$. 
If we further assume that every summand is of the same sign then    
 the sum \eqref{eq:summy}  behaves like
\begin{equation}
\label{eq:prescale}
 T^{\bar B_s \to \ga} _\perp(q^2) \propto  \sum^{{\bar n}} _{n=1}  \frac{  f^{\textrm em}_{V_n}} {m_{V_n}}  \, 
T_1^{\bar B_s \to {V_n}}(q^2)     \propto \, m_b^{-3/2} \sqrt{\bar n}\;.
\end{equation}
 It remains to determine how $\bar n$ scales with $m_b$. 
\com{ Let us imagine to stretch $m_b$ by a factor of $\la$, $m_b \to \la m_b$, 
 then $m_{V_{\bar n}} \to \la m_{V_{\bar n}}$ (but $\LaQCD$ 
 and
 $m_{V_{\bar n}}/ m_B$ fixed),  then implies 
$\bar n \to \la^2 \bar n$ and thus 
 $\bar n \propto m_b^2$. } 
Hence from  \eqref{eq:prescale} we  finally conclude our anticipated result of 
 \begin{equation}
 \label{eq:Qqscale}
 T^{\bar B_s \to \ga} _\perp(q^2) \propto m_b^{-1/2} \;.
 \end{equation}
 The argumentation for the other three FFs is completely analogous.
\com{ This is a satisfactory result in that it matches our explicit results but remains heuristic mainly since we 
 are not able to make a precise statement about  the $n > \bar n$-terms other than 
 that they are unlikely to change the scaling (cf. above).}

\section{The \texorpdfstring{$\bar B \to \gamma$}{} Form Factors from LCSR}
\label{sec:LCSR}

The method of QCD sum rules \cite{SVZ79I,SVZ79II} was originally established for 
two-point correlation functions with a local OPE.  LCSR were introduced \cite{Balitsky:1997wi} 
to describe $\Sigma^+ \to p \ga$ FFs.  It has been  understood that heavy-to-light FFs 
are to be described by LCSR rather than $3$-point functions with local OPE \cite{Braun:1997kw}. 
We refer the reader to \cite{Colangelo:2000dp} for a review on practical aspects of sum rules.

\subsection{Definition of the Correlation Function appropriate for the Sum Rule}

In the LCSR the $B$-meson is interpolated by a current, with standard choice
\begin{equation}
\label{eq:fBjB}
J_{B_q} \equiv (m_b + m_q) \bar{b} i \ga_5 q \;, \quad 
 \matel{\bar{B}_q}{J_{B_q}}{0} = m_{B_q}^2 f_{B_q} \;,
\end{equation}
and the FF is extracted from the correlation function 
 \begin{align}
\label{eq:corr}
\Corr^\type_{\mu\rho}(p_B,q)  \equiv \;&  i \int_x  e^{-ip_B \cdot x} \matel{\ga(k,\rho)}{ T J_{B_q}(x) O_\mu^\type(0)}{0} 
     \\[0.1cm]
 =\;& 
s_e \left( \pperp{\mu \rho} \, \Corr^\type_\perp(p_B^2,q^2) 
- \ppara{\mu \rho} \, \Corr^\type_\para(p_B^2,q^2)    +   \pB{\mu}   \pB{\rho} X^\type + 
 g_{\mu \rho}k\cdot p_B Z^\type +
  {\cal O}(k_\rho) \right) \;.
\nonumber
\end{align}
Above and hereafter we use the shorthand  $\int_x = \int d^4 x$ and $\type = V,T,{\cal D}$ is the index 
associated with the operators \eqref{eq:weak_op}.
The mass dimensions of the scalar functions are: $[\Corr^{V,T,{\cal D}}] =1$.
The index $\rho$ is to be contracted by $\eps^{*\, \rho}$ and the structures $X^\type$ and $Y^\type$ 
are related to the \CTs and can be unambiguously identified from the QED- and or axial-Ward identity (WI), 
to be discussed in \SEC \ref{sec:QEDWI}.

The LCSR is then obtained by evaluating \eqref{eq:corr}, in the light-cone OPE, 
and equating it to  a dispersion representation. For  $\Corr^V_\perp$ the LCSR reads 
\begin{equation}
\label{eq:disp}
\Corr^V_\perp(p_B^2,q^2) = \frac{1}{\pi} \int_{0}^\infty ds \frac{\textrm{Im}[\Corr^V_\perp(s,q^2)]}{s-p_B^2- i0} =  
\frac{  \FV(q^2) m_{B_q}^2 f_{B_q }}{m_{B_q}^2 - p_B^2 - i0}  +  \dots \;,
\end{equation}
where the narrow width approximation for the $B$-meson has been assumed and the ellipses 
correspond to heavier resonances and multiparticle states. 
 The FFs are then extracted via 
 the standard procedures of Borel transformation and approximating the heavier states  
 by the perturbative integral, which is exponentially suppressed due to the Borel transform
 \begin{equation}
 \label{eq:FFSR}
 \FV(q^2)  = \frac{1}{ m_{B_q}^2 f_{B_q }} 
  \int_{m_b^2}^{s_0} ds  \, e^{(m_{B_q}^2-s)/M^2} \rho_\perp^V(s,q^2)   \;.
 \end{equation}
 Above $ \pi \rho_\perp^V(s,q^2) = \textrm{Im}[\Corr^V_\perp(s,q^2)]$ and $M^2$ is the Borel mass, 
 with some minimal detail given in  \APP\ref{app:Borel}. If we were able to compute $\rho_\perp^V$ 
 exactly then $\FV(q^2)$, obtained from \eqref{eq:FFSR}, would be independent of the Borel mass and its variation serves as a quality measure 
 of the sum rule. The other FFs are obtained in exact analogy.  
 Hereafter we will abbreviate $m_{B_q} = m_{B}$ and $f_{B_q} = f_B$ unless otherwise stated.
 
\subsection{QED and Axial Ward Identity}
\label{sec:QEDWI}

This section can be skipped by the casual reader as it deals 
with the technical issue of how to project on the FFs and 
the appearance of the $\DelQ$-terms in  \eqref{eq:ffs}. 

Generally on-shell matrix elements can be extracted as pole 
residues from correlation functions. However for three-point functions, 
with a current insertion for the photon, this procedure is complicated by 
\CTs which  manifest themselves in the axial- and QED-WI.  
Whereas these matters have been discussed previously in the literature  
\cite{Khodjamirian:2001ga}, our presentation is more complete in that we 
discuss the derivative and tensor case as well and work purely at the level of 
correlation functions.

A general decomposition of the correlator \eqref{eq:corr} with electromagnetic current insertion  reads
 \begin{eqnarray}
\label{eq:corrPT}
\Corr_{\mu \rho} ^\type(p_B,q) 
&\equiv &\!\!  s_e e  
 \int_{x,y} e^{-ip_B \cdot x} e^{i k \cdot y} \matel{0}{ T j_\rho(y)  J_{B}(x) O_\mu^\type(0)}{0}   = s_e \left( \pperp{\mu \rho} \, \Corr^\type_\perp    \right.
- \ppara{\mu \rho} \, \Corr^\type_\para +
 \nonumber   \\[0.1cm]
 &\phantom{+}& \!\! 
    \left.   \pB{\mu}   \pB{\rho} X^\type + 
 k_\mu  \pB{\rho} Y^\type + g_{\mu \rho}  (k \Cdot p_B)  Z^\type  +
  k_\mu k_\rho W^\type + \pB{\mu} k_{\rho} U^\Gamma \right) \;,
\end{eqnarray}
where the $p_B^2$ and $q^2$  dependence on the right-hand-side (RHS) is suppressed for brevity.  
Compared to \eqref{eq:corr} the three structures $Y^\type$, $W^\type$ and $U^\type$ are introduced. 
The $U$- and $W$-terms are irrelevant after contraction with $\eps(k)^\rho$, as they vanish, and will hereafter be ignored.
It is the resolution of the ambiguity  between $Y$-$Z$ and  $\Corr^\type_\para $ that needs addressing.

One can derive a QED WI directly from the path-integral by varying $q(x) \to (1 + \omega(x) i Q_q)q(x)$ 
and demanding invariance under $\omega(x)$, similar to the derivation of the EOM given in \APP\ref{app:eomCorr}.  For the QED WI we find
\begin{alignat}{2}
\label{eq:GWI}
& k^\rho \Corr_{\mu \rho}^V &\;=\;&  s_e 
 \DelQ \left( C^A_\mu(q) - C^A_\mu(p_B) \right)  m_B
  \;,  \nonumber \\[0.1cm]
  & k^\rho \Corr_{\mu \rho}^D &\;=\;&  s_e 
\left( \DelQ \left( C^A_\mu(q) - C^A_\mu(p_B) \right) m_-
-   \SumQ   k_\mu  C^{P}(p_B^2)  \right)
  \;,  \nonumber \\[0.1cm]
& k^\rho \Corr_{\mu \rho}^T &\;=\;& 0 \;, 
\end{alignat}
   and for the axial WI 
($q(x) \to (1 + \omega(x) i \ga_5)q(x)$) one obtains
\begin{alignat}{2}
\label{eq:AWI}
  & q^\mu \Corr_{\mu \rho}^V &\;=\;&    s_e   \left(\Pi^P_\rho + 
  \DelQ {C}^A_\rho(p_B)  \right) m_{B}  \;, \nonumber \\[0.1cm]
   & q^\mu \Corr_{\mu \rho}^D &\;=\;&   s_e   \left(  ( \Pi^P_\rho + 
  \DelQ {C}^A_\rho(p_B) ) m_-
    -   \SumQ   q_\mu   C^{P}(p_B^2)  \right) \;,
  \nonumber \\[0.1cm]
& q^\mu \Corr_{\mu \rho}^T &\;=\;& 0 \;, 
\end{alignat}
where $\SumQ \equiv (Q_b+Q_q)$ and the other terms on the RHS are
\begin{alignat}{2}
\label{eq:final}
&  \Pi^P_\rho(p_B,k)  &\;=\;& 
i   
 \int_{x,y} e^{-ip_B \cdot x} e^{i k \cdot y} \matel{0}{ T j_\rho(y)  J_{B_q}(x)   \bar q \ga_5 b (0)}{0}
 \;, \nonumber \\[0.1cm]
 & C^A_\mu(p  ) &\;=\;& 
i \int_x e^{- i p \cdot x} \matel{0}{T J_{B_q}(x) \bar q \ga_\mu \ga_5(0)b}{0} = p_\mu C^A(p^2)  \;,
 \nonumber \\[0.1cm] 
&  C^{P}(p^2) &\;=\;&  i \int_x e^{- i p \cdot x} \matel{0}{T   J_{B_q}(x)  \bar q \ga_5 b (0)}{0}
  \;,
\end{alignat}
and $p^2 C^A(p^2) = (m_b+m_q) C^P(p^2) + m_b \vev{\bar b b} + m_q\vev{\bar q q}$. 
Moreover, we have used 
\begin{alignat}{1}
&m_-  \int_x e^{- i p \cdot x} \matel{0}{T J_{B_q}(x)   q \ga_\mu \ga_5(0)b}{0} = 
\int_x e^{- i p \cdot x} \matel{0}{T J_{B_q}(x) \bar q ( i \!  \overset{\leftrightarrow}{D})_{\mu} (0)\ga_5 b }{0} \;, \nonumber 
\\[0.1cm]
& \int_y e^{ i k \cdot y} \matel{0}{T j_\rho(y)  (\bar q q + \bar b b)(0)}{0}=0 \;.
\end{alignat}
where the former follows from \eqref{eq:EOMxspace} upon omission of the QED current and the latter
follows from gauge invariance.
Note that for the tensor part all  lower point correlation function vanish since there are not enough  vectors to 
impose the antisymmetry in the Lorentz vectors. 
 The correction term proportional to the sum of charges is due to the derivative term in the $O^D_\mu$-operator cf. \eqref{eq:EOMxspace}.
Consistency, $q^\mu k^\rho \Pi^V_{\mu\rho}|_{\mbox{\eqref{eq:GWI}}} =  q^\mu k^\rho \Pi^V_{\mu\rho}|_{\mbox{\eqref{eq:AWI}}}$, follows
 upon using $k^\rho \Pi^P_\rho  = \DelQ (C^A(q^2) - C^A(p_B^2))$.

From the decomposition \eqref{eq:corrPT} and the QED WI one infers that
\begin{alignat}{3}
\label{eq:hence}
& X^T = 0 \;,  & &   Y^T = 0 \;, & &  \quad Z^T = 0  \;,  \\[0.1cm]
& X^V =   \frac{ \DelQ}{k \cdot p_B} (C^A(q^2)- C^A(p_B^2)) m_B  \;, \quad 
& &  Y^V  + Z^V  &\;=\;&  - \frac{ \DelQ}{k \cdot p_B} \, C^A(q^2) m_B \;,  \nonumber \\[0.1cm]
& X^D =   \frac{ \DelQ}{k \cdot p_B} (C^A(q^2)- C^A(p_B^2)) m_- \;, \quad 
& &  Y^D  + Z^D  &\;=\;&  - \frac{ \DelQ}{k \cdot p_B} \, C^A(q^2) m_- + \frac{  \SumQ}{k \cdot p_B}  C^P(p_B^2)   \;. 
 \nonumber
 \end{alignat}
The following three points are worth clarifying:
 \begin{itemize}
\item 
The choices  $k \Cdot p_B (Y^V, Z^V) = (0, -  \DelQ \, C^A(q^2))$ and 
$k \Cdot p_B (Y^D, Z^D) = (0, -  \DelQ \, C^A(q^2) +  \SumQ C_P(p_B^2) )$ are compatible with 
the WI \eqref{eq:hence}  and would not contribute to the  $\|$-FFs.
\item 
The $X^V$-term is  the analogue of the Low-term in \eqref{eq:ffs}.
This can be seen by rewriting the correlator 
\begin{alignat}{2}
& \eps^{*\rho}   \Corr_{\mu \rho} ^V(p_B,q)   &\;=\;& - s_e \big( \ppara{\mu } \, \Corr^V_\parallel(p_B^2,q^2)  
\;+\;   i  \plow{\mu}   \frac{  \DelQ }{k \cdot p_B} (C^A(q^2) - C^A(p_B^2))m_B   + \dots\; \big) \nonumber \\[0.1cm]
& &\;=\;&  - s_e \frac{   m_B^2 f_{B}}{m_B^2- p_B^2}  \big(  \ppara{\mu } \,   (\FA(q^2) + \DelQ \FBlowno )   \;+\;   \plow{\mu} \DelQ  \FBlowno
   + \dots \big)  \;,
\end{alignat}
using  the dispersion representation  \eqref{eq:disp} and
\begin{equation}
C^A(p_B^2  ) = \frac{  i m_B^2 f_{B}^2}{m_B^2- p_{B}^2} +  \dots \;.
\end{equation}
This derivation becomes exact, reproducing \eqref{eq:ffs}, when applying the  LSZ-limit  

${\lim_{p_B^2 \to m_B^2}(p_B^2 -m_B^2)}$ to the correlation function.
 \item 
The analogous  $X^{\cal D}$-term can be treated in the very same way. 
The missing information is the matrix element 
\begin{equation}
\matel{0}{O^D_\mu}{\bar B_q} = f^D_B  \pB{\mu}  \;,  
\end{equation}
which is related to $f_B$  by the EOM \eqref{eq:eom} implying  $f^D_B = - f_B$.
When factors are arranged one reproduces the \CT in \eqref{eq:ffs}.
\end{itemize}
In practice, $Y^\type =0$,   meaning that 
\begin{equation}
\label{eq:recipe}
  \eps^{*\rho}  \Corr_{\mu \rho} ^\type =  s_e ( ( i \eps^* \Cdot p_B)
 \Corr^\type_\para k_\mu + A \eps_\mu^* + B \pB{\mu}  )   \;,
 \end{equation}
 is the correct recipe for projecting on $\Corr^\type_\para$. 
 It should be added that the entire discussion of these \CTs 
 can be avoided  if one were to consider directly a physical amplitude, as then 
 invariance under $\eps \to \eps +k$ holds cf. \APP\ref{app:GIrestored}.  
 The final recipe to obtain  $\FA$, with the point-like term subtracted, is to use
 \begin{equation}
\pi  \rho_\para^V(s,q^2)  = \textrm{Im}[\Corr^V_\para(s,q^2) - i X^V(s,q^2)] \;,
 \end{equation}
  in \eqref{eq:FFSR}, and similarly for $\cal{D}_\|$.  The expression $ iX^V$ corresponds to $\DelQ \FBlowno$. 
 From  \eqref{eq:hence} and  the corresponding WI below \eqref{eq:final}
 one can match to the standard pseudoscalar correlator  \cite{Pullin:2021ebn}.
 In the notation of this reference  the imaginary part of $i X^V$ reads
 \begin{equation}
 \label{eq:sub}
 \textrm{Im}[i X^V(s,q^2)]  = \pi  \frac{2 \DelQ}{s-q^2}  \, m_B (m_b+m_q)^2 \frac{\rho_{f_B}(s) }{s}\;.
 \end{equation}
For convenience we quote 
 $ \rho_{f_B}(s)|_{\textrm{PT-LO}}   =  \frac{N_c}{8 \pi^2} s ( (1-\frac{m_b^2}{s})^2   + \ORD(m_q))$ 
 which we have  explicitly verified to emerge  in our computation as in \eqref{eq:sub}.

\section{The Computation}
 \label{sec:comp}
 
 The aim of this section is to describe the computation of the
  correlation function \eqref{eq:corr} which contain both perturbative and light-cone OPE contributions, 
 as discussed in \SECs \ref{sec:PT} and  \ref{sec:twist} respectively.
These computations have been performed using 
 the packages  of
 \texttt{FeynCalc} \cite{FeynCalc1,FeynCalc2},  LiteRed \cite{litered:2013}, \texttt{Kira} \cite{kira:2017}, 
 \texttt{pySecDec} \cite{pySecDec1,pySecDec2}  and
 \texttt{PolyLogTools} \cite{polylogtools}. 
 
 First we discuss 
 the treatment of $\ga_5$ in dimensional regularisation (DR). The emergence of DR as the standard tool for multi-loop calculations has led to a number of prescriptions for $\gaf$ to allow for its consistent use. In this work we adopt two different prescriptions depending on the number of $\gaf$'s that appear in the traces. Our case includes either one or two $\gaf$'s. 
 For a single $\ga_5$ we adopt  the so-called Larin scheme  \cite{Larin:1993tq}, whereby 
\begin{equation}
\label{eq:Larin}
\gaf=\frac{1}{4!}i\,\varepsilon_{\alpha\beta\gamma\delta}\,\gamma^{\alpha}\gamma^{\beta}\gamma^{\gamma}\gamma^{\delta},
 \end{equation}
 and the Levi-Civita tensor $\varepsilon_{\alpha\beta\gamma\delta}$ is interpreted 
 as a $d$-dimensional object, in the sense that $\varepsilon_{\alpha\beta\gamma\delta}\varepsilon^{\alpha\beta\gamma\delta} = 
 d (d-1) (d-2)(d-3)$ for example.  This scheme is equivalent to the 't Hooft-Veltman scheme \cite{tHooft:1972tcz}, up to $\ORD(d-4)$, when 
 \eqref{eq:Larin} is applied to the bare as well as the counterterm graph.
 Breaking of $d$-dimensional Lorentz invariance, as only $4$ indices are present in  \eqref{eq:Larin}, has to 
 be restored by  a finite renormalisation  $Z^{P,s}_5=1-8\,C_F\frac{\alpha_s}{4\pi}+\ORD(\alpha_s^2)$ at NLO 
 \cite{Larin:1993tq}. In practice this amounts to the NLO replacement
 \begin{equation} 
 J_{B_q}\rightarrow J^{\textrm{Larin}}_{B_q}\equiv -
 Z^{P,s}_5  (m_b + m_q)/4! \,\varepsilon_{\alpha\beta\gamma\delta} \;\bar{b} \,\gamma^{\alpha}\gamma^{\beta}\gamma^{\gamma}\gamma^{\delta} q  \;,
 \end{equation}
  in \eqref{eq:corr}.
 When two $\ga_5$'s are present we use naive DR, whereby  $\ga_5$ is treated as an anticommuting 
 object with regards to all $\ga_\mu$ matrices  ($\mu = 1 ...d$). 
 We have checked that this is equivalent to replacing each one with \eqref{eq:Larin}.

 As the handling of the $\ga_5$ is delicate it is good practice to subject it to consistency tests. To this end we have checked the EOM at LO and NLO which constitutes a non-trivial test (cf. \SEC\ref{sec:EOMtest} for more detail).
 We have also verified the axial WI at LO; however since we project directly on scalar function at NLO the same test is not easily 
 performed, but given the EOM test we can be confident of our results. A further test  is that the
  FF relation \eqref{eq:algebraic} is explicitly obeyed, which is particularly valuable as it tests $\ga_5$'s 
  of even and odd number against each other.

Before going into some detail on the calculation let us comment on   light quark 
mass  $m_q$-contributions  which can be relevant for the strange quark. These effects are included 
everywhere except at NLO twist-$1$ which would necessitate master integrals beyond the current state of the art. These effects are rather small and the inclusion of $m_s$-effects in the magnetic vacuum susceptibility, that we propose, is a more significant effect.

 \subsection{Perturbation Theory and the Local OPE: Twist-1,4 }
 \label{sec:PT}
 
 In the perturbative and local OPE contribution the photon is treated as a perturbative vertex with
 \begin{equation}
 \matel{\ga(k,\la) }{A_\mu(x) }{0} = \eps^*_\mu(k,\la) e^{i k \cdot x}\; .
\end{equation} 
 In the literature this has been referred to as the hard contribution \cite{Ball:2003fq}.
 On grounds of  the $1/m_b$ power behaviour, the perturbative part and the condensate part are counted
 as twist-$1$ and -$4$ respectively. We will make this clearer when discussing the light-cone OPE where
 the notion of twist arises.
 It is helpful to distinguish the diagrams where the photon is emitted from the 
$q$- or the $b$-quarks, proportional to the $Q_q$- and $Q_b$-charge respectively,  
which we refer to as A- and B-type. It seems worthwhile to point out that the soft contributions are all 
proportional to $Q_q$.

 \subsubsection{Leading Order}
 \label{sec:LO}
 
\begin{figure}[!h]
 \centering
 	\includegraphics[width=\textwidth]{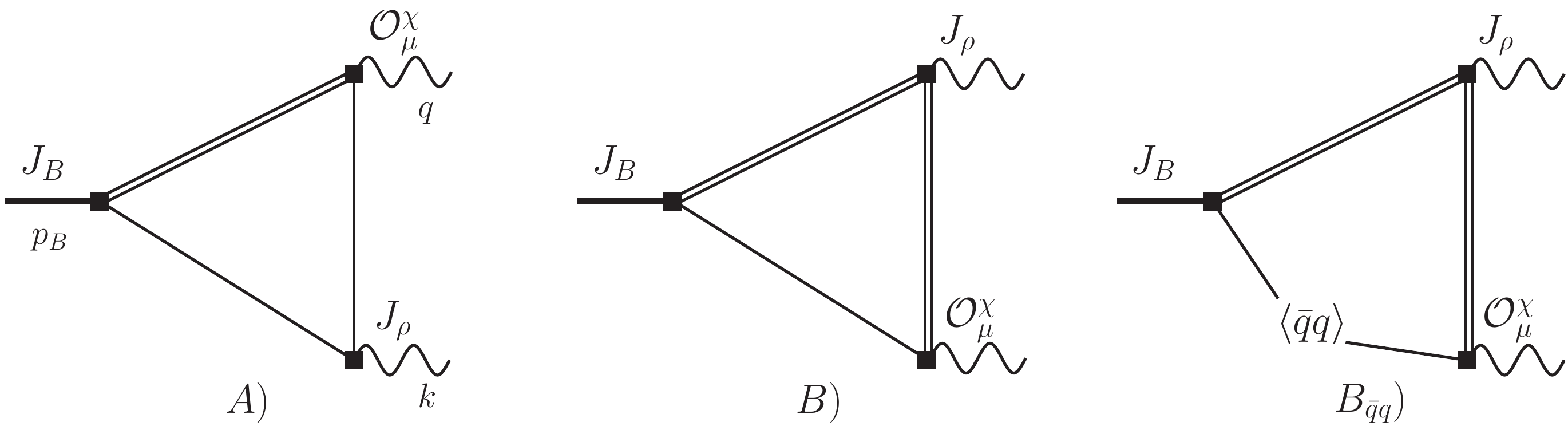}
	\caption{\small Perturbative diagrams of the A- and B-type as well as the B-type quark condensate 
	diagram. The absence of the A-type quark condensate diagram, due its singular $1/k^2$-behaviour,  
	is commented on in the main text.
	Throughout, single and double line denote 
	light $q = u,d,s$- and a heavy   $b$-quark respectively.}
	\label{fig:PT-graphs}
\end{figure}

The LO diagrams shown in  \FIG\ref{fig:PT-graphs} are straight forward to compute. 
Besides the perturbative graph we include the quark condensate but neglect the 
gluon condensate since $-m_b \vev{\bar qq} \gg \vev{G^2}$. 
Explicit analytic results for the correlation function can be found in \APP \ref{app:results}.
A relevant and interesting aspect is that the A-type quark condensate diagram behaves as 
$Q_q \vev{\bar q q }/k^2$ and diverges for an on-shell photon.  
This can be seen as a heuristic motivation for the photon DA which replaces these diagrams.

 \subsubsection{Next-to-leading Order}
 \label{sec:NLO1}
 
The most labour-intensive task of this work is the computation of 
the radiative corrections to the perturbative part. This computation is performed off-shell, $k^2 \neq 0$, as
it provides a regularisation to  IR-divergences and we can make use
of the NLO basis of master integrals \cite{DiVita:2017xlr}. 
With three external momenta and one internal mass, intended for 
 $X^0\rightarrow W^+ W^-$ with $X^0=\{H,Z,\gamma^*\}$, they are 
 state-of-the-art if the  method of differential equations
 in massive 2-loop integrals.
   \begin{figure}[tbp]
   \centering
   		\includegraphics[width=\textwidth]{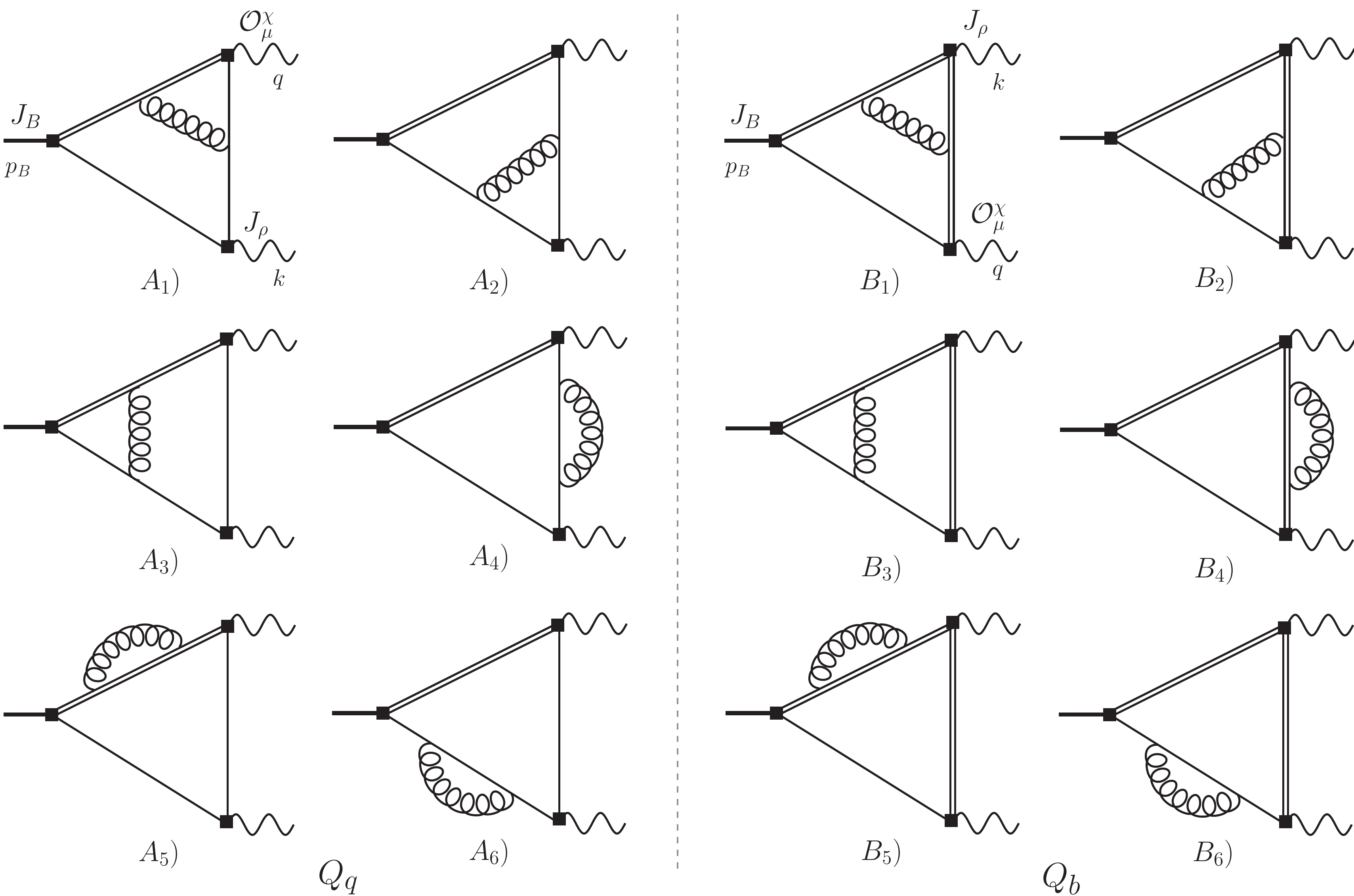}
   		\caption{\small Radiative corrections to the perturbative contribution to the form factors. Diagrams $A_{1-6})$ and  $B_{1-6})$ are proportional to the charges $Q_q$  and  $Q_b$  respectively.}
   		\label{fig:PT_2L}
   \end{figure}
The diagrams are depicted in \FIG\ref{fig:PT_2L} with the additional derivative diagrams in \FIG\ref{fig:PT_2L_Der}. After evaluating the traces in \texttt{FeynCalc} \cite{FeynCalc1,FeynCalc2} scalar products of momenta are efficiently replaced with propagator denominators via the \texttt{LiteRed} \cite{litered:2013} package, and then reduced to master integrals  
by \texttt{Kira} \cite{kira:2017}. The master integrals (MIs) are expressed in terms of 
multiple-polylogarithms (MPLs), cf.  \APP\ref{app:mpl}. The $k^2 \to 0$ limit is taken and thereafter
the discontinuities  for the dispersion representation \eqref{eq:disp}
are found using the \texttt{PolyLogTools} \cite{polylogtools} package.  
We extended the basis of MIs in  \cite{DiVita:2017xlr} to 
 include those that are given by the swapping of external legs.  
 Whereas technically this leads to an over-complete basis, this was necessary in order
 to perform both the renormalisation and the $k^2 \to 0 $ limit efficiently where cancellation of the IR-divergent  $\ln k^2$-terms takes place.
 We have included these MIs as an ancillary file  
 to the arXiv version.\footnote{Of course these additional integrals follow by exchanging the external momenta, namely $p_B^2\leftrightarrow q^2$ or $p_1^2\leftrightarrow p_2^2$, in the MPLs of their original basis counterparts. 
 However, this would lead to lengthy results and
 make the explicit cancellation of ultraviolet-(UV)  and IR-divergences ($k^2 \to 0$ limit) 
 difficult as the MPLs contain many hidden relations. 
 Hence we found it easier to construct the additional basis elements using the procedure in \cite{DiVita:2017xlr}.}
  All integrals, including the new MIs,  have been numerically verified against \texttt{pySecDec} \cite{pySecDec1,pySecDec2} and \texttt{PolyLogTools} \cite{polylogtools}.

Following this executive summary we give some more detail below. 
With the help of \texttt{LiteRed}, scalar products of loop and external momenta appearing in the numerators of the integrals depicted in the diagrams of \FIG\ref{fig:PT_2L} are written in terms of propagator denominators such that all integrals are  reduced to the following form
   \begin{equation}
   		\text{MI}[T,\{n_1,n_2,n_3,n_4,n_5,n_6,n_7\}]=\int  d^d\ell_1\int  d^d\ell_2 \frac{1}{D_{T1}^{n_1}\,D_{T2}^{n_2}\,D_{T3}^{n_3}\,D_{T4}^{n_4}\,D_{T5}^{n_5}\,D_{T6}^{n_6}\,D_{T7}^{n_7}}\;,
   \end{equation}
   with $n_i\in\mathbb{Z}$ and $T\in\{A,B\}$.  For the A-type basis, corresponding to diagrams proportional to the light-quark charge $Q_q$, the propagator denominators  are
      \begin{align}
   \nonumber
		D_{A1}&=\ell_1^2,     & D_{A2}&=\ell_2^2,    & D_{A3}&=(\ell_1+p_B)^2-m_b^2, \qquad D_{A4}=(\ell_2+p_B)^2-m_b^2, \\
		D_{A5}&=(\ell_1+k)^2, & D_{A6}&=(\ell_2+k)^2, & D_{A7}&=(\ell_1-\ell_2)^2,
   \end{align} 
   and for the  B-type basis, corresponding to diagrams proportional to the heavy-quark charge $Q_b$, we have,
   \begin{align}
   \nonumber
		 D_{B1}&=\ell_1^2-m_b^2\;,      &D_{B2}&=\ell_2^2-m_b^2\;, &D_{B3}&=(\ell_1-p_B+k)^2\;, \nonumber\\
		 D_{B4}&=(\ell_2-p_B+k)^2\;,  &D_{B5}&=(\ell_1+k)^2-m_b^2\;,  &D_{B6}&=(\ell_2+k)^2-m_b^2\;,     \nonumber \\
     D_{B7}&=(\ell_1-\ell_2)^2\;,&&&&
   \end{align} 
   with the usual  $i0$-prescription  implied.  
  We note that the A-type and B-type diagrams 
  correspond to the  $(a)$-$(b)$ and $(c)$-$(d)$ topologies in \cite{DiVita:2017xlr}.
   The corresponding MPLs are of weight $1$ to $4$ as is appropriate for a 2-loop computation. The additional MIs that we compute are:\footnote{We note that the MIs for the B-type basis, including the additional MIs \eqref{eq:Add_MI}, were also recently computed in \cite{Ma:2021cxg} in the context of mixed QCD-EW corrections to the $HW^+W^-$ vertex.}
   \begin{align}\label{eq:Add_MI}
   \text{MI}[B,\{0,2,1,0,1,1,1\}]\;,\qquad\text{MI}[B,\{0,1,2,0,1,1,1\}]\;,\qquad\text{MI}[B,\{0,1,1,0,1,1,1\}]\;,\nonumber\\
   \text{MI}[B,\{0,2,2,0,1,1,1\}]\;,\qquad\text{MI}[B,\{1,0,1,1,2,2,0\}]\;,\qquad\text{MI}[B,\{0,0,1,1,2,2,0\}]\;.
   \end{align}
The bare $\Pi^{\para,\perp}$ are subject to renormalisation. Besides the standard renormalisation 
of the $b$-quark mass in connection with the self-energy graphs, one need to renormalise 
the operators of the correlation function.\footnote{There is no need to discuss the renormalisation of overall 
local $1/\eps^2_{\textrm{UV}}$ and $1/\eps_{\textrm{UV}}$ terms due to the three operators coinciding as they are free of discontinuities in $p_B^2$. Such terms do not contain any physical information and consequently do not contribute to the dispersion integral.}
The electromagnetic current does not renormalise, $Z_{J_b} = Z_{m_b}$ and the 
renormalisation of the weak operators \eqref{eq:weak_op} is discussed in \APP\ref{app:renorm}. 
In particular the renormalisation of $O^D_\mu$ provides a rather non-trivial test of the computation. After this we perform 
the $k^2 \to 0$ limit and take the discontinuities aided by  \texttt{PolyLogTools}.
 In order to do this efficiently we use  shuffle algebra relations, such as
   \begin{equation}
   G(a,b,z)\,G(c,z)=G(a,b,c,z)+G(a,c,b,z)+G(c,a,b,z),
   \end{equation}
    to reorganise the result into a linear sum of MPLs. The task of computing the discontinuity of the correlation function then becomes one of computing the discontinuity of each individual MPL 
   with  details given in \APP\ref{app:mpl}.
It should be emphasised 
 that the $Q_b$-type  MIs are much more involved than the $Q_q$-ones,  
 which is also mirrored in their alphabet length of   $18$ and $10$ respectively \cite{DiVita:2017xlr}.

   The final step in the computation of the perturbative densities is to translate the MPLs into the more familiar classical polylogs.\footnote{Alternatively one could use the \texttt{HandyG} package \cite{Naterop:2019xaf} to perform the numerical evaluation.}
    In doing so we move from complicated functions with relatively simple branch cut structures, to simple functions with complicated branch cut structures. Consequently, the representation of a given MPL in terms of polylogs depends on the regions in which its arguments lie, meaning that a single expression may not be sufficient to span all possible values of the arguments. An explicit example of this can be seen in Eq.$(4.2)$ of \cite{Frellesvig:2016ske}. We have numerically verified that the perturbative densities written in terms of classical polylogs are equivalent to those given in terms of MPLs in the region of interest, $q^2<m_b^2<p_B^2$.

   \begin{figure}[ht]
   \centering
   		\includegraphics[width=0.8\textwidth]{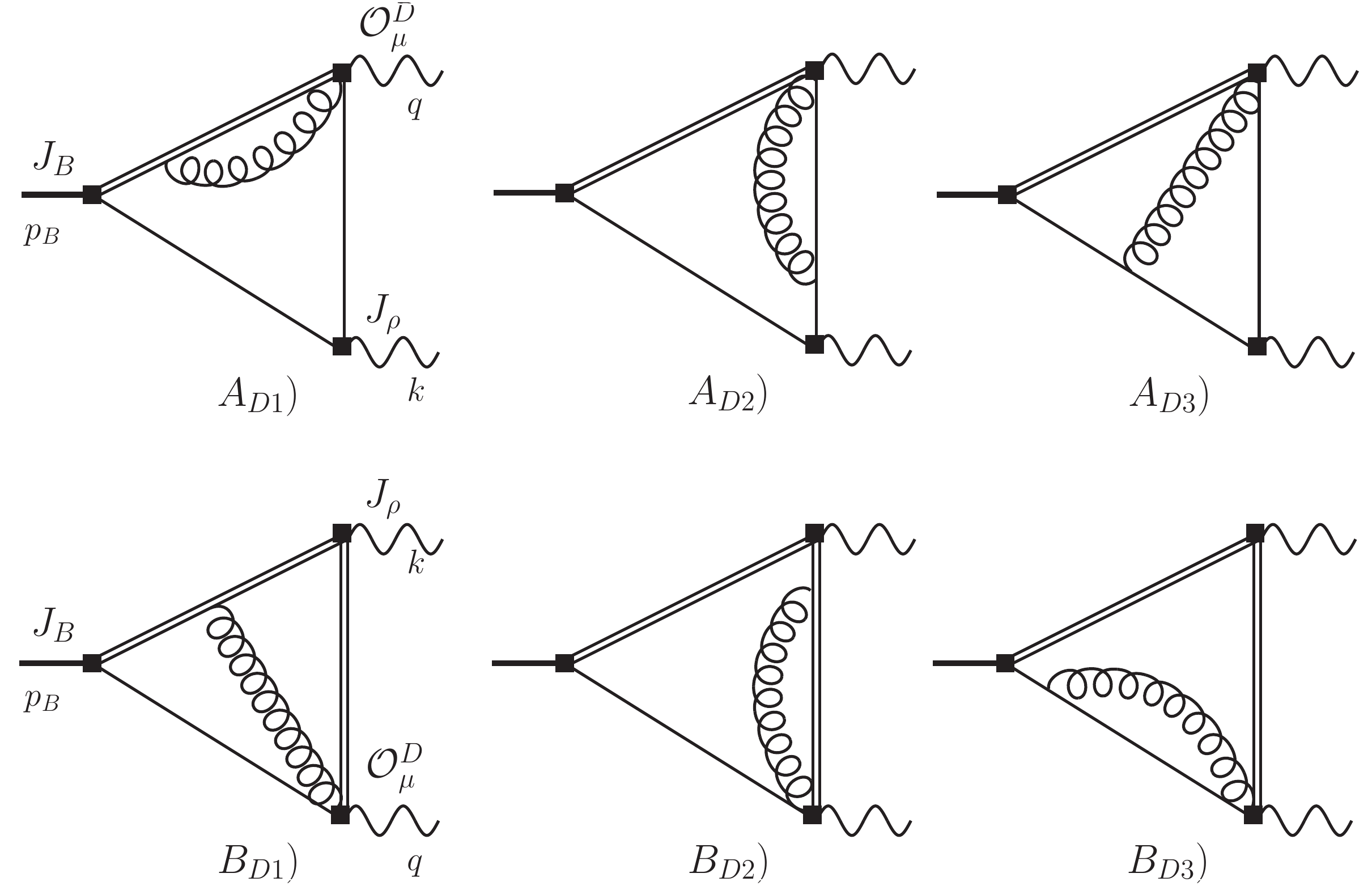}
   		\caption{\small Additional diagrams required when computing the correlation function $\Corr_\mu^{\mathcal{D}}$.}
   		\label{fig:PT_2L_Der}
   \end{figure}
 \subsection{The light-cone OPE diagrams of Twist-\texorpdfstring{$2$}{},\texorpdfstring{$3$}{} and \texorpdfstring{$4$}{}}
 \label{sec:twist}
 
 In phenomenology the light-cone OPE has its roots in deep inelastic scattering, which 
 is an inclusive measurement  where the operators on the light-cone are the parton distribution functions 
 $\matel{P(p)}{\bar q(z) \Gamma q(0)}{P(p)}$.
 The Fourier transform of the light-cone direction  $p^- z^+$  describes the probability of finding a constituent parton of a certain momentum fraction. 
 The application of the light-cone OPE to hard exclusive processes \cite{CZ1984} is a different matter, as the 
 light-cone operators correspond to  DAs, 
e.g.  $\matel{0}{\bar d(z) \Gamma u(0)}{\pi^+(p)}$ and the Fourier transform in the light-cone direction, denoted 
by the variable $u$ in this work, describes the amplitude for each parton having momentum fraction $u$.
 We refer the reader 
 to the technical review  \cite{Braun:2003rp} where the connection to conformal symmetry is presented in depth and detail.

 In a concrete setting an OPE is associated with factorisation. The light-cone OPE corresponds 
 to collinear factorisation and is often schematically presented (in momentum space) as 
 \begin{equation}
 \label{eq:collfac}
 \Pi(q^2,m_b) = \sum_i   T^{(i)}_H(q^2,m_b,\mu_F) \otimes \phi_i(\mu_F)  \;,
 \end{equation}
 where $\Pi$ is the correlation function, $T^{(i)}_H$  a perturbatively calculable hard scattering kernel, 
 $\phi_i$  a DA,  $\otimes$ denotes the integration over the collinear momentum fractions (suppressed in the formula), $\mu_F$ is the factorisation scale and the sum runs over DAs of increasing twist. 
A necessary condition for factorisation to hold is that the $\mu_F$-dependence cancels 
 to the given  order  in a  computation. This corresponds to the absorption of 
 collinear IR-divergences of the hard kernel into the DAs. We will discuss this explicitly in \SEC\ref{sec:NLO}.

 \subsubsection{Quark Mass Correction to the Photon Distribution Amplitude}
 \label{sec:mqDA}
 
 This section serves to illustrate the leading twist-$2$ photon DA and we present a method to 
 incorporate quark mass corrections,  which are sizeable in the case of  the strange quark. 
 The leading twist-2 photon DA is given by \eqref{eq:t2}
 \begin{equation}
 \label{eq:g-DA}
 \vev{\gamma(k,\epsilon)|\bar q (0) \sig_{\al \be} q (z)|0} \; =\;      i s_e e Q_q  \vev{\bar q q }  \chi
 \eps^*_{[\al} k_{\be]}
\int_0^1 du e^{i  \bar uk z}   \phi_\gamma(u,\mu) + \textrm{higher twist} \;,
\end{equation}
 where the Wilson line is omitted for brevity and 
 $\chi$ is known as the magnetic vacuum susceptibility, which serves as a normalisation of the photon DA 
 in the sense that $f^\perp_{\ga } = \vev{\bar qq } \chi  $ is the analogue of the transverse decay constant $f_\rho^T$ for a $\rho$-meson.
 Its value without the inclusion of quark mass effects (and thus no quark subscript) 
 is $\chi  \approx  3 \GeV$, at the scale $\mu = 1 \GeV$,   known from QCD sum rules 
 \cite{Ioffe:1983ju,Ball:2002ps} 
 and lattice computations \cite{Bali:2012jv,Bali:2020bcn}. 
 
 For finite quark mass $m_q$ the \emph{local} matrix element is UV-divergent. 
 More precisely, it mixes with the identity (additive renormalisation) 
 which is symptomatic of any OPE.
 Thus including finite quark mass  corrections 
 requires a specific prescription which we propose. 
 Our starting point is the non-local matrix element for an off-shell photon $\ga^*(k^2 \neq 0)$ 
 \begin{eqnarray}
 \label{eq:nonloc}
  \vev{\ga^*(k,\epsilon)|\bar q (0) \sig_{\al \be} q (x)|0} &\;=\;&  - i e  \eps_\mu^* \int e^{i k y} 
  \matel{0}{T j^\mu(y) \bar q (0) \sig_{\al \be} q (x)}{0}  =
 \nonumber \\[0.1cm]
&\;=\;&  i s_e e Q_q  
 \eps^*_{[\al} k_{\be]}    \int_0^1 du e^{i  \bar uk x}  \Phi_{\ga^*}(u,k^2,x^2) \;,
 \end{eqnarray}
 where  we parameterise ($\mu$-dependence suppressed)
 \begin{eqnarray}
 \Phi_{\ga^*}(u,k^2,x^2) =    \vev{\bar qq  }  \chi  \phi_\gamma(u,k^2)  + \frac{N_c m_q}{4 \pi^2}  \phimq_{\ga^*} (k^2,x^2)  \;,    \nonumber 
 \end{eqnarray}
and remind the reader that, by definition, $\chi$  does not include quark mass corrections.
The LO contribution is straightforward to compute 
 \begin{eqnarray} 
 \phimq_{\ga^*} (k^2,x^2) &\;=\;& -2 K_0\left( \sqrt{ (m_q^ 2-  u \bar u k^2 )(-x^2+i0) } \,\right) + \ORD(\al_s)
 \nonumber \\[0.1cm]
&\;=\;& 
\ln \left( \frac{ m_q^2 - \bar u u k^2}{4 \mu^2} \right) + \ln (-x^2+i0)\mu^2  +2 \ga_E + \ORD(\al_s,x^2) \;,
 \end{eqnarray} 
 where $K_0$ is the modified Bessel function of the second kind.
 The $\ln (-x^2+i0)$-term signals the UV-divergence which is regulated 
 by $x^2 \neq0$.  Upon taking the local 
 limit from the start and using DR $d = 4 - 2\eps$, we identity  at LO
 $1/\eps_{\textrm{UV}} + \ln 4 \pi - \ga_E  \leftrightarrow  - \ln \frac{(-x^2+i0)\mu^2}{4} 
   - 2 \ga_E$.
   
  Before discussing the  LO-prescription to remove the UV-divergence,
   let us digress on the sign of the $\ln (-x^2+i0)$-term.
   Some time ago Ioffe and Smilga \cite{Ioffe:1983ju}  analysed the local matrix element 
 using a UV cut-off $\Lambda$ and a constituent quark mass. Taking into consideration their 
 sign convention  of $\chi < 0$, we expect an increase of $   \Lambda^2\leftrightarrow 1/x^2$
 to effectively enhance $\chi$ (recall $\vev{\bar qq} < 0$ for $m_q >0$). The same remark applies to the evaluation of this quantity in the background
 of a constant magnetic field via its Dirac eigenvalues cf. \APP B in 
 \cite{Bali:2012jv}.  
Hence our sign is in accordance with these treatments.\footnote{The 
relative sign, between the $\chi \vev{\bar qq}\phi_\ga$,  and 
the $\ln(-x^2)$-term differs from  \cite{Balitsky:1997wi} (eq A.2) (and \cite{Ball:2002ps} eq (2.20,3.17)),
when $\chi > 0$ is assumed as stated in those papers. This could be due to $\chi<0$ convention used in previous publications \cite{Balitsky:1986st}.
Note that there is a  typo in that paper. Namely 
$g_\phi^2/4 \pi = 9 m_\phi^2/( 4 \pi f_\phi^2) \approx 11.7$ and not the inverse of what is quoted in that reference.}

We would like to motivative the LO-prescription.  
 The Fourier transform of $\ln (-x^2)$
 \begin{equation}
\label{eq:FT}
\mu^{d-4} \int d^d x e^{i p x}   \ln (-x^2 \mu^2)  =   i  (4 \pi)^2 \frac{1}{p^4}  +\ORD(\eps) 
\end{equation} 
corresponds to a zero momentum insertion approximation.
We have checked by explicit computation that this leads to the same result
when one uses a mass-insertion approximation in perturbation theory.\footnote{
This is not a good choice per se as it is IR-divergent. 
Moreover, if one were to use  the 
$ \ln (-x^2 \mu^2)$-term directly, via \eqref{eq:FT}, this would lead to an IR-divergence. 
Of course these issues are resolved  if one drops the $ \ln (-x^2 \mu^2)$-term 
and uses perturbation theory with quark masses in the denominator. 
In particular,  the IR-divergence becomes  regular $m_q \ln m_q$.}
Hence at LO in $\al_s$  dropping the $ \ln (-x^2 \mu^2)$-term is the correct prescription 
in order to avoid double counting. 
Whether dropping solely the  $ \ln (-x^2 \mu^2)$-term is the correct prescription for higher order or not is an open question and  deserves further exploration.

Now that the UV-divergent part is separated it remains to be seen what we can do with the 
DA piece
\begin{equation}
\phimq_{\ga^*}(k^2) =  \phimq_{\ga^*} (k^2,x^2)|_{\textrm{twist--2}}  = \ln (  - \bar u u k^2 /(4 \mu^2) )  + 2 \ga_E  \;.
\end{equation}
If one could trust the $k^2 \to 0$ limit then   one could simply add it to the twist-$2$ DA 
$\Phi^{\twist}_\ga$ with  $\phimq_{\ga^*}(0)$ in place. Clearly  this is not the case since
perturbation theory is not applicable for $k^2 \to 0$ as illustrated by the singularity. 
A well-known  solution to this problem is to invoke a  dispersion relation \cite{Ball:2002ps}. 
We can build on their work for determining $\chi$ in the 
massless quark limit and add our contribution. 
The only relevant conceptual difference is that we have to use 
a once-subtracted dispersion relation because of the  logarithmic divergence. 
 
 It was found in  \cite{Ball:2002ps}  that the sum rules for higher moments are unstable 
 and we thus focus on the zeroth moment $\chi_q$ (recall $f^\perp_{\ga q} = \vev{\bar qq} \chi_q$).
 The once subtracted dispersion representation reads  
\begin{equation}
   f^\perp_{\ga^*q}(k^2) = \int_0^1 du \Phi^{(m_q)}_{\ga^*}(u,k^2) 
   = f^\perp_{\ga^* q}(\ksub) +   \int_{\textrm{cut}}^\infty \frac{d r \,  \rho^{f^\perp_{\ga^* q}}(r)}{r-k^2-i0} R(r)   \;,
\end{equation}
with $R(r) = \frac{ k^2 - \ksub}{r - \ksub}$ and   $\pi \rho^{f^\perp_{\ga^* s}} = \Ima  f^\perp_{\ga^* s}$.

Assuming the quark content 
($| \rho_0[\omega]  \rangle \simeq ( | \bar u u \rangle \mp |\bar d d\rangle)/\sqrt{2}$ and $|\phi \rangle \simeq | \bar ss\rangle $) and adopting the zero width approximation we may rewrite the dispersion relation as follows
\begin{alignat}{2}
\label{eq:dispi}
& f^\perp_{\ga^* s}(k^2)  &\;=\;& f^\perp_{\ga^* s}(\ksub) +  \frac{m_\phi f_\phi f_\phi^T}{ m_\phi^2 - k^2-i0}R(m_\phi^2) + 
\int_{r^{(\phi)}_0}^\infty \frac{d r \, \rho^{f^\perp_{\ga^* s}}(r)}{r-k^2-i0}R(r)  \;, \nonumber \\[0.1cm]
& f^\perp_{\ga^* q}(k^2)  &\;=\;&  f^\perp_{\ga^* q}(\ksub) +  \frac{m_\rho f_\rho f_\rho^T}{ m_\rho^2 - k^2-i0}R(m_\rho^2) + 
\int_{r^{(\rho)}_0}^\infty \frac{d r \, \rho^{f^\perp_{\ga^* q}}(r)}{r-k^2-i0}R(r)  \;,
\end{alignat}
where  $r_0$ is the hadronic threshold  due to higher states 
and we have adopted the isospin limit $ f^\perp_{\ga^* q} \equiv  f^\perp_{\ga^* u} \approx 
 f^\perp_{\ga^* d}$. 
Now, in order to evaluate the RHS, besides the decay constants 
 $(f_\rho, f_\rho^T) = (0.220(5),0.160(7)) \GeV$, 
$(f_\phi, f_\phi^T) = (0.233(4),0.191(4)) \GeV$   \cite{BSZ15} and continuum thresholds
$(r_0^{(\rho)},r_0^{(\phi)}) = (1.5(1),2.0(1)) \GeV^2$  \cite{Ball:2002ps}, one needs 
the subtraction constant and the spectral density $\rho(r)$. Both of these can be obtained from    
the local OPE   $f^\perp_{\ga^* s}(k^2) $.
Using the results in \APP B in \cite{Ball:2002ps} and adding our 
$m_q$-correction we find 
\begin{equation}
\label{eq:above}
 f^\perp_{\ga^* q}(k^2) = \frac{2 \vev{\bar q q}}{k^2}  \left(-1 +  C_F \frac{\al_s}{\pi} (1 - \ln  \frac{-k^2}{\mu^2} )  - \frac{1}{3}\frac{m_0^2}{k^2}   \right) +
  \frac{N_c  m_q}{4 \pi^2}\left( 2 ( \ga_E - 1) -\ln 4 +\ln  \frac{-k^2}{\mu^2}  \right) \;,
\end{equation}
where $\vev{\bar q \sig \Cdot Gq } = m_0^2 \vev{qq}$ with $m_0^2 = 0.8(2)\GeV^2$.  
 The spectral function   
 \begin{equation}
 \rho^{f^\perp_{\ga^* s}}(s) = \frac{2  C_F \al_s}{\pi}  \frac{\vev{\bar qq }}{s} - \frac{N_c   m_q}{4 \pi^2} \;,
 \end{equation}
  is obtained from  \eqref{eq:above}.
 Using $\chi_q =  f^\perp_{\ga q}/\vev{\bar qq } $ and adopting the notation
$f^\perp_{\ga q} = f^\perp_{\ga^* q}(0)$  with reference to \eqref{eq:dispi} we find (with $\mu = 1 \GeV$)
\begin{alignat}{4}
\label{eq:fga}
&   f_{\ga q}^\perp &\;=\;&  - 48(3) \MeV   \quad &\Leftrightarrow& \quad  \chi_q &\;=\;&  3.21 (15) \GeV^{-2} \;,
 \quad  \nonumber \\[0,1cm]
& f_{\ga s}^\perp &\;=\;& - 61(3) \MeV   \;, \quad &\Leftrightarrow& \quad \chi_s &\;=\;&   3.79  (17) \GeV^{-2} \;, 
\end{alignat}
where we have kept $- k_0^2 > 3 \GeV^2$ in the deep Euclidean region as appropriate for 
a short distance expansion. 
In addition we quote an $SU(3)$-ratio 
\begin{equation}
\label{eq:SU3}
r_{SU(3)} = \frac{f_{\ga s}^\perp }{f_{\ga q}^\perp} = 1.28(4) \;.
\end{equation}
All results are highly stable with respect to the subtraction point $k_0^2$ which 
is a sign of consistency in both the formula and its input, as the individual parts show sizeable variation 
 (for $k_0^2 \in [2,10]\GeV^2$ the results vary less than a percent).
The value of \eqref{eq:SU3} corresponds to a typical $SU(3)$-ratio in the non-perturbative realm. The $m_s$-contribution in perturbation theory 
 can be estimated to be of the order of $m_s/m_b \approx 0.02$ and can therefore be dropped.

We wish to emphasise that whereas  $\chi_q (f_{\ga q})$ merely reproduces the result in \cite{Ball:2002ps} (within minor variation of the input),  $\chi_s (f_{\ga s})$ is the first   determination of this quantity.
 The quantity $\chi_s$ in \cite{Bali:2012jv,Bali:2020bcn} should not be compared to the $\chi_s$ above 
 as it differs in the  renormalisation prescription. 
 We re-emphasise that our prescription emerges naturally from the twist expansion. 
 The fact that a correction to a non-perturbative matrix element can be computed with 
 perturbative methods is somewhat exceptional and so let us comment. 
 A key difference to QCD condensates or matrix elements is that $f^\perp_{\ga q}$ is of 
 mass dimension one which is taken by the quark mass. In QCD this would involve $\ORD(\LaQCD^n)$-terms.
 The remaining logarithmic divergence $\ln \LaQCD$ can be absorbed into the perturbative part (twist-$1$).

  \subsubsection{Leading Order}

\begin{figure}[H]
 \centering
 	\includegraphics[width=0.75\textwidth]{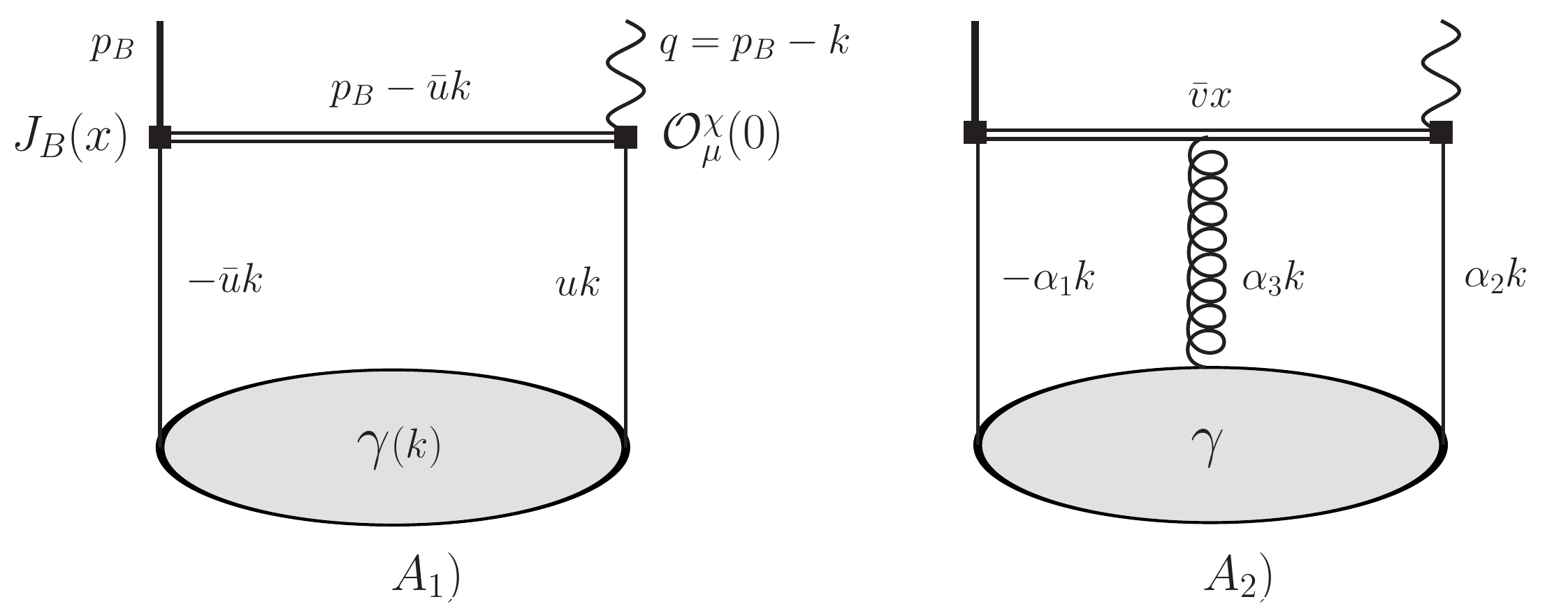}
	\caption{\small LO  contributions in the DAs. $A_1)$  $2$-particle DA $A_2)$ $3$-particle DA. 
	The variable $u \in [0,1]$ and $\bar u = 1-u$ are the momentum fractions carried by quarks.   
	The point $vx$, $ v \in [0,1]$, corresponds to the gluon in the background field gauge propagator \eqref{eq:BGFprop}.
	The $\al_i$ denote the momentum fractions of the quarks and gluon with $\al_1\!+\!\al_2\!+\!\al_3\!=\!1$.}
	\label{fig:DAtree-graphs}
\end{figure}

   The LO graphs, corresponding to the light-cone DA contributions to the FFs are depicted in \FIG\ref{fig:DAtree-graphs}. 
 The expansion is performed up to twist-$4$, including $3$-particle DAs, with the
 definition of the DAs given in \APP\ref{app:DAs}.  As mentioned above 
the $Q_q \vev{\bar qq}$-term is part of this expansion cf. Eq.~\eqref{eq:qqfun} to see this explicitly.

\subsubsection{Next-to-leading Order}
\label{sec:NLO}

\begin{figure}[ht]
 \centering
 	\includegraphics[width=0.8\textwidth]{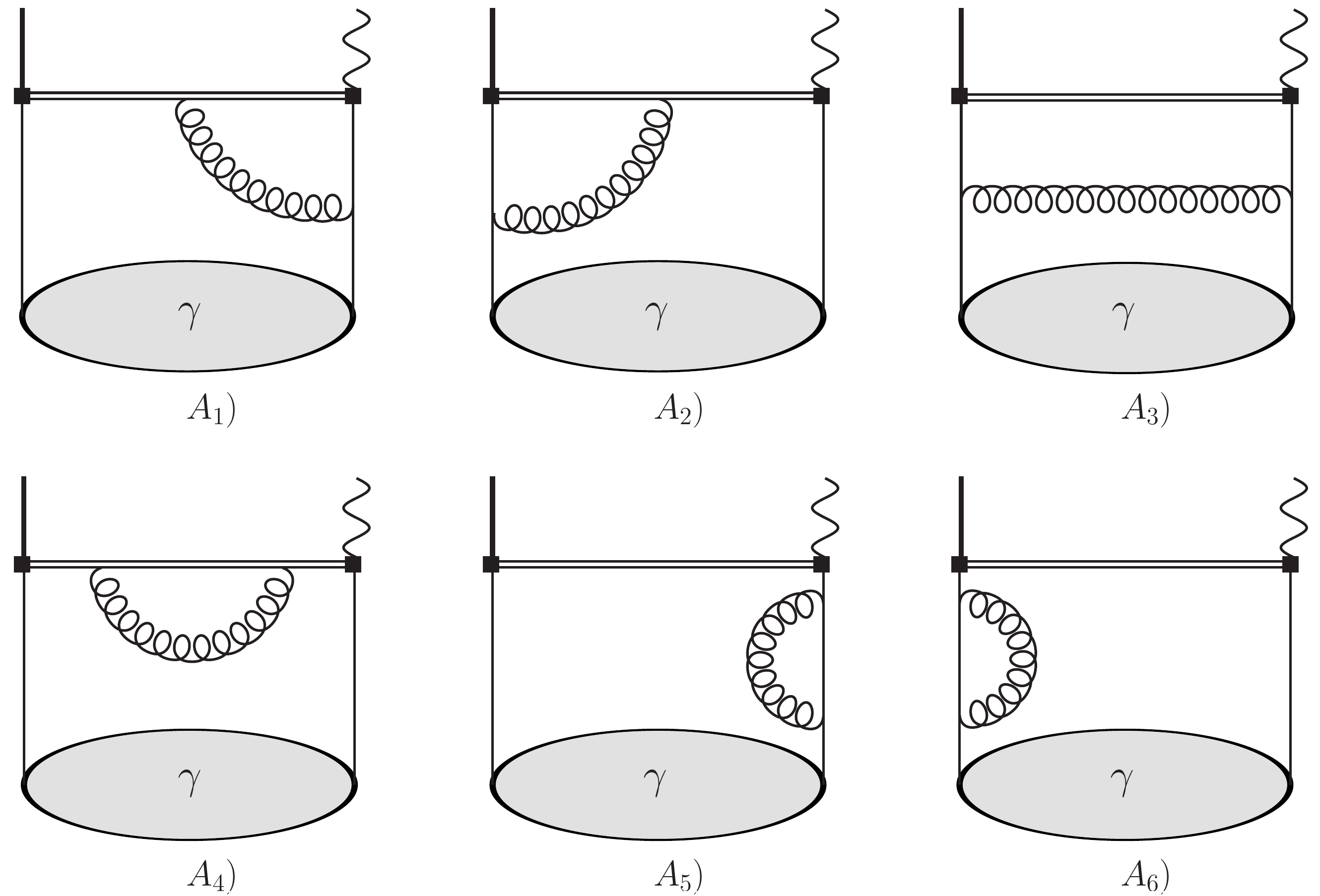}
	\caption{\small $\mathrm{O}(\al_s)$ corrections to twist-$2$ DA contribution. 
	$A_1$) vertex-correction $A_2$) $J_B$-correction $A_3$) box-diagram $A_{4-6}$) $b$- and $q$-quark self energy corrections.}
	\label{fig:DAloop-graphs}
\end{figure}

\begin{figure}[ht]
 \centering
 	\includegraphics[width=0.76\textwidth]{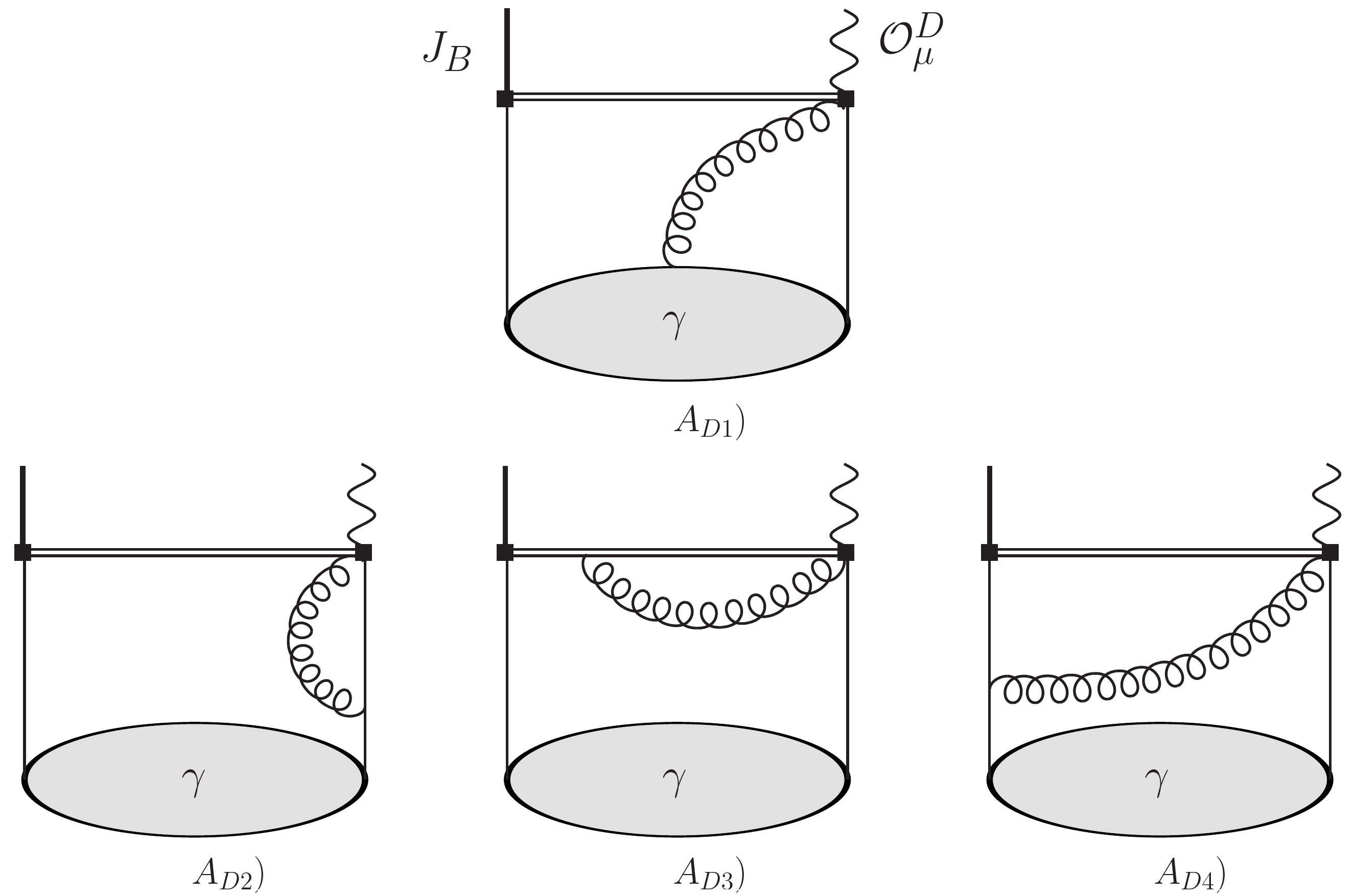}
	\caption{\small Additional diagrams originating from  the emission of the photon from 
	the  $\mathcal{O}_{\mu}^{D}$-operator. In principle there are two additional graphs, one for perturbation theory and
	for $\SumQ \vev{\bar qq}$, where the photon is emitted from the covariant derivative 
	of the $O^D_\mu$-vertex. However, these graphs turn out to be proportional to $\eps^*_\mu$ and are orthogonal 
	to our projection prescription \eqref{eq:recipe}, and we therefore do not include them. 
	Further not that if they were truly missing the EOM would not close.}
	\label{fig:DA-deriv-graphs}
\end{figure}

Further to the LO graphs we have also computed the twist-$2$ $\ORD(\al_s)$ corrections in the photon DA 
$\phi_\ga(u)$.
The relevant diagrams are depicted 
in \FIG\ref{fig:DAloop-graphs}, with additional diagrams for  $\Pi^D_{\mu\rho}$ shown  in \FIG\ref{fig:DA-deriv-graphs},  and the total results given in \APP\ref{app:resultsNLOt2} 
in terms of Passarino-Veltman (PV) functions. 
A noticeable feature is that in the Feynman gauge the box diagram 
$A_3)$ vanishes.
For the radiative corrections there are issues with regard to IR- and UV-divergences 
to be discussed. A particularly clear discussion of these matters can be found in 
\cite{B83}.
First, the necessary conditions for collinear factorisation to hold are (i) that there 
are  no divergences on integration over the collinear momentum fraction $u$ 
and (ii) that the collinear IR-divergences can be absorbed into the 
bare DA, $\phi^{\textrm{bare}}_\ga$, which is not observable.  
We find that our integrals converge when integrated over $u$ therefore satisfying the first
condition.  The collinear IR-divergences are absorbed into the LO DA.  
For this to be possible the divergences have to assume the following form
$T_{H,1}^{\textrm{div}} = c\, T_{H,0} \otimes V_0 $. In this equation  $T_{H,0(1)}$ are the LO and NLO 
hard scattering amplitudes,  $V_0$ is the LO Efremov-Radyushkin-Brodsky-Lepage 
evolution kernel \cite{Efremov:1978rn,Efremov:1979qk,Lepage:1980fj}
and $c$ is a constant or simply the counterterm \cite{B83}. 
One may use the property that the  eigenfunctions and eigenvalues of $V_0^\perp$
 are known  \cite{Shifman:1980dk}
\begin{equation}
 V^\perp_0 \otimes \varphi_n = \int_0^1 V^\perp_0(u,v) \varphi_n(v)  =   - \ga_n^{(0)\perp} \varphi_n(u) \;,
 \end{equation}
  with  
\begin{equation}
\varphi_n(u)  =  6 u \bar u \, C^{(3/2)}_n(2u - 1) \;, \quad \ga_n^{(0)\perp} =  2 C_F \left(1 +4  \sum_{j= 2}^{n + 1} \frac{1}{j} \right) \;,
\end{equation}
 where $C_F = ( N_c^2-1)/(2 N_c)$ and $C^{(3/2)}_n(x)$ is the $n^{\textrm{th}}$ Gegenbauer polynomials of degree $3/2$. The divergencies can then be absorbed into
\begin{equation}
\phi_\ga^{\textrm{bare} } = 6 u \bar u \sum_{n\geq 1} (a_n^\perp)^{\textrm{bare} } C^{(3/2)}_n(2u - 1) \;, \quad 
\end{equation} 
where
\begin{equation}
(a_n^\perp)^{\textrm{bare}} = (\mu/\mu_0)^{2 \eps} a_n^\perp(\mu) \left(1 + c \frac{g^2}{ (4 \pi)^2} \ga_n^{(0)\perp}
\frac{1}{\eps} \right) \;.
\end{equation}
  Requiring 
 \begin{equation}
 0 = \frac{d}{d \ln \mu} (a_n^{\perp})^{\textrm{bare}} =  a_s \ga_n^{(0)\perp} ( -  1  +  2 c) \;, \quad a_s = g^2/(4 \pi)^2 \;,
 \end{equation}
 determines $c = \frac{1}{2}$. This is indeed the constant that we find in our explicit computation 
 and thus shows $\mu_F$-independence  and cancellation of the collinear IR-divergences.
  This is not a new 
 result as the computation is equivalent to the $\phi^\perp$-amplitude in $\bar B \to V$ FFs 
 where it was verified at the twist-$2$ level in \cite{BB98b,BZ04b} and for the $V^{\bar B_u \to \gamma}$  in 
 \cite{Ball:2003fq}.\footnote{The difference to these references is that we give explicit results. In 
 \cite{BB98b} results were given with Dirac traces to be evaluated from where the cancellation can easily be verified.}

Obtaining the dispersion representation \eqref{eq:disp} of the twist-$2$ NLO contribution 
is not straightforward from the representation given in \eqref{eq:t2NLO}, which we write schematically 
 \begin{equation}
 \label{eq:schema}
\Pi_\perp(p_B^2) = \int_0^1 du \pi_\perp(u, p_B^2) \;,
\end{equation}
with $\pi_\perp(u, p_B^2) =   T_\perp(u,p_B^2)  \phi_\ga(u)$.
Whereas the function $\Pi_\perp(p_B^2)$ has a cut starting at $m_b^2$, 
 $\pi_\perp(u, p_B^2)$ has additional cuts in the $(u,p_B^2)$-variables.
 Two natural choices present themselves. Firstly, one commits to a certain form 
of the twist-$2$ photon DA and integrates over $du$ in \eqref{eq:schema} such that the (poly)logs are  functions
 of the external kinematic 
variables only and one can then obtain the discontinuity in the standard way. The other possibility is that one uses the fact that a FF's discontinuity is equivalent to its imaginary part.   
 One may write, working with a single subtraction term at $p_B^2 = \bar{s}$ in order to regulate the logarithmic UV-divergence, 
\begin{eqnarray}
\Pi^V_\perp(p_B^2) - \Pi^V_\perp(\bar s) &\;=\;& \frac{(p_B^2- \bar s)}{\pi} \int_{m_b^2}^\infty \frac{ds}{s - \bar{s}} 
\frac{ \textrm{Im}[ \int_0^1 du \pi^V_\perp(u,s)] }{s-p_B^2- i0}  \nonumber \\[0.1cm]
&\;\stackrel{p_B^2 < m_b^2}{=}\;& \textrm{Im}\left[ \frac{(p_B^2- \bar s)}{\pi} \int_{\Gamma_1} \frac{ds}{s - \bar{s}} 
\frac{  \int_{\Gamma_2}  du \pi_\perp^V(u,s)}{s-p_B^2} 
    \right] \;,
\end{eqnarray}
where $\Gamma_{1,2}$ are paths distorted into the upper complex planes $\mathbb C_{s,u}$ to avoid 
singularities.
We have validated this procedure numerically. 

In the final dispersion integral, after Borel transformation, the recipe reads
\begin{equation}
V_\perp(q^2) =  \frac{1}{f_B m_B^2} \textrm{Im}\left[\frac{1}{\pi} \int_{\Gamma_{m_b^2,s_0}} ds e^{(m_B^2-s)/M^2}
 \int_{\Gamma_2}  du \pi_\perp^V(u,s,q^2) \right] \;,
\end{equation}
where the integration endpoint of $\Gamma_1$, which is $\infty$, is moved to 
the continuum threshold $s_0$. This integral is numerically very stable.

\subsection{Comparison of Form Factor computation with the Literature}
\label{sec:CC}

The twist-$1$ and $2$ LO results agree with \cite{Khodjamirian:1995uc,Ali:1995uy}.
At  twist-$3$ and $4$  we can compare 
with \cite{Ball:2003fq} and find that the $ \vev{\bar qq}$, $\one$, $\mathbb{A}$, $S_{\ga}$ \&  $T_{4\ga}$  contributions are missing  and our result  in $\tilde{S}$  is larger by a factor of six.  At NLO solely  the twist-$2$ result has been reported in \cite{Ball:2003fq,Wang:2018wfj}.  In the first reference it is presented only in numeric form 
and the agreement is reasonable taking into account the slightly different numerical input.
In   \cite{Wang:2018wfj} analytic results are given but  
there is  a sign  difference between the twist-$1$ and the twist-$2$ term
(cf. their figure 5) in contrast to our result.\footnote{We would like to thank Yuming Wang for confirming this sign error.} 
To continue the discussion at this point it is useful to already take into account 
some of our numerical results of the proceeding section. 
Correcting for the mentioned sign, and taking into account the global sign difference \eqref{eq:above2}, 
focusing on twist-$1$ and $2$ one
gets  $V^{\bar{B}_u \to \ga}_\perp|_{\mbox{\cite{Wang:2018wfj}}}  \approx - (0.28_{t=1} + 0.08_{t=2}) = -0.36$
whereas we find $V^{\bar{B}_u \to \ga}_\perp \approx - ( 0.124_{t=1} + 0.082_{t=2}) = -0.206$ 
(cf. \TAB\ref{tab:twist_breakdown}). The discrepancy is too large in view of our 
$\ORD(10\%)$-uncertainty. In addition  
the value $V_\perp = -0.36$ is  too large in view of
the Belle exclusion limit (cf. \SEC\ref{sec:rate}).
The breakdown suggests that the culprit  is the twist-$1$ contribution 
for which two sources can be identified. 
First, the twist-$1$ result in  \cite{Wang:2018wfj} is only to leading power in $1/m_b$ and this suggests that the non-leading part is sizeable. Second, the not too well-known $B$-meson DA parameters  and the assumption of a specific model for the DA itself (cf. the discussion in \SEC\ref{sec:BDA}). To conclude, our analysis suggests that a posteriori 
the hybrid approach in  \cite{Wang:2018wfj} is numerically not sound because it does not include the full next-leading power corrections in the $1/m_b$-expansion.

\subsection{Equations of Motion as a Test of the Computation}
\label{sec:EOMtest}

Generally EOM give rise to relations between correlation functions, cf. \APP\ref{app:eom}
which often involve CTs. For simple local FFs these \CTs are absent 
and imply, in the case at hand, the following relations
 \begin{alignat}{2}
\label{eq:EOMFF}
&   T_\perp(q^2) +\sD  {\cal D}_\perp(q^2)  &\;=\;& \frac{(m_b + m_q  )}{m_B} V_\perp(q^2)  \;,  \nonumber \\[0.1cm]
&  T_\parallel(q^2)  + \sD {\cal D}_\parallel(q^2)   &\;=\;&
 \frac{(m_b - m_q  )}{m_B}  V_\parallel(q^2)     \;.
\end{alignat}
We wish to stress that these relations are completely general 
 and thus have to be obeyed by any approach. We further note that the point-like part 
 $\FBlowno$ \eqref{eq:ffs}  cancels between the LHS and RHS in the $\para$-equation.
  
Reassuringly, we find that \eqref{eq:EOMFF} holds in all parameters for which we have  complete results: 
perturbation theory (twist-$1$), $\{ \chi \vev{\bar qq},a_2(\ga) ,a_4(\ga)\} $ (twist-$2$), $f_{3\ga}$ at LO in $1/m_b$ (twist-$3$) 
and $Q_b \vev{\bar q q}$ (twist-$4$ local OPE). For twist-$1$ and $2$ this includes the NLO correction in $\al_s$.  
There are some comments to be made on the other cases. The parameter 
$f_{3\ga}$ is defined from a $3$-particle matrix element of twist-$3$
but could mix into a $4$-particle matrix element. This effect is $1/m_b$-suppressed, which 
we have checked analytically, and fortunately numerically negligible.
The only true twist-$4$ term that we include is $Q_b \vev{\bar q q}$ and now turn to the reason why,  
which has to our knowledge not been thoroughly discussed in the literature. 

It concerns the  consistent inclusion of twist-$4$ $3$-particle DA parameters. 
Our finding is that this demands $4$-particle DAs and is based
on the following two observations. 
 Firstly,  $4$-particle DAs of the 
form $\bar q(x) G_{\al \be}(vx) G_{\ga \de}(ux) \sigma_{\rho \sigma}q(0)$, which 
do appear in the light-cone expansion of the propagator cf. eq.~A.16 in  \cite{Balitsky:1987bk}, 
allow for twist-$4$ contributions. 
Secondly, these terms would mix via the EOM, generalisations of \EQ4.40 \cite{Ball:2002ps}, 
into the $3$-particle DAs of twist-$4$.\footnote{For example, 
for  vector meson DAs the $3$-particle parameter
$\zeta_3 \omega^T$, of  twist-$3$, mixes into the  $2$-particle DA $h_3$  twist-$4$, e.g. \cite{Ball:1998ff}, due to the EOM. In the same way $3$- and $4$-particle DAs are coupled through the EOM and lower twist parameter can always mix into a higher 
twist parameter.}  The point  is that the EOM at the FF-level 
need to close in each hadronic parameter separately, to be used in the numerics. 
In fact we find that of the $3$-particle twist-$4$ parameters only $\zeta_2^+$ closes and $\{\kappa, \kappa_+, \zeta_1, \zeta_1^+,\zeta_2\}$ including the $\vev{\bar q Fq}$-type contributions \eqref{eq:F3} do not.\footnote{\label{foot:charged} For the charged transitions 
further twist-$4$ contributions are needed, namely the photon parton of the photon DA connecting to the 
charged lepton.  
Strictly speaking, it is then better to consider the amplitude rather than the FF (beyond the rather trivial issue of the $\DelQ$-CTs in \eqref{eq:ffs}). }
As there is no argument as to whether any of the FF-combination are to be preferred over others 
we are led to conclude that one needs to 
drop those parameters altogether.
Fortunately quite a few of them are zero (\TAB\ref{tab:input_params}) 
and the numerical impact of the others is well below $10\%$ (\TAB\ref{tab:twist_breakdown}). 
Whereas the numbers in the table seem higher for the charged  $\bar B_u$, footnote \ref{foot:charged} is relevant in that it probably means that the numbers are overestimated. 
Note that for the photon DA the omission of twist-$4$ parameters for the $3$-particle case then 
implies that the $2$-particle twist-$4$ DAs are effectively zero as their only contribution comes from the
EOM with the $3$-particle DA of twist-$4$.  This is different to the vector mesons where
the $m_V^2$-contributions appear as $2$-particle ones of twist-$4$ and are sizeable for the $\phi$-meson 
for example  \cite{BSZ15}.

\section{Numerics} 
\label{sec:numerics}

In this section we turn to the numerics for which we first discuss the input, the sum rule parameters, 
correlations of parameters due to the EOM 
and the fit-ansatz in \SEC\ref{sec:inputs}.  
 In  \SEC\ref{sec:plots}  
FF-plots, fit parameters   and  values for $q^2=0$ and are given in   \FIG\ref{fig:FFplots}, \TAB\ref{tab:FFval}
 \TAB\ref{tab:fitparams} respectively 
The correlation matrix  can be found in an ancillary file to the arXiv version.  
The prediction of the $\bar B_u \to \ga \ell^- \bar \nu$ rate with bins and correlation matrix is given in \SEC\ref{sec:rate}
In \SEC\ref{sec:BDA} we present the extraction of the $B$-meson DA parameter $\la_B$ by comparing 
to a set of three SCET-computations.

\subsection{Input Parameters, Sum Rule Parameters and Fits}
\label{sec:inputs}

The input parameters, including the values for the photon DAs, are collected in \TAB\ref{tab:input_params}.
An important aspect is the choice of the mass scheme. 
The   $\MSbar$- and the  pole-scheme give rise to large  effects in either 
higher twist or  radiative corrections  which  suggests that neither  is optimal (cf. \FIG\ref{fig:polekinetic} and \TAB\ref{tab:kin_vs_msbar}).
The  kinetic scheme, originally developed for the OPE in inclusive decays \cite{Bigi:1994em}, 
can be regraded as a compromise of these two schemes and we have found 
that it does indeed give rise to stable results. 
From \FIG\ref{fig:FFscale} it is seen that 
this leads to stability in $\mukin$ with uncertainties in the $1$-$2\%$ range. 
To our knowledge this is the first time this scheme
is used in the context of LCSR and we also adapt it in the closely related work on 
the $g_{BB^*\ga}$-type on-shell couplings \cite{Pullin:2021ebn}.  
The  value of the kinetic mass, shown in \TAB\ref{tab:input_params}, has been obtained using the two-loop relation between the $\MSbar$ and kinetic mass given in \cite{Gambino:2017vkx}. 
The uncertainty on the kinetic mass has been estimated by adding the error on the $\MSbar$ mass 
and the  one from the conversion formula (variation of scale $\mu_m \in [m_b/2,2m_b]$) in quadrature. 
In line with other applications we set the scale of the kinetic scheme to  $\mukin\!=(\!1\pm 0.4 )\GeV$.

\begin{table}[btp]
\addtolength{\arraycolsep}{2pt}
\renewcommand{\MeV}{\,{\textrm{MeV}}}
\renewcommand{\GeV}{\,{\textrm{GeV}}}
\renewcommand{\arraystretch}{1.3}
\resizebox{\columnwidth}{!}{
\begin{tabular}{C|C|C|C|C|C} 
\multicolumn{6}{C}{\mbox{Running coupling parameters}}\\\hline
\alpha_s(m_Z) \mbox{~\cite{PDG}} & m_Z\mbox{~\cite{PDG}}\\\hline
0.1176(20) & 91.19 \GeV \\\hline 
\multicolumn{6}{c}{\mbox{Scales}}\\\hline
\multicolumn{2}{C}{\mu_F^{B_{d,u}}=\mu_{\text{cond}}^{B_{d,u}} = \mu_{\al_s}^{B_{d,u}} } & \multicolumn{2}{|C|}{\mu_F^{B_{s}}=\mu_{\text{cond}}^{B_{s}} = \mu_{\al_s}^{B_{s}} } & \muUV          & \mukin        \\\hline
\multicolumn{2}{C}{2.24(1.00)  \GeV }                                                    & \multicolumn{2}{|C|}{2.44(1.00)  \GeV }                                              & 4.78(1.00)\GeV & 1.00(40) \GeV  \\\hline
\multicolumn{6}{C}{J^P =0^- \mbox{ Meson masses~\cite{PDG}}}\\\hline
\mBz        & \mBu       & \mBs       &            &            & \\\hline
5.280  \GeV & 5.280 \GeV & 5.367 \GeV &            &            & \\\hline
\multicolumn{6}{C}{J^P =1^\pm \mbox{ Meson masses~\cite{PDG}}}\\\hline
\mBsz       & \mBsu      & \mBss      & \mBoz      & \mBou      & \mBos   \\\hline
5.325  \GeV & 5.325 \GeV & 5.415 \GeV & 5.726 \GeV & 5.726 \GeV & 5.829 \GeV \\\hline

 \multicolumn{6}{C}{\mbox{Quark masses~\cite{PDG}}}\\\hline
m_{s\lscale{2}}                                                  & m_b(m_b)     &     m_b^{\textrm{pole}}  & m_b^{kin}(1\GeV)^{\dagger}&&\\\hline
92.9(7)  \MeV                                                    & 4.18(4) \GeV &  4.78(6)\GeV                & 4.53(6)\GeV&&  \\\hline
\multicolumn{6}{C}{\mbox{Condensates}} \\\hline
\qbarq_\lscale{2} \mbox{~\cite{Bali:2012jv}} & \sbars \mbox{~\cite{McNeile:2012xh}} & \vev{G^2}\mbox{~\cite{SVZ79I,SVZ79II}} & m_0^2 \mbox{~\cite{Ioffe:2002ee}} & & \\\hline
-(269(2) \MeV)^3                             & 1.08(16) \qbarq                      & 0.012(4)\GeV^4                         & 0.8(2)\GeV^2                      & & \\\hline
\multicolumn{6}{c}{\mbox{Photon distribution amplitude parameters}} \\\hline
f_{\gamma q\lscale{1}}^{\perp}\mbox{~\eqref{eq:fga}}  & f_{\gamma s\lscale{1}}^{\perp}\mbox{~\eqref{eq:fga}}  
   & f_{3\gamma \lscale{1}}\mbox{~\cite{Ball:2002ps}}  & \omega_{\gamma \lscale{1}}^A \mbox{~\cite{Ball:2002ps}} & \omega_{\gamma\lscale{1}}^V \mbox{~\cite{Ball:2002ps}} & \\\hline
-48(3)\MeV                             & -61(3)\MeV                           
            & -0.004(2)^*\GeV^{2}                                       &
3.8(1.8)^*                                    & -2.1(1.0)^*   &      \\  \hline
\kappa_\lscale{1}  \mbox{~\cite{Ball:2002ps}} & \kappa^+_\lscale{1} \mbox{~\cite{Ball:2002ps}} & \zeta_{1 \lscale{1}} \mbox{~\cite{Ball:2002ps}} & \zeta_{1 \lscale{1}}^+ \mbox{~\cite{Ball:2002ps}} & \zeta_{2 \lscale{1}} \mbox{~\cite{Ball:2002ps}}         & \zeta_{2 \lscale{1}}^+ \mbox{~\cite{Ball:2002ps}} \\ \hline
 
0.2(2)                                      & 0(0)                                           & 0.4(4)                                        & 0(0)                                              & 0.3(3)                                            & 0(0)
\end{tabular}
}
\caption{\small Summary of input parameters. 
An asterisk, *, indicates that the quantity has been evaluated in the vector meson dominance approximation.
Recall that $f_{\gamma q}=\chi_q\CondQQ{q}$ where $\chi_q$ is the magnetic vacuum susceptibility of the quark condensate. The Gegenbauer moments of the photon DA, absent in this table, are discussed after \eqref{eq:phigamma} and set to $a_2(\ga) = 0(0.1)$ and $a_4(\ga) = 0(0)$. 
For the meson masses we have not indicated an error as they are negligible for this work.
Although we do not use the twist-$4$ in the numerics, for reasons discussed in \SEC\ref{sec:EOMtest} 
we include their values in the table for completeness.
The values of the $\kappa$ and $\zeta$-parameters are taken from 
the paper where the photon DA was first introduced \cite{Balitsky:1997wi},  and estimated those 
matrix elements to be correct within a factor of two and note that the ones set to zero are systematically smaller. 
We have  $\mu_{\al_s}^2 = \mu_F^2\!=\!m_B^2\!-\!(m_b^{pole})^2$ and the renormalisation scale of the tensor current is set to $\muUV=m_b^{pole}$.
}\label{tab:input_params}
\end{table}

The sum rule specific parameters, given in \TAB\ref{tab:FFSRparams}, are determined
by imposing a series of consistency conditions. Whilst we refrain from assigning a $q^2$ dependence to the SR parameters we consider the constraints set out below to apply for $q^2\in[0,7]\GeV^2$. For the effective threshold 
$s_0$ it is required  that the daughter SR, which  is formally exact,  
\begin{equation}\label{eq:daughterSR}
m_B^2=M^4\frac{d}{dM^2}\ln\int_{m_b^2}^{s_0}ds\; e^{-s/M^2}\rho(s,q^2)\;,
\end{equation}
 reproduces the known $B$-meson mass to within $1$-$2\%$. The $SU(3)$ ratio of effective thresholds should approximately reproduce that of the meson masses $s_0^{B_s}/s_0^{B_{d,u}}\approx m_{B_s}^2/m_{B_{d,u}}^2$, which we use as an additional loose constraint. The Borel mass is obtained by 
 considering two competing factors. A large Borel mass leads to faster convergence of the LC-OPE  at the cost of greater contamination from the continuum states, whereas a small value of the Borel mass achieves exactly the opposite. We balance these two factors by requiring both that the continuum contributes no more than $30\%$ and that the highest twist contribution remains below $10\%$ of the result.
Varying the Borel mass to $\pm2\GeV$ results in   $1$-$2\%$ changes in the FF-value  which is rather
satisfactory.  Next, we turn to correlations imposed by the EOM.

\begin{table}[t]
	\centering
	\begin{tabular}{l|c|c|c}
                                                         & $\bar B_d$                & $\bar B_u$                & $\bar B_s$ \\ \hline\rule{0pt}{1.2em}
$\!\!\{s_0,M^2\}^{V_{\perp},T_{\perp},T_{\parallel}  } $ & 35.2(2.0),\;8.0(2.0) & 35.2(2.0),\;8.0(2.0) & 36.2(2.0),\;8.0(2.0)  \\
$\{s_0,M^2\}^{V_{\parallel}}$                            & 35.8(2.0),\;8.5(2.0) & 35.8(2.0),\;8.5(2.0) & 36.8(2.0),\;9.5(2.0) \\ \hline \hline\rule{0pt}{1.2em}
$\!\!\{s_0,M^2\}^{f_B}$                                  & 34.3(2.0),\;5.6(2.0) & 34.3(2.0),\;5.6(2.0) & 35.5(2.0),\;6.4(2.0)
	\end{tabular}
	\caption{\small Summary of the sum rule parameters. The reason for not correlating the continuum thresholds for $T_{\parallel}$ and  $V_{\parallel}$ is given 
in \SEC\ref{sec:EOMFF}.}
	\label{tab:FFSRparams}
\end{table}

\subsubsection{Equations of Motion as Constraints on Sum Rule Parameters}
\label{sec:EOMFF}

Besides providing a non-trivial check of both the computation and the formalism of the OPE and DAs 
itself cf. \SEC\ref{sec:EOMtest}, 
the EOM serve an additional purpose in that they 
 correlate the vector and tensor FF as the derivative FF turns out to be small. 
This observation was first made for the  $B \to V$-FFs, with $V=\rho,\phi,...$ a light meson, in \cite{Hambrock:2013zya} and more systematically exploited in \cite{BSZ15}.  
Numerically,  the derivative FFs in \eqref{eq:EOMFF}  are suppressed by an order of magnitude  as compared to the vector and tensor ones.
This means that the sum rule specific parameters, of Borel mass and continuum threshold, 
ought to be approximately equal  $s_0^V \simeq s_0^T$ and $M_V \simeq M_T$; otherwise 
this would imply an unprecedented violation of quark-hadron duality for the derivative FF.  

Does this still hold for the $B \to \ga$ FFs?
Inspecting  \TAB\ref{tab:FFval} we note that this pattern persists for  the  $\perp$-direction but not 
for the $\para$-direction.  
As the computation is a 1-to-1 formal map with the $B \to V$ FFs for higher twist-$2$ and above 
(cf. \TAB\ref{tab:dict})
this suggests that there is something special going on in the perturbative diagrams (twist-$1$ and $Q_b\vev{qq}$).

\begin{table}
\begin{center}
    \begin{tabular}{c|c|c|c|c|c|c}
        FF            & $Q_q$ PT$^{\text{ LO [NLO]}}$ & $Q_b$ PT$^{\text{ LO [NLO]}}$ & $\chi\langle\bar qq\rangle^{\text{ LO [NLO]}}$ & $f_{3\gamma}$ & $Q_b \langle \bar q q \rangle$ & $Q_q \langle \bar q q \rangle$ \\
        \hline
        $V_\perp$     & \m{1/2[-1/2]}                 & \m{3/2[-3/2]}                 & \m{3/2[-1/2]}                                  & \m{3/2}       & \m{3/2}                        & \m{5/2} \\
        $V_\parallel$ & \m{1/2[-1/2]}                 & \m{3/2[-3/2]}                 & \m{3/2[-1/2]}                                  & \m{3/2}       & \m{3/2}                        & \m{5/2} \\
        $T_\perp$     & \m{1/2[-1/2]}                 & \m{3/2[-3/2]}                 & \m{3/2[-1/2]}                                  & \m{5/2}       & \m{3/2}                        & \m{5/2} \\
        $T_\parallel$ & \m{1/2[-1/2]}                 & \m{3/2[-3/2]}                 & \m{3/2[-1/2]}                                  & \m{5/2}       & \m{3/2}                        & \m{5/2} \\
        $D_\perp$     & \m{3/2[-1/2]}                 & \m{5/2[-3/2]}                 & 0\;[\m{3/2}]                                     & \m{5/2}       & 0                              & 0 \\
        $D_\parallel$ & \m{3/2[-1/2]}                 & \m{3/2[-3/2]}                 & 0\;[\m{3/2}]                                     & \m{5/2}       & \m{3/2}                        & \m{3/2}
    \end{tabular}
    \caption{\small Heavy quark scaling of the form factors according to the rules in \eqref{eq:HQ}. 
    The breaking of the hierarchy $V,T \gg {\cal D}$ at NLO is compatible with the findings in 
    \cite{Beneke:2000wa}. The $T,V_{\perp,\para} \propto Q_q m_b^{-1/2}$  and  $T,V_{\perp,\para} \propto Q_b m_b^{-3/2}$ are discussed from a
    dispersion relation approach in \SECs\ref{sec:scaling} (\EQ\eqref{eq:Qqscale}) and \ref{sec:EOMFF} respectively.} 
\label{tab:HQ}
\end{center}
\end{table}

In order to understand the origin  let us briefly recall how the FF-hierarchy can be understood 
in $B \to V$. 
We proceed by investigating the heavy quark limit of the FF. 
Whereas sum rules do not lend themselves to a heavy quark expansion, since some $m_b$ dependence 
is hidden in hadronic parameters, it is nevertheless possible to extract the leading $1/m_b$-behaviour 
by making the following well-known substitutions  
\begin{equation}
\label{eq:HQ}
m_B \to m_b + \bar \Lambda  \;, \quad   s_0 \to (m_b+\omega_0)^2 \;,  \quad M^2 \to 2 m_b \tau \;,
\end{equation}
where $\bar \Lambda,\omega_0$ and $\tau$ are all parameters of $\ORD(\LaQCD)$.  For $B \to V$ it was observed 
 that the derivative FFs   are $1/m_b$-suppressed  at $\ORD(\al_s^0)$ \cite{Hambrock:2013zya,BSZ15}. This pattern is broken at NLO where they show the same scaling.\footnote{This is without doubt related 
to the large energy limit FF relations \cite{Charles:1998dr}, subsequently 
systematically integrated into  QCD factorisation \cite{Beneke:2000wa} and soft-collinear effective theory 
\cite{Bauer:2000yr}. Such corrections were dubbed symmetry breaking corrections in \cite{Beneke:2000wa} 
and investigated at NNLO in \cite{Beneke:2005gs}. 
The FF relations in \cite{Beneke:2000wa} are compatible with our  finding from the EOM at LO in $1/m_b$.} 
The $B \to \ga$ scaling are collected 
in \TAB\ref{tab:HQ}.  The twist-$2$ parameter $\chi \vev{\bar qq}$, of course, confirms the earlier findings, whereas for twist-$1$ it is found that the hierarchy is obeyed in $Q_q$ but not $Q_b$. 
The former is in complete accord with the heavy quark scaling argument, presented in \SEC\ref{sec:scaling}
where the $Q_q$-part is inferred as a sum of $B \to V_n$ FFs. 
As for the latter we can offer a similar argumentation as  in \SECs\ref{sec:scaling}, \EQ\eqref{eq:Qqscale},  by using 
the dispersion representation in the limit of large colours
\begin{equation}
\label{eq:summy2}
T^{\bar B_s \to \ga} _\perp(q^2)|_{Q_b}  =  
\sum_{V_n = \Upsilon,\Upsilon',\dots }  \frac{  f^{\textrm em}_{V_n}} {m_{V_n}}  \, 
T_1^{\bar B_s \to V_n  }(q^2)   
+ \textrm{s.t.} \;,
\end{equation}
where we focus on the resonances for simplicity. The heavy quark scaling  
\begin{equation}
(m_{V_n} , f^{\textrm em}_{V_n},T_1^{V_n \to \bar B_s}(0) ) \propto ( m_b, m_b^{-1/2}, m_b^0) \;,
\end{equation}
of the hadronic parameters are all trivial.
In fact each generic term in  \eqref{eq:summy2} scales like $m_b^{-3/2}$.
Contrary to the case for the $Q_q$-charge, there is no obvious significant alteration 
for the sum when scaling in $m_b$ and we  conclude 
$T^{\bar B_s \to \ga} _\perp(q^2)|_{Q_b}  \propto m_b^{-3/2}$,
which is indeed consistent with our findings cf. \TAB\ref{tab:HQ}.  This explains the scaling  but also makes it clear that 
the $Q_b$-part has nothing to do with light vector mesons and 
the ideas behind the large energy limit are thus not applicable.\footnote{Charmonium states 
can be neglected since  $\matel{0}{\bar b \ga_\mu b}{\psi_{\bar cc}}$ is rather small.}  Hence there is no reason to believe that such a hierarchy is in place and the pattern in \TAB\ref{tab:FFval} is now qualitatively understood.  We consider it instructive to give a numeric breakdown for the LO contributions
up to twist-$2$ 
\begin{alignat}{5}
\label{eq:breakdown}
& V^{\bar B_q \to \ga}_\para(0) &\;\simeq\;&  (- 0.01 \,Q_b    &\;-\;& 0.09\,  Q_q)|_{\text{PT}} &\;-\;&   0.11\, Q_q|_{\text{twist-2}}      &\;+\;&  \textrm{higher twist} \;, \nonumber \\[0.1cm]
& T^{\bar B_q \to \ga}_\para(0) &\;\simeq\;&  (-0.03\, Q_b    &\;-\;& 0.17\,  Q_q )|_{\text{PT}}&\;-\;&   0.09\, Q_q|_{\text{twist-2}}       &\;+\;& \textrm{higher twist} \;, \nonumber \\[0.1cm]
& {\cal D}^{\bar B_q \to \ga}_\para(0) &\;\simeq\;&  (\phantom{-}   0.02\, Q_b    &\;+\;& 0.10 \, Q_q)|_{\text{PT}} &\;+\;&  \phantom{0.0} 0\, Q_q|_{\text{twist-2}}       &\;+\;&  \textrm{higher twist}  \;,\nonumber \\[0.1cm]
& V^{\bar B_q \to \ga}_\perp(0) &\;\simeq\;&  (- 0.03 \,Q_b    &\;-\;& 0.17\,  Q_q)|_{\text{PT}} &\;-\;&   0.11\, Q_q|_{\text{twist-2}}      &\;+\;&  \textrm{higher twist} \;, \nonumber \\[0.1cm]
& T^{\bar B_q \to \ga}_\perp(0) &\;\simeq\;&  (-0.03\, Q_b    &\;-\;& 0.17\,  Q_q )|_{\text{PT}}&\;-\;&   0.09\, Q_q|_{\text{twist-2}}       &\;+\;& \textrm{higher twist} \;, \nonumber \\[0.1cm]
& {\cal D}^{\bar B_q \to \ga}_\perp(0) &\;\simeq\;&  (\phantom{-}   0.00\, Q_b    &\;+\;& 0.03 \, Q_q)|_{\text{PT}} &\;+\;&  \phantom{0.0} 0\, Q_q|_{\text{twist-2}}       &\;+\;&  \textrm{higher twist}  \;,
 \end{alignat}
where $q=u,d$. Above we see that whilst the derivative FF is small in the $\perp$-direction it provides a considerable contribution in the $\|$-direction, and is the reason we allow the central values of the continuum thresholds of $V_\para^{\bar B \to \ga}$ and $T_\para^{\bar B \to \ga}$ to differ, in contrast to the other FFs. Moreover the numerical suppression of ${\cal D}^\perp$ in perturbation theory might 
or might not be accidental. 
In view of these findings we  apply a moderate (smaller than in $B \to V$ \cite{BSZ15})   correlation of the tensor versus vector thresholds.
Further to the correlations induced by the EOM,  the effective thresholds of the FFs and the decay constant 
 are correlated, which is warranted as both thresholds are associated with the $B$-meson state 
\begin{equation}
\label{eq:corrs0}
\text{corr}(s_0^{V^q_{\|[\perp]}},s_0^{T^q_{\|[\perp]}})=0.5 \;, \quad \text{corr}(s_0^{\type^q_{\|[\perp]}},s_0^{f_B})=0.5 \;,
\end{equation}
where $q=u,d,s$ and $\type=V,T$.

\subsubsection{Fit-ansatz}
\label{sec:fit}

A generic FF, say $F_n$,  is fitted to a $z$-expansion ansatz, 
following the same setup as in previous work \cite{BSZ15,Albrecht:2019zul}, 
where the leading pole is factored out 
\begin{equation}
\label{eq:Fon}
    F_n^{\bar B \to \ga}(q^2) = \frac{1}{1 -q^2/m_R^2}\left( \alpha_{n0}  + \sum_{k=1}^N \alpha_{nk}(z(q^2)-z(0))^k \right) \;.
\end{equation}
The variable $z$ describes the following map into the complex plane: $ z(t) = \frac{\sqrt{t_+ - t} - \sqrt{t_+ - t_0}}{\sqrt{t_+ - t} + \sqrt{t_+ - t_0}} $ 
($t_0\equiv t_+(1-\sqrt{1-t_-/t_+})$, $t_\pm \equiv (m_{B_q} \pm m_\rho)^2$ and
 $m_\rho = 770\MeV$).  It is noted that  $t_\pm \to m_{B_q} \pm 2 m_\pi, 
 m_{B_s} \pm 2 m_K$ is formally correct but we use $t_\pm$ as above with no impact on our results in the physical region. 
  We note that whilst $t_+$ corresponds to the multiparticle production threshold, the choice of $t_0$ (and therefore $t_-$) is arbitrary and does not impact our results.
The masses of the  lowest lying $J^P = 1^{-(+)}$ states, relevant to the $\perp(\parallel)$-projection, are  $m_R=m_{B^*_q(B_{1q}) }$. 

Relations between the FFs provide constraints on the fit parameters. For example \eqref{eq:algebraic} implies
\begin{equation}
T^{\bar B \to \ga}_{\perp}(0) = T^{\bar B \to \ga}_{\parallel}(0) \;\; \Leftrightarrow  \;\; \al_{T_{\perp}0} = \al_{T_{\parallel}0} \;.
\end{equation}
Further constraints arise from the  dispersion representation. Considering the vector FF for concreteness, 
\begin{equation}
\label{eq:VFF}
V^{\bar B \to \ga}_{\parallel[\perp]} (q^2) =   
\frac{1}{\pi} \int^\infty_{\textrm{cut}}  dt \frac{\Ima[ V^{\bar B \to \ga}_{\parallel[\perp]} (t)]}{t - q^2 - i0} =  
\frac{r^{V}_{\parallel[\perp]}}{1 - q^2/m_{B^*[B_1]}^2}  + \dots \;,
\end{equation}
one can infer the residue of the lowest-lying resonance\footnote{For $B \to \pi$  a two pole ansatz  fitted to the $f_+(q^2)$ FF
in LCSR within the $0 < q^2 < 14 \GeVs^2$-range reproduces a residue, consistent with the best determination of the these hadronic parameters \cite{BZ04} ($g_{DD^*\pi}$ from experiment and heavy quark scaling).}
\begin{alignat}{1}
\label{eq:r}
 r^{(V,T)}_{\perp}      \;=\; \left(  \frac{m_B \fBst{} }{m_{B^*}}, \fBst{^T} \right) \gonemi   \;, \quad 
 r^{(V,T)}_{\parallel}  \;=\;   \left( \frac{m_B \fBone{} }{m_{B_1}} ,  \fBone{^T} \right)   \gonepl \;,
\end{alignat}
with some more detail in \cite{Pullin:2021ebn} along with the definition of the decay constants $f_{B^*,B_1}^{(T)}$ 
and the on-shell matrix elements $g_{BB^*(B_1)\gamma}$ describing the 
$B^*(B_1) \to B \ga$ transition.
This relation enforces the constraint 
\begin{equation}
r_{V_\perp} =   \al_{V_\perp 0}  + \sum_{k=1}^N \al_{V_\perp k}(z(m_{B^*_q}^2)-z(0))^k \;,
\end{equation}
and similar constraints for other FFs. 
The main idea of the fit is that the four FFs with $15$ $q^2$-points (ranging from $[0,14]\GeV^2)$ 
and $N_{\text{samples}}=2500$ samples are subjected to a single fit providing correlations between the fit parameters of the FFs. This is the same Markov Chain Monte Carlo methodology as used in  \cite{BSZ15}.   Data points are generated from normal distributions  correlated
as in \eqref{eq:corrs0}, with the 
exception of  $\{\mu_F,\mu_\text{cond},\mu_{\alpha_s},\mu_\text{kin},M^2, M_{f_B}^2\}$ for which a 
skew normal distribution is applied. The latter is necessary since clearly not all parameters have a symmetric validity-range.  The skew normal distribution is defined by three parameters, which we take to be the  mean, 
the standard deviation and a \emph{shape parameter}, $\alpha_\text{shape}$. The latter is chosen such that the probability of  generating a sample below a certain cut-off is less than $10^{-4}$. 
For the scales we impose $\{\mu_F,\mu_\text{cond},\mu_{\alpha_s}\}>0.8\GeV$ and $\mu_\text{kin}>0.2\GeV$ whilst for the Borel parameters we require $M^2>4\GeV^2$ and $M^2_{f_B}>2\GeV^2$, below which the twist expansion is non-convergent.
The fitted FFs reproduce the mean values of generated FF data sets to within $\approx0.5\%$ over the fitted region. The values of the fit parameters are given in \TAB\ref{tab:fitparams}, with 
some more description in the caption. Covariance 
and correlation matrices along with the central values of the fit-parameters are given as ancillary file appended to the arXiv version of this paper. The \texttt{JSON} formatted file is ordered according to charge, statistical quantity and FF(s) of interest and can be readily used in \texttt{Mathematica}. For example, after loading the data\\

\begin{table}[t]
  \centering
\resizebox{\columnwidth}{!}{
  \begin{tabular}{| C|L L L L L L|} \hline \hline\rule{0pt}{1.2em}
          & V_{\perp}^{\bar B_d\to\gamma}        & V_{\perp}^{\bar B_s\to\gamma}        & V_{\perp}^{\bar B_u\to\gamma}      & T_{\perp}^{\bar B_d\to\gamma}        & T_{\perp}^{\bar B_s\to\gamma}        & T_{\perp}^{\bar B_u\to\gamma}\\\hline
 \alpha_0 & \phantom{\text{-}}0.133 (17)    & \phantom{\text{-}}0.158 (18)    & \text{-}0.213 (26)            & \phantom{\text{-}}0.133 (13)    & \phantom{\text{-}}0.151 (13)    & \text{-}0.213 (19) \\
 \alpha_1 & \text{-}0.123 (66)              & \text{-}0.459 (200)             & \phantom{\text{-}}0.278 (155) & \text{-}0.093 (362)             & \text{-}0.284 (357)             & \phantom{\text{-}}0.240 (545) \\
 \alpha_2 & \phantom{\text{-}}1.280 (913)   & \phantom{\text{-}}1.798 (1.303) & \text{-}1.760 (1.486)         & \phantom{\text{-}}1.075 (3.277) & \phantom{\text{-}}2.489 (3.356) & \text{-}1.346 (4.707) \\
 \alpha_3 & \phantom{\text{-}}3.373 (2.485) & \phantom{\text{-}}6.603 (3.283) & \text{-}4.940 (4.016)         & \phantom{\text{-}}2.764 (6.493) & \phantom{\text{-}}7.288 (6.915) & \text{-}3.848 (9.281) \\  \hline       & V_{\|}^{\bar B_d\to\gamma} & V_{\|}^{\bar B_s\to\gamma} & V_{\|}^{\bar B_u\to\gamma} & T_{\|}^{\bar B_d\to\gamma} & T_{\|}^{\bar B_s\to\gamma} & T_{\|}^{\bar B_u\to\gamma}\\[.1cm]\hline
 \alpha_0 & \phantom{\text{-}}0.081 (9)     & \phantom{\text{-}}0.104 (9)     & \text{-}0.151 (18)            & \phantom{\text{-}}0.133 (13)    & \phantom{\text{-}}0.151 (13)    & \text{-}0.213 (19) \\
 \alpha_1 & \text{-}0.138 (53)              & \text{-}0.177 (63)              & \phantom{\text{-}}0.312 (129) & \text{-}0.071 (377)             & \text{-}0.237 (359)             & \phantom{\text{-}}0.230 (535) \\
 \alpha_2 & \phantom{\text{-}}0.803 (769)   & \phantom{\text{-}}1.882 (649)   & \text{-}1.749 (1.260)         & \phantom{\text{-}}1.241 (3.207) & \phantom{\text{-}}1.841 (3.139) & \text{-}1.397 (4.340) \\
 \alpha_3 & \phantom{\text{-}}1.986 (1.408) & \phantom{\text{-}}3.951 (1.220) & \text{-}4.263 (2.314)         & \phantom{\text{-}}2.746 (4.768) & \phantom{\text{-}}4.180 (4.684) & \text{-}3.472 (6.397) \\\hline
  \end{tabular}
  }
  \caption{\small Fit parameters. To obtain the fits we generate $2500$ random samples of the input parameters, normally distributed such that for a given parameter the mean of the total sample set corresponds to the central value and the standard deviation to the uncertainty.  The fit parameters $\al_n$ are then produced by fitting the data set generated by evaluating the form factors at integer values of $q^2 \in [0,14]\GeV^2$ for each sample to the ansatz \eqref{eq:Fon} with $N=3$.}
  \label{tab:fitparams}
\end{table}

\texttt{data}$=$\texttt{Import[\dots/BtoGam\_FF.json]}\\

\noindent all information can be accessed via the \texttt{OptionValue} command. For example, the central value of $\al_{T_{\perp}0}^{\bar B_u}$ is obtained with \\

\texttt{OptionValue[data, "Bu"$\to$"central"$\to$"Tperp"$\to$"a0"]} .\\ 

\noindent Similarly the correlation between $\al_{V_{\|}0}^{\bar B_u}$ and $\al_{T_{\|}2}^{\bar B_u}$  can be accessed via\\

\texttt{OptionValue[data, "Bu"$\to$"correlation"$\to$"VparaTpara"$\to$"a0a2"]} .\\

\noindent The \texttt{uncertainty} and \texttt{covariance} matrix are accessed in an analogous manner.

\subsection{Results and Plots}
\label{sec:plots}

The central values of the FFs \eqref{eq:ffs}  at $q^2 =0$ and the residues \eqref{eq:r} are given in \TAB\ref{tab:FFval}. 
Plots of the neutral  $\bar B_d$-, $\bar B_s$- and charged $\bar B_u$-modes   are shown in \FIG\ref{fig:FFplots}. 
Together with the fit-values given in \TAB\ref{tab:fitparams} of the previous section
and the correlation matrix given in an ancillary file this constitutes the main practical results of our paper.

A breakdown of  contributions, split according to twist and DAs, is given in \TAB\ref{tab:twist_breakdown}.
We remind the reader that  all  twist-$4$ contributions, other than the $Q_b\CondQQ{q}$ condensates, are dropped as they  
require the inclusion  of $4$-particle DAs cf. \SEC\ref{sec:EOMtest}. This effect which 
is fortunately well below $10\%$ as can 
 inferred from  \TAB{\ref{tab:FFval}.

\begin{table}[H]
  \centering
  \setlength\extrarowheight{-3pt}
  \begin{tabular}{| l|r|r|r|r |} \hline \hline
              & \mc{$V_{\perp}(0)$}    & \mc{$V_{\|}(0)$}       & \mc{$T_{\perp,\|}(0)$} & \\  \hline     \rule{0pt}{1.2em}
    $\!\!\bar B_d$ & $0.130(17)$            & $0.079(9)\phantom{0}$  & $0.129(17)$            & $-$ \\
    $\bar B_s$     & $0.160(18)$            & $0.105(10)$            & $0.153(16)$            & $-$ \\
    $\bar B_u$     & $-0.209(26)$           & $-0.148(18)$           & $-0.209(26)$           & $-$ \\ \hline \hline
    \rule{0pt}{1.2em} 
              & \mc{$r_{\perp}^V$}     & \mc{$r_{\parallel}^V$} & \mc{$r_{\perp}^T$}     & \mc{$r_{\parallel}^T$}\\     \hline\rule{0pt}{1.2em}
    $\!\!\bar B_d$ & $\phantom{-}0.179(19)$ & ${0.076}(16)$          & ${0.171}(20)$          & ${0.104}(15)$\\
    $\bar B_s$     & $\phantom{-}0.235(25)$ & ${0.114}(18)$          & ${0.224}(24)$          & ${0.146}(17)$\\
    $\bar B_u$     & $-0.300(34)$           & ${-0.159}(31)$         & ${-0.284}(33)$         & ${-0.199}(28)$\\\hline
  \end{tabular}
  \caption{\small Values of the form factors \eqref{eq:ffs} at $q^2=0$ along with the residues \eqref{eq:r} evaluated in the kinetic scheme, cf. \cite{Pullin:2021ebn}. Values for the threshold and Borel mass can be found in 
    \TAB\ref{tab:FFSRparams}. All other input parameters are given in \TAB\ref{tab:input_params}.}
  \label{tab:FFval}
\end{table}

Whereas formally the fit-parameter $\alpha_0$ equals the  FF-value at $q^2=0$, on comparison with \TAB\ref{tab:FFval} we see a deviation of $\ORD(2\%)$ between the two. This comes as the FFs are evaluated analytically whereas 
the fit-values follow from the  Monte Carlo Markov-chain analysis ($N_{\text{samples}}=2500$ random samples) and the deviations are due to slightly asymmetric distribution of the uncertainty.

From the plots one infers the typical behaviour of a FF. 
Its value at $q^2 =0$ is 
below the normalisation of a current, e.g.  the electromagnetic pion FF $f_+^{\pi \to \pi}(0)=1$
and it raises 
as a result of  hadronic poles  (or multiquark thresholds in partonic QCD). 
One sees that the constraint from the first pole is well in line with the raise reproduced in our computation 
for $q^2 \approx 14 \GeV^2$ and would become progressively worse.
In the tensor case the $\perp$- and $\para$-directions only show significant deviations for large $q^2$, with the coincidence at $q^2=0$ enforced by \eqref{eq:algebraic}. For the vector FFs this pattern would be reproduced if the derivative FFs were small. However, this is not the case in the $\|$-direction, as the vector form factor differ more notably  (cf. \eqref{eq:breakdown} and the discussion of \SEC\ref{sec:EOMFF}). Furthermore, we find that for both the twist-2 and leading $1/m_b$ twist-1 contribution (proportional to $Q_q$ only) the $\perp$- and $\para$-directions give identical results. Hence any degree of asymmetry between $\perp$ and $\para$ must arise from sub-leading perturbative and/or higher twist contributions. Our decision to drop hadronic parameters that do not close under the EOM from the analysis (cf. \SEC\ref{sec:EOMtest}) therefore acts to increase the symmetry between the two directions.

\begin{table}[H]
  \centering
    \resizebox{\columnwidth}{!}{
  \begin{tabular}{c| c | l |r|r|r|r|r|r|r}
{\small twist} & {\small pa} & DA                                & $V_{\perp}^{\bar B_u}(0)$ & $V_{\perp}^{\bar B_d}(0)$ & $V_{\|}^{\bar B_u}(0)$ & $V_{\|}^{\bar B_d}(0)$ & $T_{\perp,\|}^{\bar B_u}(0)$ & $T_{\perp,\|}^{\bar B_d}(0)$ & $T_{\perp,\|}^{\bar B_s}(0)$ \\\hline
 
1              & -           & $\textrm{PT}\,\ORD{(\al_s^0)}$             & $-0.104$             & $0.065$              & $-0.054$          & $0.032$           & $-0.105$                & $0.065$                 & $0.086$\\
1              & -           & $\textrm{PT}\,\ORD{(\al_s)}$               & $-0.024$             & $0.019$              & $-0.010$          & $0.006$           & $-0.027$                & $0.021$                 & $0.021$\\\hline
2              & 2           & $\phi_{\ga}(u)\,\ORD{(\al_s^0)}$  & $-0.072$             & $0.036$              & $-0.073$          & $0.037$           & $-0.062$                & $0.031$                 & $0.037$\\
2              & 2           & $\phi_{\ga}(u)\,\ORD{(\al_s)}$    & $-0.010$             & $0.005$              & $-0.011$          & $0.006$           & $-0.011$                & $0.005$                 & $0.007$\\\hline
3              & 2           & $\Psi_{(a)}(u)$                   & $-0.002$             & $0.001$              & --                & --                & $-0.001$                & $<\!10^{-3}$            & $-0.003$ \\
3              & 2           & $\Psi_{(v)}^{(1)}(u)$             & --                   & --                   & $-0.003$          & $0.001$           & $-0.002$                & $0.001$                 & $-0.001$ \\
3              & 3           & $\mathcal{A}(\underline{\alpha})$ & --                   & --                   & --                & --                & $-0.002$                & $0.001$                 & $<\!10^{-3}$\\
3              & 3           & $\mathcal{V}(\underline{\alpha})$ & --                   & --                   & --                & --                & $-0.001$                & $<\!10^{-3}$            & $<\!10^{-3}$ \\\hline
4              & 2           & $h_{\ga}^{(2)}(u)$                & --                   & --                   & $0.011$           & $-0.005$          & $0.004$                 & $-0.002$                & $-0.002$\\
4              & 2           & $\mathbb{A}(u)$                   & $0.015$              & $-0.007$             & $0.014$           & $-0.007$          & $0.011$                 & $-0.005$                & $-0.005$ \\
4              & 3           & $\sSum$                           & $<\!10^{-3}$         & $<\!10^{-3}$         & $<\!10^{-3}$      & $<\!10^{-3}$      & $-0.001$                & $<\!10^{-3}$            & $<\!10^{-3}$ \\
4              & 3           & $\tSum$                           & $-0.002$             & $0.001$              & $-0.002$          & $0.001$           & $-0.004$                & $0.002$                 & $0.002$\\
4              & 3           & $S_{\ga}$                         & $0.003$              & $0.003$              & $-0.002$          & $-0.002$          & $0.003$                 & $0.001$                 & $0.003$\\
4              & 3           & $T_4^{\ga}$                       & $<\!10^{-3}$         & $<\!10^{-3}$         & $<\!10^{-3}$      & $<\!10^{-3}$      & $<\!10^{-3}$            & $<\!10^{-3}$            & $<\!10^{-3}$ \\
4              & -           & $Q_q\VEV{\bar q q}$               & --                   & --                   & $0.010$          & $0.002$           & --                      & --                      & -- \\
4              & -           & $Q_b\VEV{\bar q q}$               & $0.004$              & $0.004$              & $0.003$          & $-0.003$          & $0.003$                 & $0.003$                 & $0.003$ \\\hline
               &             & Total$^*$                             & $-0.194$             & $0.127$              & $-0.118$          & $0.067$           & $-0.197$                & $0.127$               & $0.151$
  \end{tabular}
  }
  \caption{\small A breakdown of contributions for a representative selection of form factors according to associated twist, ``pa" = number of partons and DAs.  The asterisk  in total is a reminder that it includes twist-$4$ contributions  that do not close under the EOM (cf.~\SEC\ref{sec:EOMtest}).
  The definitions of the DAs can be found in \APP\ref{app:DAs}. Results are given in the kinetic scheme with the kinetic scale $\mukin=1\GeV$.}
  \label{tab:twist_breakdown}
\end{table}

\begin{figure}[btp]
\centering
    \begin{minipage}{0.5\textwidth}
        \centering
        \includegraphics[width=1.\textwidth]{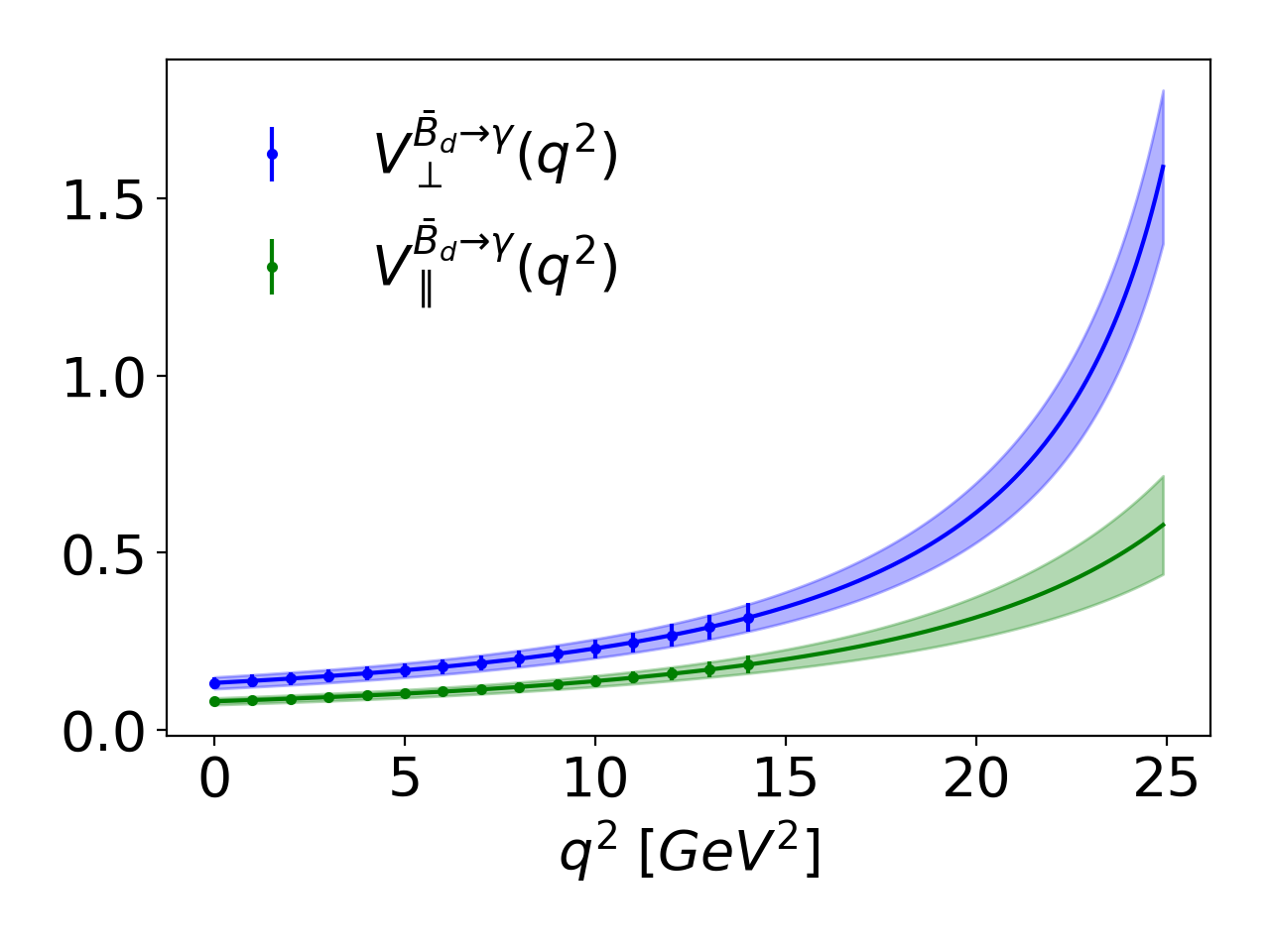}
    \end{minipage}\hfill
    \begin{minipage}{0.5\textwidth}
        \centering
        \includegraphics[width=1.\textwidth]{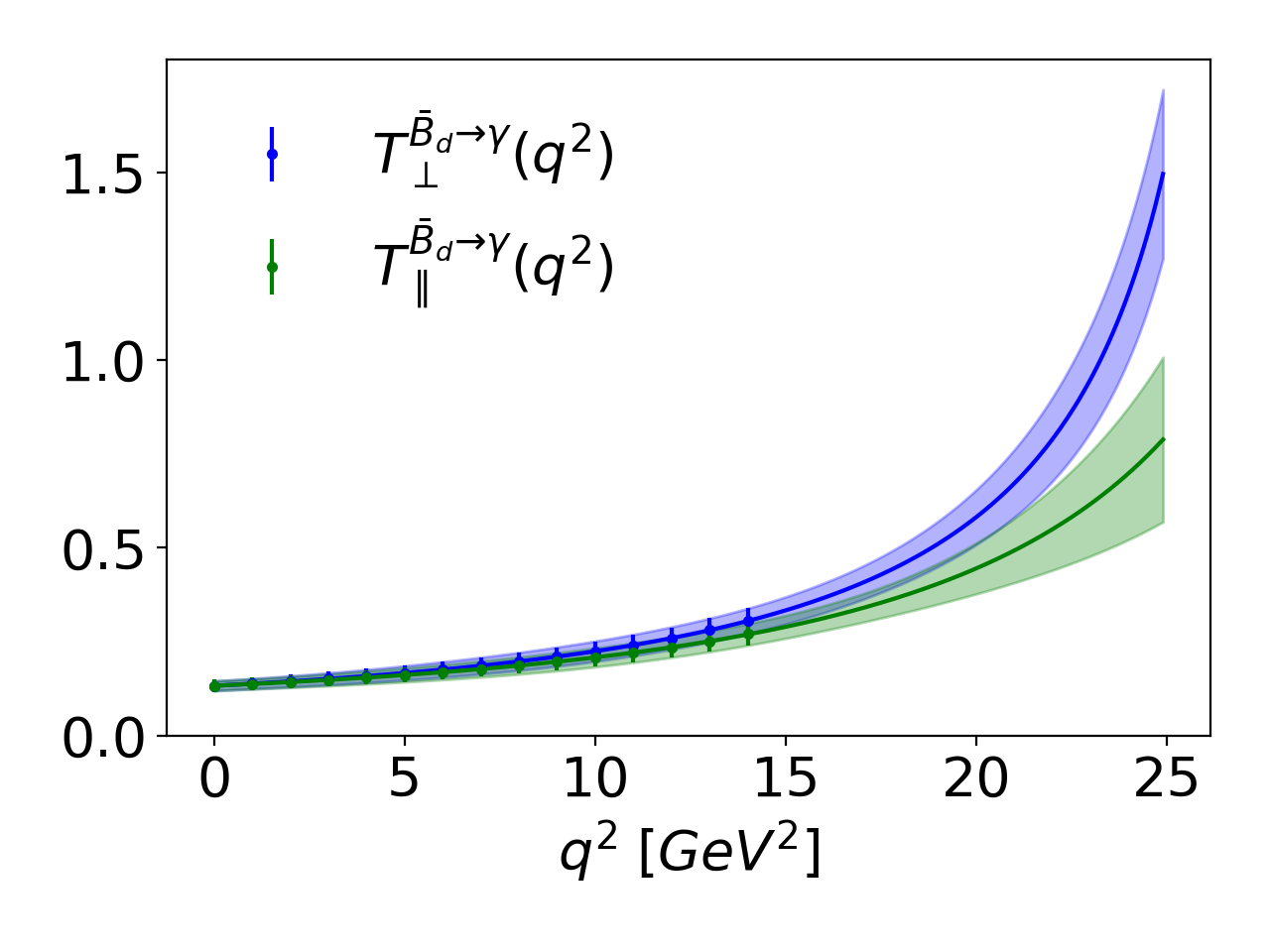}
    \end{minipage}\vspace{5mm}
      \begin{minipage}{0.5\textwidth}
        \centering
        \includegraphics[width=1.\textwidth]{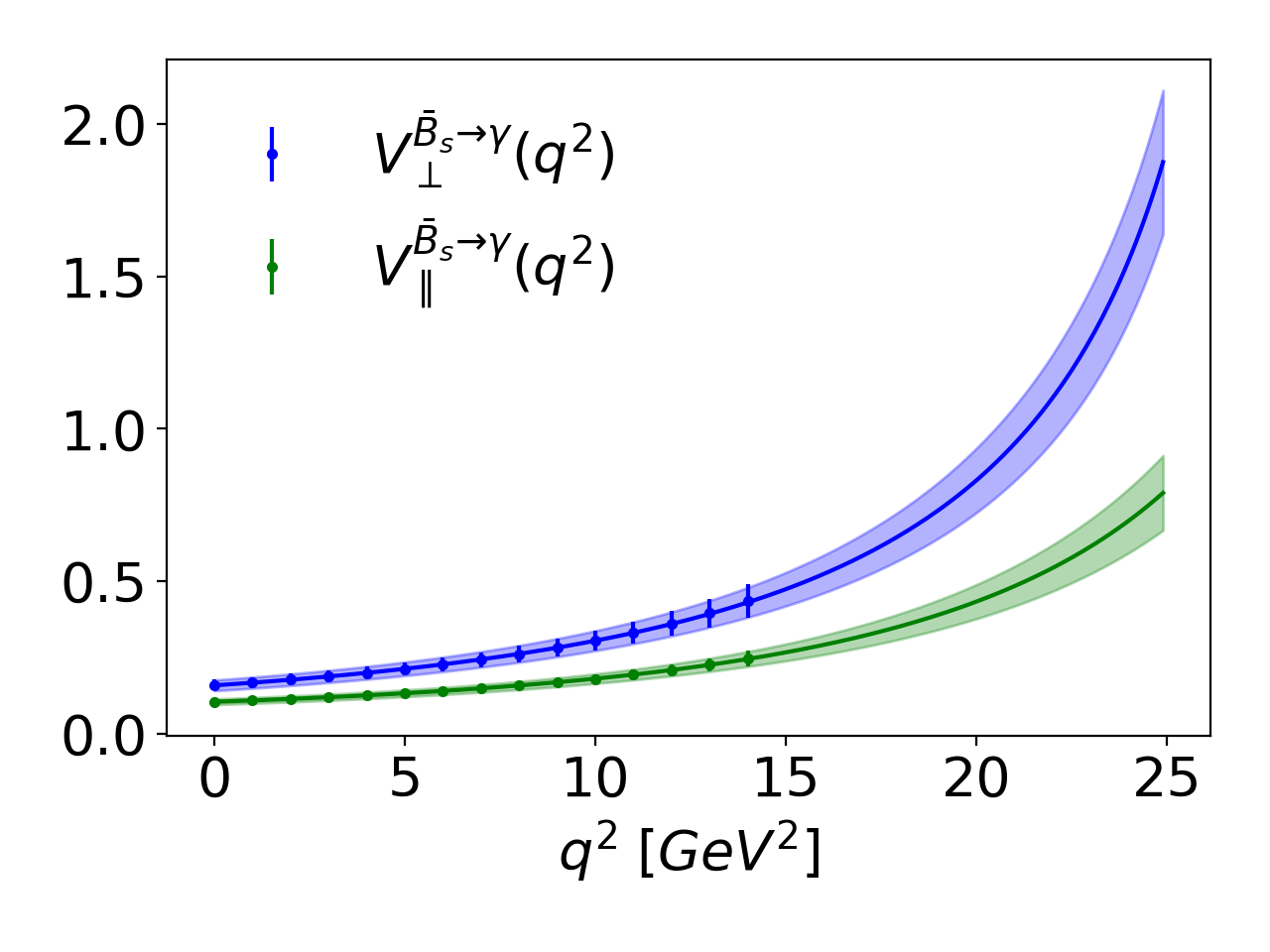}
    \end{minipage}\hfill
    \begin{minipage}{0.5\textwidth}
        \centering
        \includegraphics[width=1.\textwidth]{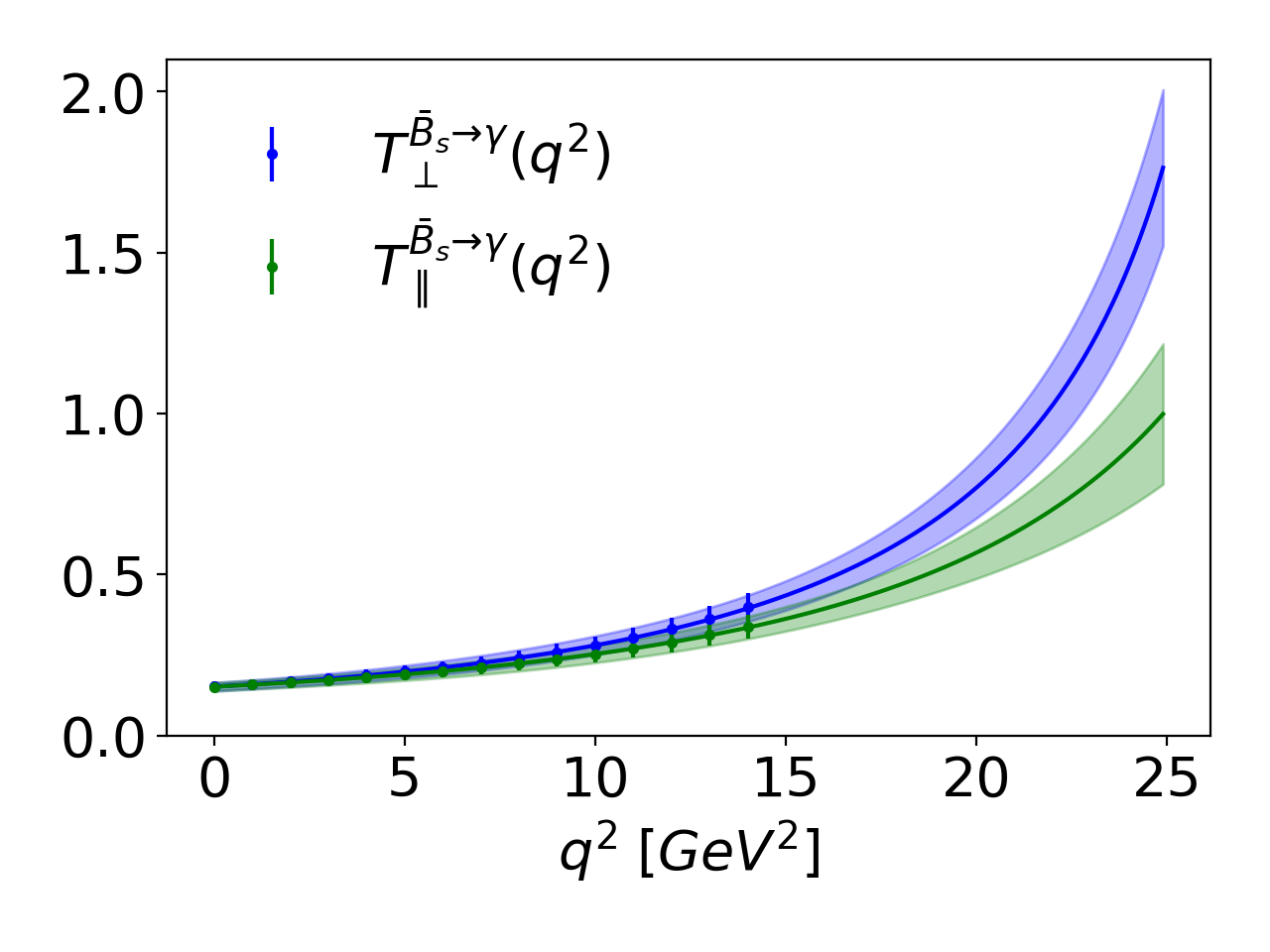}
    \end{minipage}
    \begin{minipage}{0.5\textwidth}
        \centering
        \includegraphics[width=1.\textwidth]{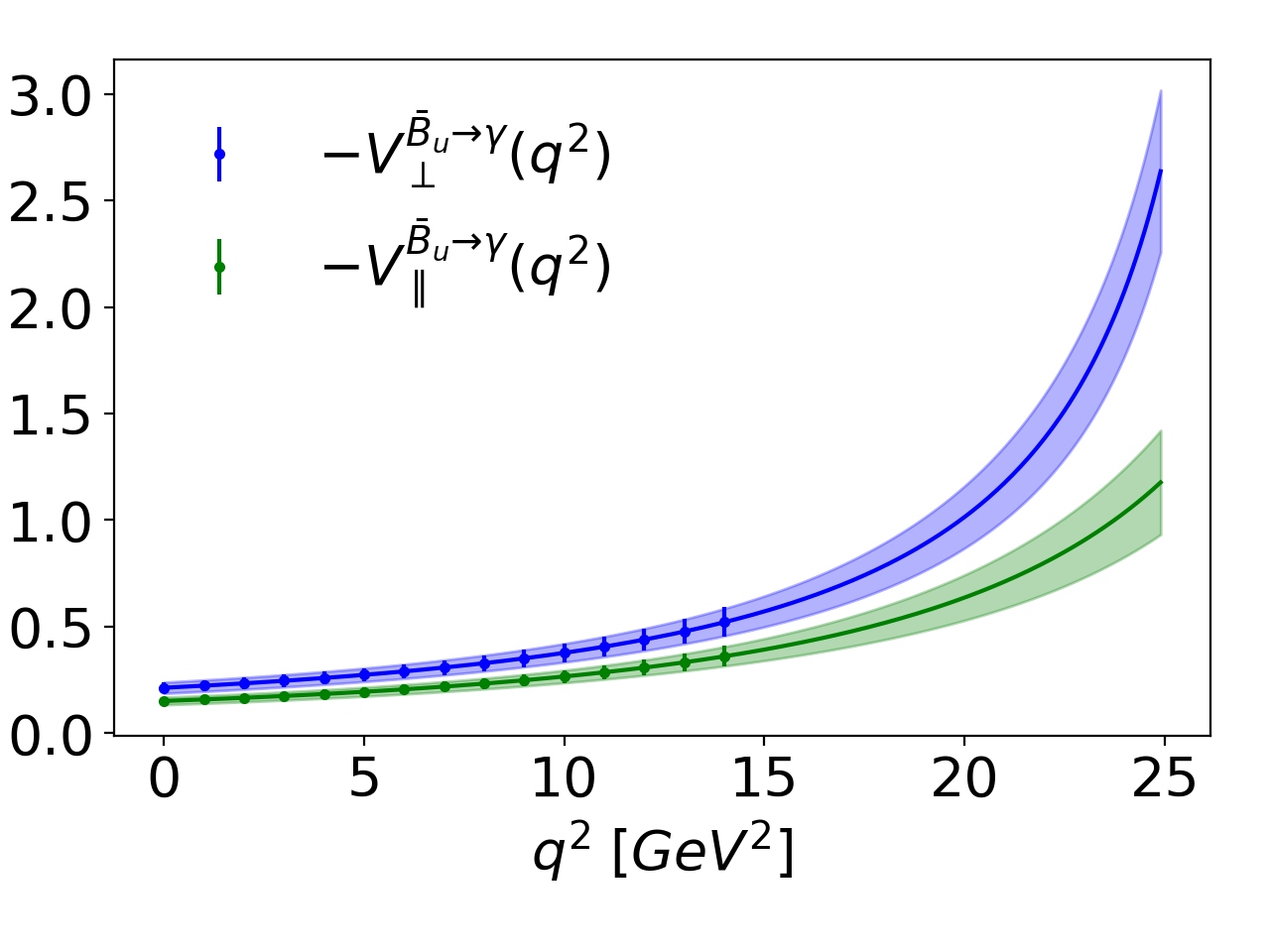}
    \end{minipage}\hfill
    \begin{minipage}{0.5\textwidth}
        \centering
        \includegraphics[width=1.\textwidth]{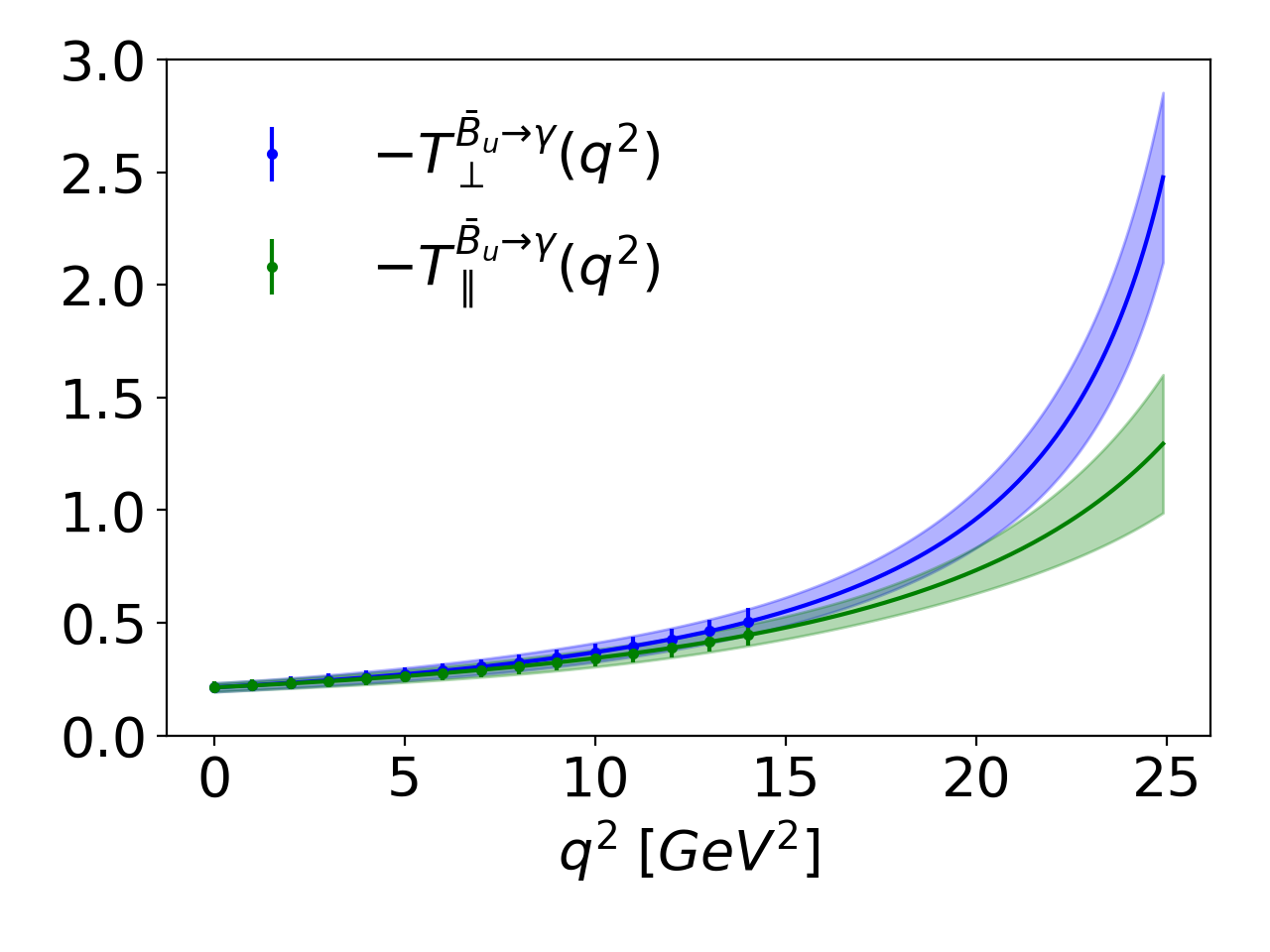}
    \end{minipage}\vspace{5mm}
    \caption{\small Form factors for the $\bar B_{d,s,u} \to \ga$ transition where the data points (with solid bars 
    as respective uncertainties) 
    have been generated 
    by a $N_{\text{samples}}=2500$ sample and fitted to the ansatz \eqref{eq:Fon}.
    The plots are reproduced from the fit-values   in \TAB\ref{tab:fitparams} and the 
    uncertainty bands stem from the correlation matrix given as an ancillary file cf. \SEC\ref{sec:inputs} 
    for further details. Whereas we consider the LCSR computation valid up to $q^2 < 14 \GeV^2$, 
    the ansatz \eqref{eq:Fon} with the pole residue from \cite{Pullin:2021ebn} allows one to extend the form factor 
    beyond this region. The sign of the charged form factors is reversed so that they are positive. We refer 
    the reader to \APP\ref{sec:FFconvLit} for discussion of conventions.}
    \label{fig:FFplots}
\end{figure}

\subsubsection{Comparison with the Literature}
\label{sec:comparison}

In this section we compare our FFs to some of the results in the literature. 
Comparison with SCET computations of these FFs \cite{Beneke:2020fot} is not straightforward 
as the  $B$-meson DA parameters are not known with a lot of certainty. 
It is therefore advisable to  turn the tables and use our predictions to set bounds 
as done in  \SEC\ref{sec:BDA}. Comments on  comparison in terms of 
analytic computations can be found  
at the beginning of  \APP\ref{app:results}. 
Lattice results are available for $D_{d,s} \to \ga$  and $K,\pi \to \ga$ modes in \cite{Desiderio:2020oej}  and in preparation by another group \cite{Kane:2019jtj}.  At low photon recoil such computations are of importance for 
QED-corrections.  Whereas we have chosen not to present $D \to \ga$ FFs, the pole 
residues of the vector FFs,  $g_{DD^*\ga}$, are computed in our other work \cite{Pullin:2021ebn} and 
are consistent with experiment in the measured modes. A notable aspect 
is that the shape in  \cite{Desiderio:2020oej}  does not seem to be compatible with the $D^*$-pole playing a
significant role close to the kinematic endpoint of $D \to \ga \ell \bar \nu$ ($q^2 = m_D^2$).

We can, however, compare to the quark model computations \cite{MN04,Kozachuk:2017mdk} 
(with $T^{\bar B_d \to \ga}_{ \perp,\para} = F^{\bar B_d \to \ga}_{T_{V,A}}$ dictionary).
Two representative plots of the neutral modes are given in \FIG\ref{fig:FFcomparison}.
There is rather good agreement at low $q^2$ but we can clearly see that our FFs 
are larger for higher $q^2$ which ought have a significant impact on phenomenology.

\begin{figure}[H]
\centering
    \begin{minipage}{0.5\textwidth}
        \centering
        \includegraphics[width=0.9\textwidth]{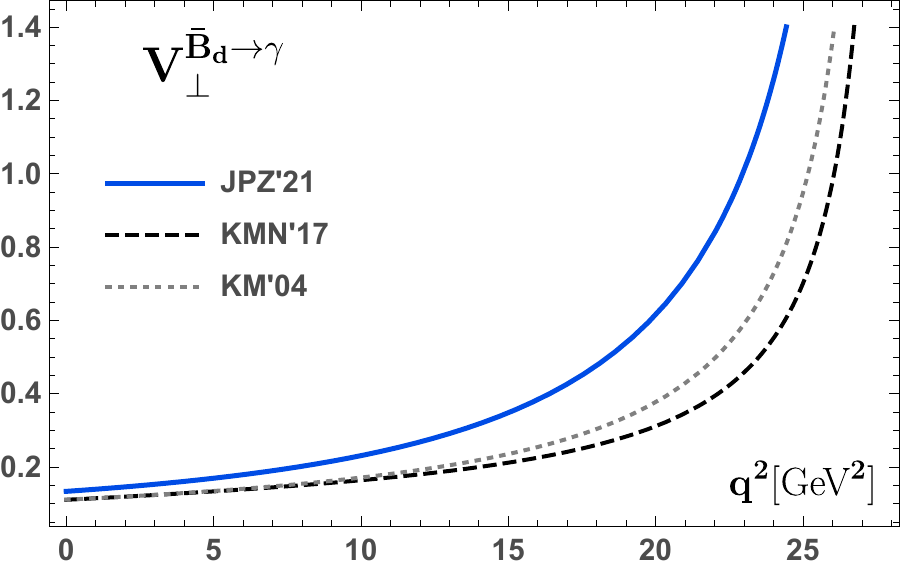}
    \end{minipage}\hfill
    \begin{minipage}{0.5\textwidth}
        \centering
        \includegraphics[width=0.9\textwidth]{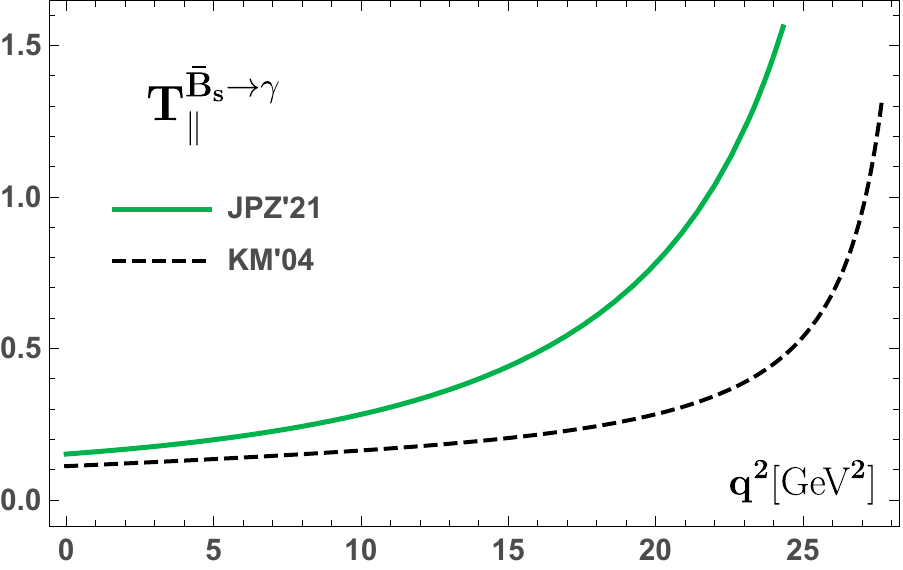}
    \end{minipage}\vspace{5mm}
     \caption{\small Representative comparison plots for neutral modes.  Our plots (solid,blue/green) ,
     denoted by JPZ,  compared against the quark model computation \cite{MN04} (short-dashed,light grey) from 2004
     and an improved version \cite{Kozachuk:2017mdk} (dashed,black) with 
     the modified pole version.  Note that there are $\bar B_s \to \ga$ results in the  2004 work.
      Our results are larger and, in particular, show an earlier rise. In view of this 
     the numerical agreement of the central value at $V^{\bar B_d \to \ga}_\perp(0)$ seems somewhat accidental.}
    \label{fig:FFcomparison}
\end{figure}
For illustrative purposes we have added \FIG\ref{fig:pt-like} comparing 
the form factor with and without the point-like contribution added.
\begin{figure}[H]
\centering
        \centering
        \includegraphics[width=0.47\textwidth]{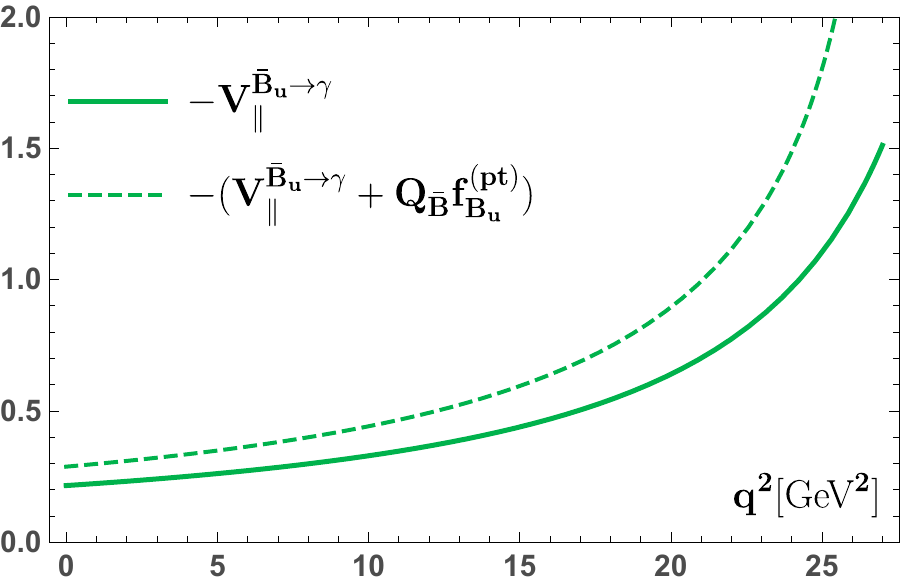}
     \caption{\small  Plot comparing the $\FA^{\bar B_u \to \ga}$ form factor,  as defined in \eqref{eq:ffs}, with and without the inclusion of the  point-like contribution  $ Q_{\bar B} \FBlowu$.  The value of the decay constant $f_B = 190 \MeV$ is taken 
     from \cite{Pullin:2021ebn} (cf. references therein).
     The point-like term provides a significant contribution, most notably towards the $q^2 = m_B^2$ pole.}
    \label{fig:pt-like}
\end{figure}

\subsection{Predictions for the  \texorpdfstring{$\bar B_u \to \ga \ell^- \bar \nu$}{} Rate}  
}
\label{sec:rate}

The $\bar B_u \to \ga \ell^- \bar \nu$ decay proceeds via a charged current and depends on the local FF only. 
Neglecting the lepton mass, anticipating a measurement in  the electron mode, the differential rate after integrating over the lepton angle is given by
\begin{equation}
\frac{d \Ga}{ d E_\ga} = \frac{G^2_F \al |V_{\textrm{ub}}|^2 }{6 \pi^2} m_B E_\ga^3\left( 1- \frac{2 E_\ga}{m_B}\right) \left( |V_\perp|^2 + |V_\para  |^2 \right) \;,
\end{equation}
where $E_\ga = (m_B^2 - q^2)/(2 m_B) \in [0,\frac{m_B}{2}]$ is the photon energy in the $B$-meson restframe, 
such that  $\frac{d \Ga}{ d q^2} = - \frac{1}{2 m_B} \frac{d \Ga}{ d E_\ga} $.
We refer the reader to 
\APP\ref{app:GIml0} for an explanation of why the point-like part, proportional to $f_B$, disappears 
in the $m_\ell \to 0$ limit. 

\begin{figure}[H]
	\centering
	\begin{minipage}{0.5\textwidth}
   		\includegraphics[width=0.9\textwidth]{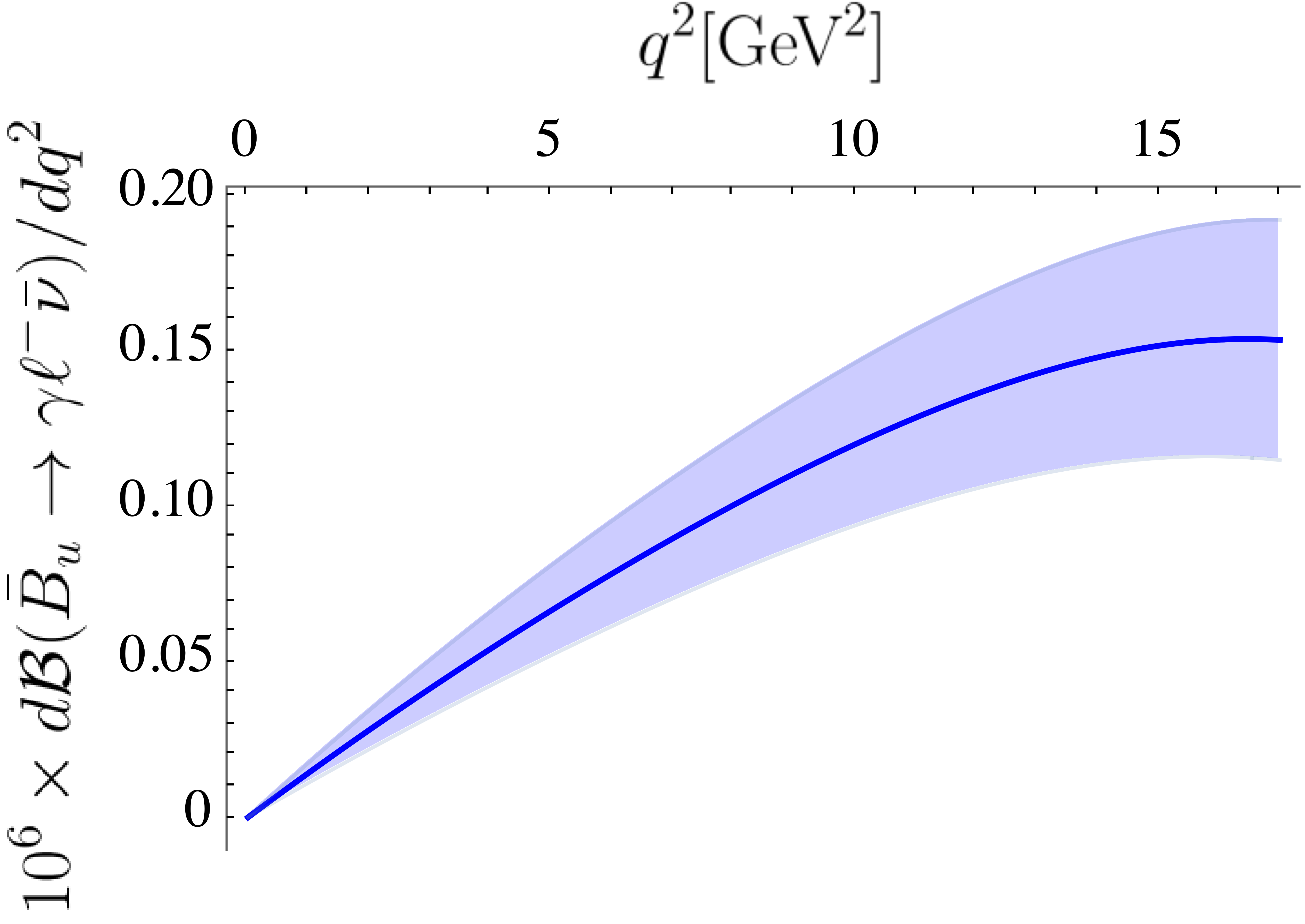}
   		\centering
    \end{minipage}\hfill
    \begin{minipage}{0.5\textwidth}
        \centering
        \includegraphics[width=0.9\textwidth]{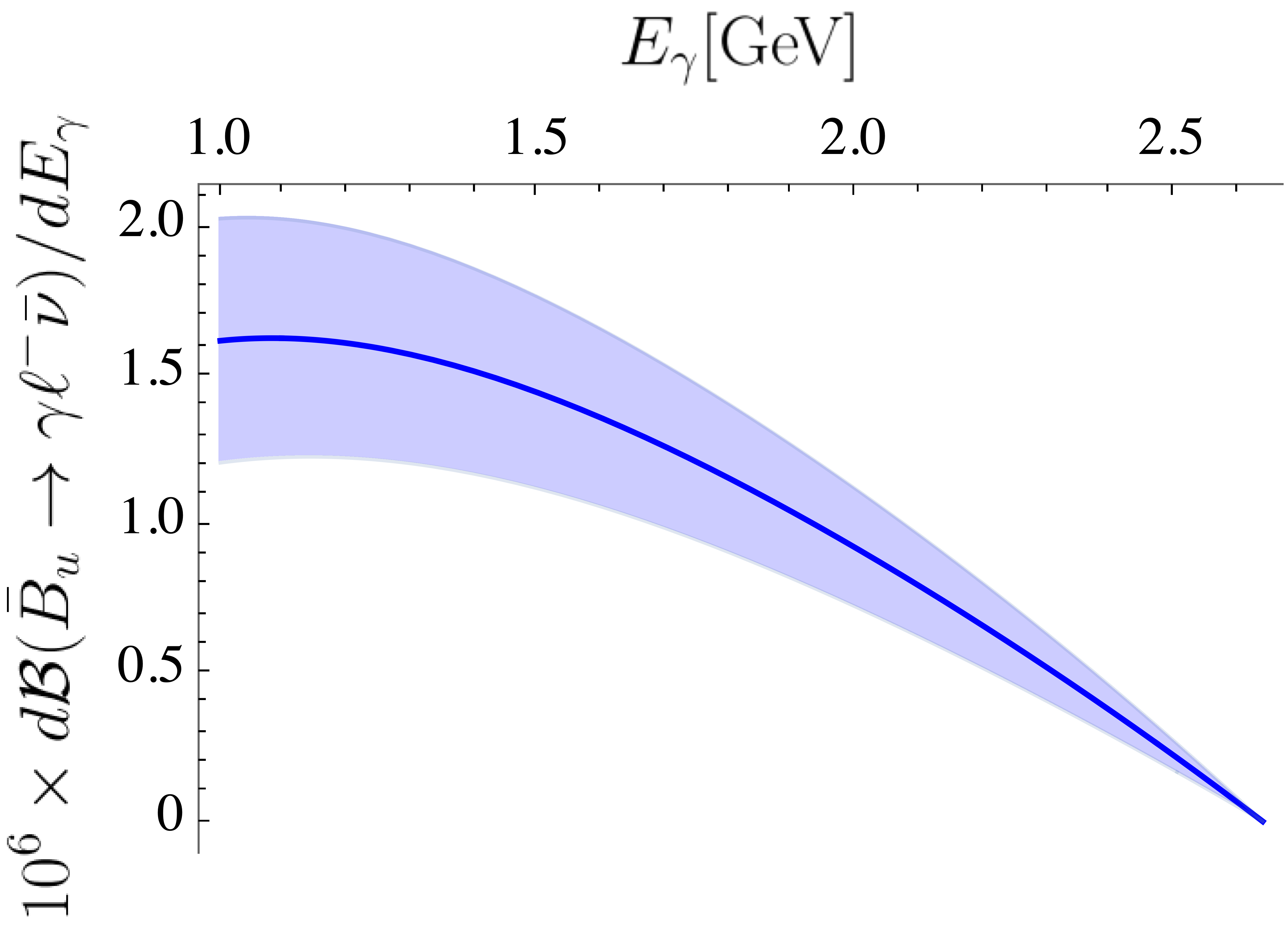}
    \end{minipage}
        \caption{\small Plots showing the differentiated branching fraction with respect to the di-lepton momentum transfer $q^2$ (left) and the photon energy $E_{\ga}$ (right). The solid blue line indicates the central value with the correlated uncertainty (shaded region)  on both the fit-parameters and the CKM matrix element $|V_{\textrm{ub}}|$. The contribution to the total uncertainty from the remaining input parameters is negligible. 
    In the evaluation of the branching ratio we use $G_F = 1.166378\cdot10^{-5} \GeV^{-2} $, $|V_{\textrm{ub}}| = 3.82 (24) \cdot 10^{-3} $, $\tau_B=1.638\cdot 10^{-12}s$ \cite{PDG} and the fine structure constant at the $B$-meson scale 
    $\al = 1/129$.}
     \label{fig:BR}
\end{figure}

Plots of the branching fraction are shown, in both the $q^2$ and the $E_{\gamma}$ variables, in \FIG\ref{fig:BR}. 
The $q^2$ and $E_\gamma$ covariance matrices between the bins of \TAB\ref{tab:BR} are respectively given by
\begin{equation}
\text{cov}_{\text{bin}}^{q^2}\!=\!\!\left(\!
\begin{array}{cccc}
 0.0534 & 0.0502 & 0.0459 & 0.0393 \\
 0.0502 & 0.0471 & 0.0431 & 0.0369 \\
 0.0459 & 0.0431 & 0.0394 & 0.0337 \\
 0.0393 & 0.0369 & 0.0337 & 0.0289 \\
\end{array}
\!\right),\quad
\text{cov}_{\text{bin}}^{E_\gamma}\!=\!\left(\!\!
\begin{array}{cccc}
 0.5537 & 0.4521 & 0.3188 & 0.1787 \\
 0.4521 & 0.3736 & 0.2675 & 0.1523 \\
 0.3188 & 0.2675 & 0.1954 & 0.1133 \\
 0.1787 & 0.1523 & 0.1133 & 0.0669 \\
\end{array}
\!\right).
\end{equation}
The $\bar B_u \to \ga \ell^- \bar \nu$ rate will, finally,  be measured at Belle II \cite{Kou:2018nap}. 
A $90\%$ confidence level (CL), 
 ${\cal B} (\bar B_u \to \ga \ell^- \bar \nu)_{E_\ga > 1 \GeV}  <3.0 \cdot 10^{-6} \;@90\% \textrm{ CL}$,  
with Belle  data has been reported two years ago \cite{Gelb:2018end}. This limit is about a factor of two 
above our predictions of $1.60(37) \cdot 10^{-6}$ given in \TAB\ref{tab:BR}.
\begin{table}[H]
  \centering
  \setlength\extrarowheight{-2pt}
  \resizebox{\columnwidth}{!}{
  \begin{tabular}{| l |r|r|r|r |} \hline \hline
	$q^2$ bin $[\GeVs^2]$                                  & $[0,6]\GeVs^2$           & $[2,6]\GeVs^2$         & $[3,6]\GeVs^2$           & $[4,6]\GeVs^2$  \\\hline
	$10^{6}\times\mathcal{B}(\bar B_u \to \ga \ell^- \bar\nu)$ & 0.243(53)         & 0.214(47)       & 0.180(39)         & 0.132(29)  \\\hline\hline
	$E_\gamma$ bin $[\GeVs]$                               & $[0.5\GeVs,\frac{m_B}{2}]$ & $[1.0\GeVs,\frac{m_B}{2}]$ & $[1.5\GeVs,\frac{m_B}{2}]$ & $[2.0\GeVs,\frac{m_B}{2}]$  \\\hline
	$10^{6}\times\mathcal{B}(\bar B_u \to \ga \ell^- \bar\nu)$ & 2.224(554)        & 1.604(374)      & 0.883(195)        & 0.305(67)         \\\hline
  \end{tabular}
  }
  \caption{\small  Integrated branching fraction for various $q^2$ and $E_\gamma$ bins.  
  The photon energies to the left are denoted  by $E_\ga^{\textrm{min}}$ elsewhere in the text.
  The last three branching fraction predictions are overlaid  (rescaled by  $(3.70/3.82)^2 $ since 
  $|V_{\textrm{ub}}|_{\mbox{\cite{Beneke:2018wjp} }} = 3.70(16)\cdot 10^{-3}$)  in \FIG\ref{fig:lamdaB} in order to  extract $\la_B$.}
  \label{tab:BR}
\end{table}

\subsection{Extraction of the Inverse Moment \texorpdfstring{$\la_B$}{} of the \texorpdfstring{$B$}{}-meson DA}
\label{sec:BDA}

As previously mentioned, besides being a toy model for factorisation,  $\bar B_u \to \ga \ell^- \bar\nu$ 
is of interest to extract the $B$-meson DA parameters from experiment. 
Here, we replace experiment by our computation 
which comes with the additional bonus that the  $\ORD(8\%)$ uncertainty of the 
 CKM matrix element  $|V_{\textrm{ub}}| = 3.82 (24) \cdot 10^{-3} $   \cite{PDG} is eliminated.
This is also useful as $|V_{\textrm{ub}}|$  suffers from some tension between exclusive and inclusive  $b \to u$ transitions  \cite{PDG} .  

Let us turn to some minimal definitions in order to clarify the context.
The $B$-meson DA of leading twist-$2$, $\phi^B_+$ was originally 
introduced in \cite{Grozin:1996pq}.
 Its inverse moment 
\begin{equation}
\label{eq:laB}
\frac{1}{\la_B(\mu)} =   \int_0^\infty d \omega  \frac{  \phi^B_+(\omega,\mu)} {\omega}  \;, 
\end{equation}
is a genuinely unknown non-perturbative parameter  $\la_B = \ORD(\LaQCD)$. Above
 $\mu$ is the renormalisation scale (or factorisation scale in processes), assumed to be 
 $\mu = 1\GeVs$ if not stated otherwise.  
Mainly due to its non-local character it has so far evaded a direct first principle determination.  
In  computations using the heavy quark limit it is often the leading uncertainty.
More precisely, at NLO  there are two further non-perturbative parameters that appear,
the inverse logarithmic moments $\hat{\sig}_{1,2}$ e.g. 
\cite{DescotesGenon:2002mw,Beneke:2011nf}
\begin{equation}
 \hat{\sig}_n(\mu) =  \lambda_B(\mu)\int_0^{\infty}\frac{ d\omega}{\omega}\ln^n\frac{ \mu_0}{\omega}\phi_+(\omega,\mu)  \;, \quad  \mu_0= \la_B e^{-\ga_E} \;,
\end{equation}
and one should be aware of different conventions due to the choice 
of $\mu_0$ resulting in different values of $\sig_n$.
Matters are further complicated by the fact that $\la_B$ and $\hat{\sig}_{1,2}$ are 
defined at  $\mu = 1\GeVs$  but in computations they are evaluated at the hard-collinear scale $\mu = \ORD(\sqrt{\LaQCD m_b}) \approx 2.2\GeVs$. 
 The evolution of the $B$-meson DA parameters involves solving an integro-differential equation \cite{Lange:2003ff,Braun:2019wyx,Galda:2020epp} already at $\ORD(\al_s)$ with  no autonomous evolution of the 
parameters (unlike for light-meson DAs where the conformal symmetry allows for a simpler picture, 
cf. \SEC\ref{sec:NLO}). 
In essence a separate (indirect) determination of these parameters is not feasible.
Thus actual studies one is forced to use a model-ansatz for  $\phi^B_+$. 
This is done in reference \cite{Beneke:2018wjp} for a $3$-parameter family,   which permits
 an analytic evolution at $\ORD(\al_s)$. This $3$-parameter model is further constrained to 
 three $2$-parameter models which give rise to the following ranges in $\hat{\sig}_{1,2}$\footnote{The ranges $(\sigma_1,\sigma_2)_{\mu_0= 1{\mbox{\tiny\text{ GeV}}}} = (1.5(1),3(2))\GeVs$ 
 (the $\hat{\sig}_1$-value  comes from non-local sum rule 
\cite{Braun:2003wx}  and the $\hat{\sig}_2$-value is the one adopted in  \cite{Beneke:2011nf}), 
 translate into ranges of $(\hat{\sig}_1 , \hat{\sig}_2) \approx  (-0.1(1.0), -4.4(3.2)(2.0)  ) \GeVs $ 
 when $\la_B = 360 \MeVs$ is assumed.} \begin{alignat}{6}
 \label{eq:models}
& \textrm{Model I} \qquad & & -0.31 < \;& \hat{\sigma}_1 & < 0.0  \;, \quad & & \phantom{-} 0.74 < \;& \hat{\sigma}_2 & < 1.64 \;, \quad & & \textrm{\color{blue} [blue]}  \;, \nonumber \\[0.1cm]
& \textrm{Model II} &  & -0.31 < \;& \hat{\sigma}_1 & < 0.69 \;, \quad & & \phantom{-} 0.74 < \;& \hat{\sigma}_2 & < 5.42 \;, \quad & & \textrm{\color{darkgreen} [green]}  \;, \nonumber \\[0.1cm]
& \textrm{Model III} &  & -0.69 < \;& \hat{\sigma}_1 & < 0 \;, \quad    & & \phantom{-}1.16 <        \; & \hat{\sigma}_2 & < 1.64 \;, \quad & & \textrm{\color{red} [red]} \;.
 \end{alignat}
 \begin{figure}[t]
  \centering
  \begin{minipage}{0.33\textwidth}
      \includegraphics[width=0.99\textwidth]{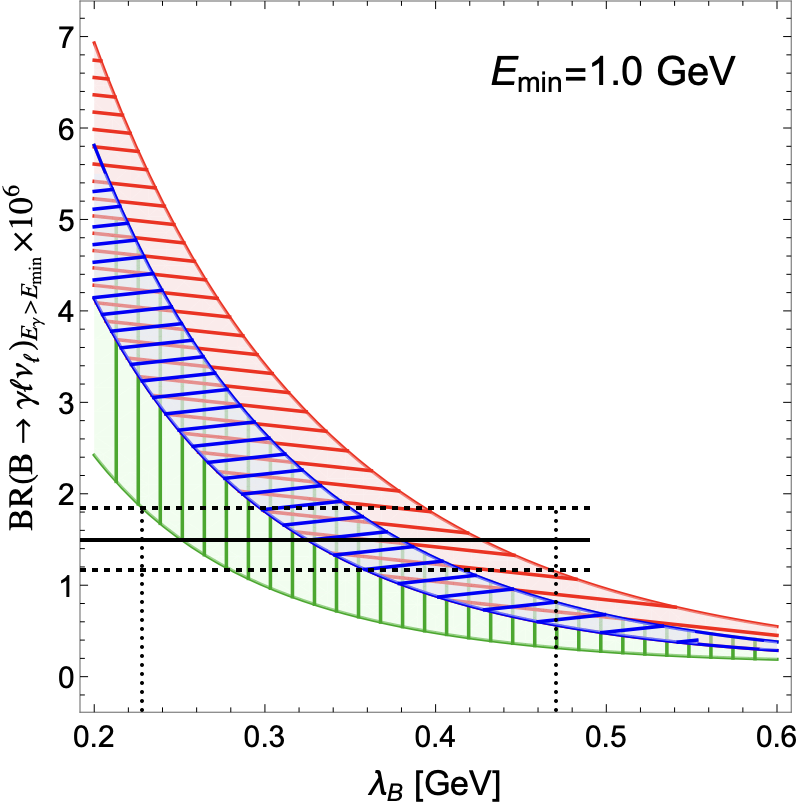}
      \centering
    \end{minipage}\hfill
    \begin{minipage}{0.33\textwidth}
        \centering
        \includegraphics[width=0.99\textwidth]{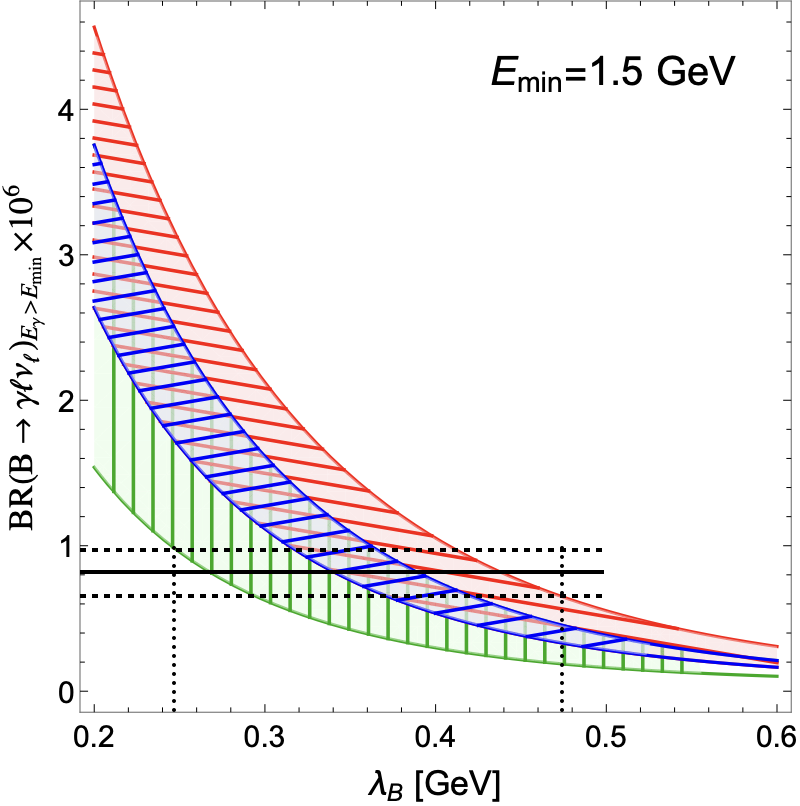}
    \end{minipage}\hfill
    \begin{minipage}{0.33\textwidth}
    \centering
    \includegraphics[width=0.99\textwidth]{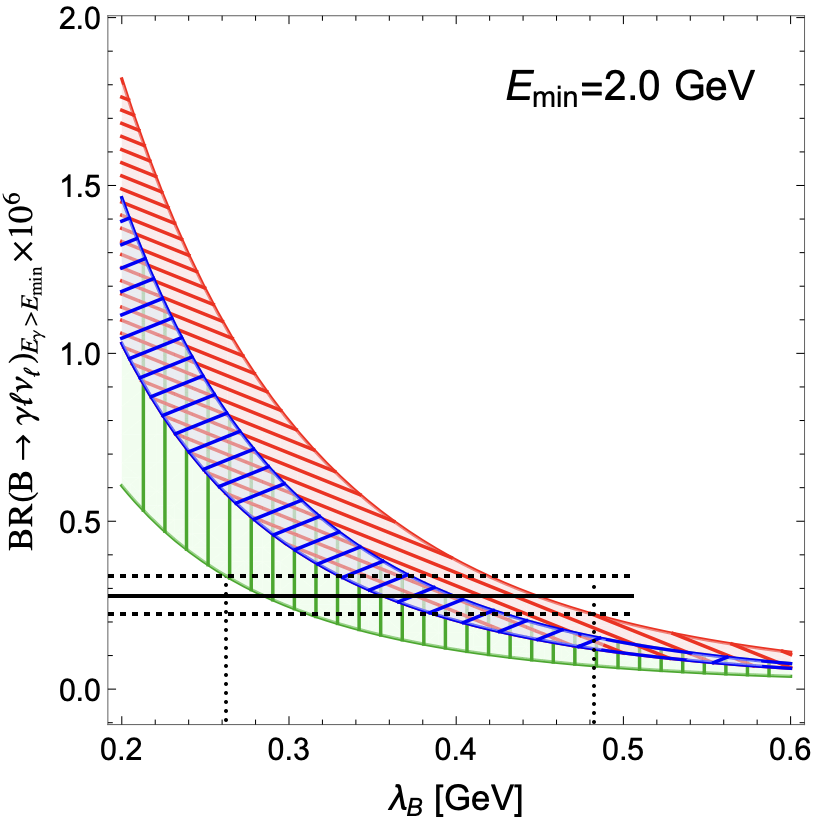}
    \end{minipage}
    \caption{\small Plots taken from \cite{Beneke:2018wjp} (coloured curves) overlaid with our predictions for the integrated branching ratios (cf. \TAB\ref{tab:BR}) scaled to account for the fact that in \cite{Beneke:2018wjp} 
 $|V_{\textrm{ub}}|_{\textrm{excl}} = 3.70(16) \cdot 10^{-3} $ is used. This results an effect of a factor of $0.938$ on the branching fraction. Solid black lines represent our central values with uncertainty bands indicated by dashed lines. The colour coding of the plots corresponds to the different models cf. \eqref{eq:models} and \cite{Beneke:2018wjp} of course.
}
     \label{fig:lamdaB}
\end{figure}
 The variation of the model and the model parameter effectively 
 accounts for the lack of knowledge of  $\phi^B_+$.

\begin{table}[btp]
  \centering 
    \begin{tabular}{|C| C C C|C|}
    \hline\hline
      E_{\gamma}^{\text{min}} & \lambda_B^{\text{Model I}} &\lambda_B^{\text{Model II}} &\lambda_B^{\text{Model III}} &\lambda_B^{\text{Model Avg.}}\\\hline
      1.0\GeV&355(60)\MeV&295(60)\MeV&415(60)\MeV&350(120)\MeV \\
      1.5\GeV&365(60)\MeV&310(60)\MeV&415(60)\MeV&360(110)\MeV \\
      2.0\GeV&375(60)\MeV&315(60)\MeV&420(60)\MeV&370(110)\MeV \\\hline 
    \end{tabular}
    \caption{\small Values of $\lambda_B$ read off from \FIG\ref{fig:lamdaB} for the three $\phi_+^B$ models, rounded to $5\MeV$. The rightmost column contains the mean value over the models for each minimum photon energy cut, rounded to the nearest $10\MeV$ and the $E_{\gamma}^{\text{min}} = 1.5\GeV$ is selected as the our final value in \eqref{eq:laBvalue}.}
    \label{tab:lambdaB}
\end{table} 
 In \FIG\ref{fig:lamdaB} we have overlaid our  branching fraction predictions, shown in \TAB\ref{tab:BR}, 
 with the plots from \cite{Beneke:2018wjp}.
 From \TAB\ref{tab:lambdaB} one sees that the values predicted for a given model are largely stable on the minimum photon energy  cut. 
  There is, as expected, some dependence on the
 specific model of $\phi_+^B$. However, averaging over the ranges of the three models (cf. right hand column of \TAB\ref{tab:lambdaB}), one again sees the consistency between the different photon cuts, which suggests good agreement in the shape for these values.
  
Whereas $E_\ga^{\textrm{min}} = 2\GeVs$ is arguably  the best result 
 because of the range of validity of the approaches, it still seems preferable to take an average 
of the three values for $\lambda_B^\text{Model Avg.}$ to deduce our final value in this work
 \begin{equation}
 \label{eq:laBvalue}
  \la_B =   360(110) {\MeV} \;.
 \end{equation} 

We now turn to discussing the value obtained in relation to other determinations. 
The previously mentioned  Belle measurement from 2018 sets a bound of 
$\la_B > 240 {\MeVs}$ $@90\% \textrm{ CL}$    using  \cite{Beneke:2018wjp} as the reference input.
A direct determination via a QCD sum rule, which however uses non-local quark condensates which are not free from model-dependence, was performed in 2003 and gave $\la_B = 460(100){\MeVs}$  \cite{Braun:2003wx} 
(and the previously quoted value for $\hat{\sig}_1$). An update \cite{Khodjamirian:2020hob} using improved numerical input 
and two models for the $B$-meson DA yields  $\la_B = 383(153){\MeVs}$, a value closer to ours. 
The same strategy as here has been applied at LO \cite{Ball:2003fq,Khodjamirian:2005ea} and NLO 
\cite{Wang:2015vgv,Gao:2019lta} with various specific model functions for $\phi_+^B$.
The $B \to \ga$  \cite{Ball:2003fq} LO-analysis has been done in the heavy quark limit and resulting in
$\la_B = 600{\MeVs}$  with no given error.   The LO $B \to \pi$ analysis with an exponential model 
gave  $\la_B = 460(160){\MeVs}$   \cite{Khodjamirian:2005ea}.
At NLO there are the works on $B \to \pi$ \cite{Wang:2015vgv} and $B \to \rho$ \cite{Gao:2019lta}
with  $\la_B = 354^{+30}_{-38}{\MeVs}$ and 
$343^{+64}_{-79} {\MeVs} $ respectively for which we quote the exponential model values (cf. those 
references for other model determinations). 

\begin{table}[btp]
  \centering 
    \resizebox{\columnwidth}{!}{
    \begin{tabular}{| l |  l  | l  |  l  l | }
    \hline\hline
      $\la_B [\MeVs]$ &  ref.  &  year &  method  & B-DA model  \\\hline
      $460(100)$ &   \cite{Braun:2003wx}  & '03   &  QCD SR (non-local condensate) & -- \\
      $383(153)$ &   \cite{Khodjamirian:2020hob} & '20   &  idem   & -- \\ \hline
      $ 354^{+30}_{-38}$ & \cite{Wang:2015vgv} &  '15   &  B-LCSR vs LCSR $B \to \pi$ \cite{BZ04b,Duplancic:2008ix}  & exponential  
       (81) \cite{Wang:2015vgv}  \\
        $368^{+42}_{-32}$ & \cite{Wang:2015vgv} &  '15   & idem   & Model-II  
        (81) \cite{Wang:2015vgv} \\
          $389^{+35}_{-28}$ & \cite{Wang:2015vgv} &  '15   & idem  & Model-II 
        (81) \cite{Wang:2015vgv}  \\
            $ 303^{+35}_{-28}$ & \cite{Wang:2015vgv} &  '15   & idem & Model-II 
        (81) \cite{Wang:2015vgv}  \\ \hline
            $343^{+64}_{-79}  $ & \cite{Gao:2019lta} & '19      &     B-LCSR vs LCSR
      $B \to \rho$  \cite{BSZ15} & exponential \\ 
      $370^{+69}_{-86}  $ & \cite{Gao:2019lta} & '19      & idem  & local duality  \\
      \hline
600 & \cite{Ball:2003fq}  & '03 &  SCET \cite{DescotesGenon:2002mw} vs LCSR $B \to \ga$    ($m_b \to \infty$) & -- \\ \hline
365(60) &  this work &  '21 &  LCSR vs SCET \cite{Beneke:2018wjp}  &  Model I  
(5.4a) \cite{Beneke:2018wjp}  \\
310(60) &  this work &  '21  & idem & Model II  (5.4b) \cite{Beneke:2018wjp} \\
415(60) &  this work &  '21 & idem &  Model III (5.4c) \cite{Beneke:2018wjp} \\ \hline
    \end{tabular}
    }
    \caption{\small Comparison-table of $\lambda_B$ 
    with respect to the literature.  
    The issue of the model-dependence of the indirect determinations by matching 
    two FF computations is clearly visible. B-LCSR stands for LCSR with a $B$-meson DA.
   Cf. final paragraph of the section for further comments.}
    \label{tab:lambdaBcomp}
\end{table} 

To conclude, from \TAB\ref{tab:lambdaBcomp} it is seen that the restriction to 
a single model does reduce the error considerably and underlines the importance of the 
QCD SR determinations.  In summary, the picture looks 
consistent but a dedicated heavy quark analysis of the NLO $B \to \ga$ analysis 
 must ultimately be the most reliable method 
 as it will not involve any photon DAs. 
 We anticipate returning to this issue  in a forthcoming publication.

\section{Summary and Discussions}
\label{sec:conclusion}

In this work we have computed  the $\bar B_{u,d,s} \to \ga$ form factors at NLO 
for twist-$1$,$2$  and LO in twist-$3$,$4$ with   light-cone sum rules. 
This involved the evaluation  of the diagrams in \FIG\ref{fig:PT_2L} which was the technically most 
challenging part of this work and involved the use of a series of tools  \cite{FeynCalc1,FeynCalc2,litered:2013,kira:2017,polylogtools}
and state-of-the-art master integrals \cite{DiVita:2017xlr}. The form factors are provided as fits to a $z$-expansion ansatz \eqref{eq:Fon} with fit-parameters
given in \TAB\ref{tab:fitparams}. The correlation matrix can be found in an ancillary file 
to the arXiv version.
Form factor values at $q^2 =0$ and plots are presented in  \TAB\ref{tab:FFval} 
and  \FIG\ref{fig:FFplots} respectively.   This constitutes our main practical results. 
Analytic results of the correlation 
functions in terms of Goncharov functions are given as an ancillary file (\texttt{Mathematica} notebook) 
to the arXiv version.
An important technical aside is the choice of the $m_b$ mass-scheme 
for which we have adapted the kinetic scheme,
which shows good scale-stability  (cf. \FIG\ref{fig:FFscale}) 
unlike the  pole- or $\MSbar$-scheme for which either the $\al_s$- or  twist-corrections are unnaturally large.

Besides the form factors themselves, we have established a few valuable theoretical results and advances. 
In \SEC\ref{sec:scaling}, starting from a dispersion representation, we motivated  the $m_b^{-1/2}$
heavy quark scaling of the $\bar B \to \gamma$ form factors from a purely hadronic picture.
Additionally, in \SEC\ref{sec:mqDA}, we proposed a scheme to add (strange) quark mass corrections 
to the magnetic vacuum susceptibility by separating the perturbative part. 
The following values have been obtained at $\ORD(\al_s^0)$:  
$ \chi_q  =  3.21 (15) \textrm{\small GeV}^{-2}  $  and $ \chi_s  =   3.79  (17) \textrm{\small GeV}^{-2} $.
Moreover, the equations of motion  \eqref{eq:EOMFF} were verified  to  hold in all hadronic 
parameters for which completeness can be expected. 
In doing so it became clear that in order to include the twist-$4$ $3$-particle 
distribution amplitude parameters, given in \TAB\ref{tab:twist}, one needs to include $4$-particle DAs of twist-$4$ which have not yet been classified. For $\bar B \to \ga$ form factors the impact is fortunately rather small. 

The \emph{charged} form factors describe the $\bar B_u \to  \ga \ell^- \bar \nu$ decay, 
anticipated to be measured at Belle II  \cite{Kou:2018nap}, for which 
our differential rate prediction, binned and including correlation matrix, are given in \SEC\ref{sec:rate}.
Comparing to SCET-computations, of three models of the $B$-meson distribution amplitude, we 
have extracted its inverse moment \eqref{eq:laB}, $\la_B = 360(110)\MeVs$.\footnote{
 It seems worthwhile 
to mention that  comparing theory computations with each other  eliminates 
the sizeable $|V_{\textrm{ub}}|$-uncertainty. Moreover, this analysis is to be regarded as a crude first approach and
 can be improved by doing a fully differential analysis, including 
the two logarithmic moments $\hat{\sig}_{1,2}$ into a $3$-parameter fit, which we might turn to in  
a future publication.}
The \emph{neutral} form factors are the input to the flavour changing neutral 
current process $\bar B_{d,s} \to \ga \ell^+ \ell^- $, which will come into focus by LHCb in the muon-mode \cite{LHCb:2021awg,LHCb:2021vsc,LHCbtalk,Ferreres-Sole:2021qxv},   alongside off-shell form factors \cite{Albrecht:2019zul} 
and genuine long distance contributions \cite{GRZ17,Kozachuk:2017mdk,Beneke:2020fot}.
The scope of applications of the $\bar B \to \ga$ form factors does not end here  as
they can serve as inputs, to related processes, in terms of subtraction 
constants of dispersion relations.
Examples are, the weak annihilation process in $\bar B \to V \ga$ e.g. \cite{Lyon:2013gba}, 
$B_s \to \mu \mu \, \textrm{axion/dark photon}$  
for which generic photon off-shell  $B \to \ga^*$ form factors  are needed   (cf. \APP A   \cite{Albrecht:2019zul}  for  
how dispersion relations can bridge the  $k^2 \approx m^2_{\rho,\omega,\phi}$  resonance region)
and  $\bar B_u \to \ell^+ \ell^- \ell^{'-} \bar{\nu}$ (requiring $q^2,k^2 \neq 0$)  as recently  investigated  in \cite{Bharucha:2021zay,Beneke:2021rjf}. 
Our form factor predictions should  be helpful  
in reducing the theory error in all of these modes and enhance the new physics sensitivity.

\subsection*{Acknowledgments}
We are grateful to  Gunnar Bali, Martin Beneke, Christoph Bobeth, Vladmir Braun, Matteo Di Carlo, Paolo Gambino, Einan Gardi, Yao Ji, Chris Sachrajda and Yuming Wang for useful discussions and to James Gratrex for careful reading of the manuscript.
We are indebted to Thomas Gehrmann and Roberto Bonciani for guidance in the master integral literature, 
Claude Duhr and James Matthews  for help with Polylog Tools 
and Stefano Di Vita, Pierpaolo Mastrolia, Amedeo Primo \& Ulrich Schubert for useful 
exchange and guidance with regard to \cite{DiVita:2017xlr}. BP is also grateful to Calum Milloy for useful discussions
and guidance regarding the method of differential equations and the subtleties therein. 
RZ acknowledges Damir Becirevic for the kind hospitality at IJCLab, P\^ole Th\'eorie  Orsay-Paris 
when this work was in its initial stages.
RZ is supported by an STFC Consolidated Grant, ST/P0000630/1. BP is supported by an STFC Training Grant ST/N504051/1.
                 
\subsection*{Ancillary files}
We include a number of ancillary files. A file with the renormalised NLO correlation functions
for twist-$1$ in terms of MPLs  (\texttt{corr\_PT\_NLO.m});  
the new set of master integrals  (\texttt{MI\_vzzb.m}) discussed in \SEC\ref{sec:NLO1} which is linearly dependent 
on the basis \cite{DiVita:2017xlr} but of value to make cancellations explicit.  
The $z$-expansion fits,  given as a \texttt{JSON} file (\texttt{BtoGam\_FF.js}on), with some description in \SEC\ref{sec:fit} and the notebook (\texttt{FF-plots\_from\_json.nb}) that exemplifies its implementation. 

\appendix

\numberwithin{equation}{section}

\section{Conventions and Additional Plots }
\label{app:conv}

We define the covariant derivative  as
\begin{equation}
\label{eq:cD}
D_\mu = \partial_\mu + s_e i e Q_f A_\mu + s_g i g_ s A^a_\mu T^a
\end{equation}
where $e > 0$ ($\al = e^2/(4 \pi))$, $Q_e = -1$, $Q_{u,c,t} = 2/3$ and $Q_{b,d,s} = -1/3$. This leads to a 
Feynman rule
$ - s_e i e Q_f \slashed{A} - s_g i g_ s \slashed{A^a}T^a $. 
Note that the sign of the FFs \eqref{eq:ffs} depends on $s_e$ which we therefore prefer to keep 
explicit. The sign of the gluon term $s_g$ is also kept which appears in the light-cone propagator 
in the background field \eqref{eq:BGFprop}   as well as the $3$-particle DAs. The final results are independent of $s_g$ 
and the rate is of course independent of $s_e$, but attention must be paid when compared or combined with 
other work.
Using the $g= \textrm{diag}(1,-1,-1,-1)$ metric the following relation 
holds,  
\begin{equation}
\label{eq:5sigma}
\sigma^{\al\be} \ga_5 = -   \frac{i}{2}  \varepsilon^{\al\be\ga\de}\sigma_{\ga\de} \;,
\end{equation}
with $ \varepsilon_{0123} =1$ and $\ga_5 = i \ga^0 \ga^1 \ga^2 \ga^3$.
Moreover we have $T^a T^a = C_F \mathbf{1}$ with $C_F = (N_c^2-1)/(2N_c)$ and $\al_s =g^2/(4 \pi)$ as usual.

\subsection{Borel Transformation}
\label{app:Borel}

The Borel transform is defined as
\begin{equation}
f_\Bor (M_\Bor ^2) = \Bor f(Q^2) =\lim_{Q^2 \to \infty, n \to \infty} \dfrac{(Q^2)^{n+1}}{n!} \left(- \dfrac{\der}{\der Q^2}\right)^{n+1} f(Q^2)
\end{equation}
where $Q^2 = - q^2$ is a Euclidean momentum and $Q^2/n = M_\Bor^2$ defines the so-called Borel mass.

For the perturbative part and the condensates, which are poles at tree level, one needs
the following Borel transformation    
\begin{align}
\label{eq:Borel}
\Bor \dfrac{1}{(Q^2+m^2)^a} &= \dfrac{1}{\Gamma(a) (M_\Bor^2)^a} e^{-m^2/M_\Bor^2} \, . 
\end{align}
The DA-part can be reduced to a form $du \frac{1}{\Delta^n}$ ($n \geq 1$) for which simple rules for Borel transformation
are given in \APP A in \cite{BSZ15}. Of course one could integrate over the DA-parameters 
$du$ or $D\underline{\al}$, then subject the result to a dispersion relation and use \eqref{eq:Borel}.
The advantage of using the former method is that one does not need to commit to a specific form/truncation 
of the DAs.

\subsection{Plots of Scale Dependence and Twist-contributions}
In this appendix we collect additional plots and tables.
\begin{figure}[H]
\centering\vspace{-3mm}
    \begin{minipage}{0.45\textwidth}
        \centering
        \includegraphics[width=1.\textwidth]{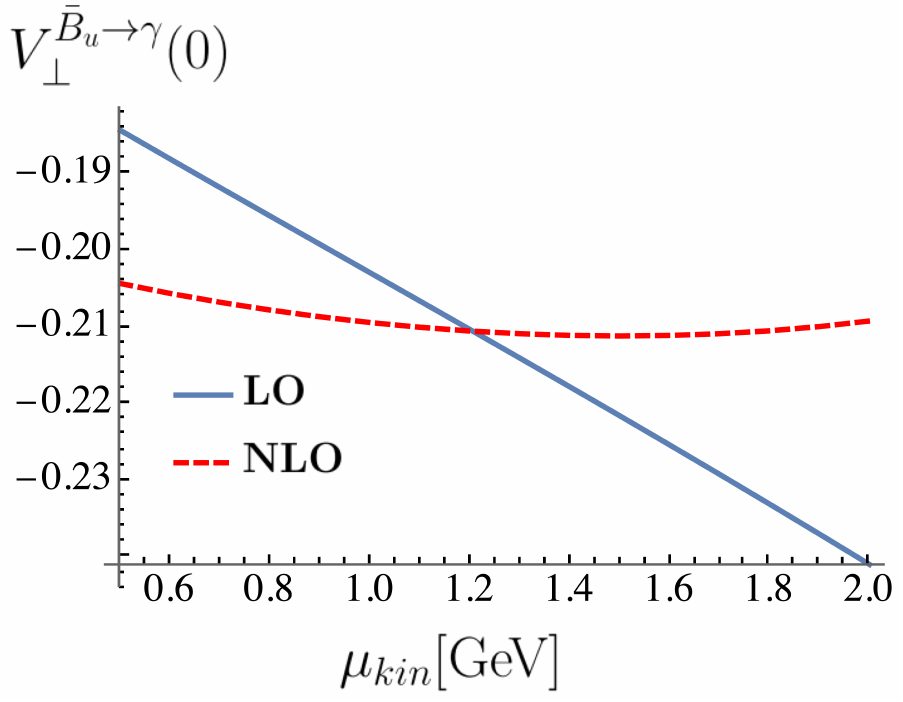}
    \end{minipage}\hfill
    \begin{minipage}{0.47\textwidth}
        \centering
        \includegraphics[width=1.\textwidth]{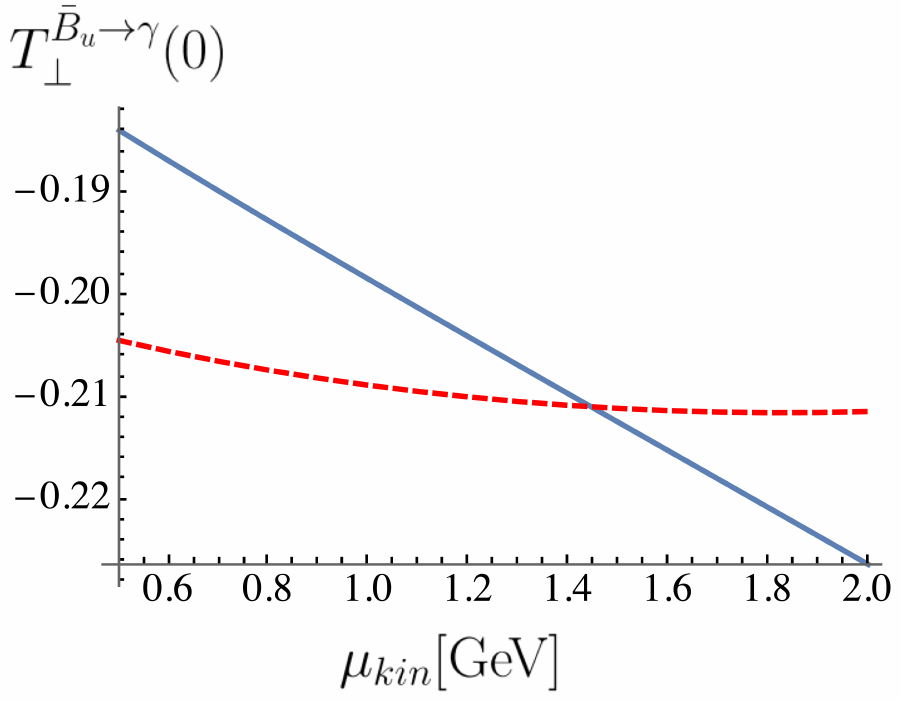}
    \end{minipage}\vspace{-1mm}
        \begin{minipage}{0.45\textwidth}
        \centering
        \includegraphics[width=1.\textwidth]{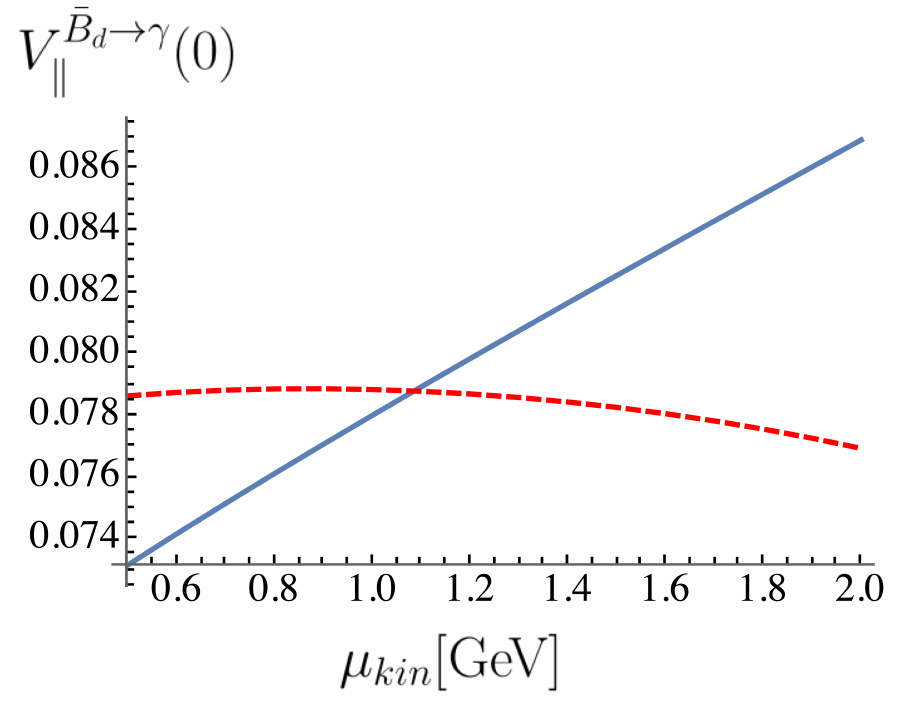}
    \end{minipage}\hfill
    \begin{minipage}{0.46\textwidth}
        \centering
        \includegraphics[width=1.\textwidth]{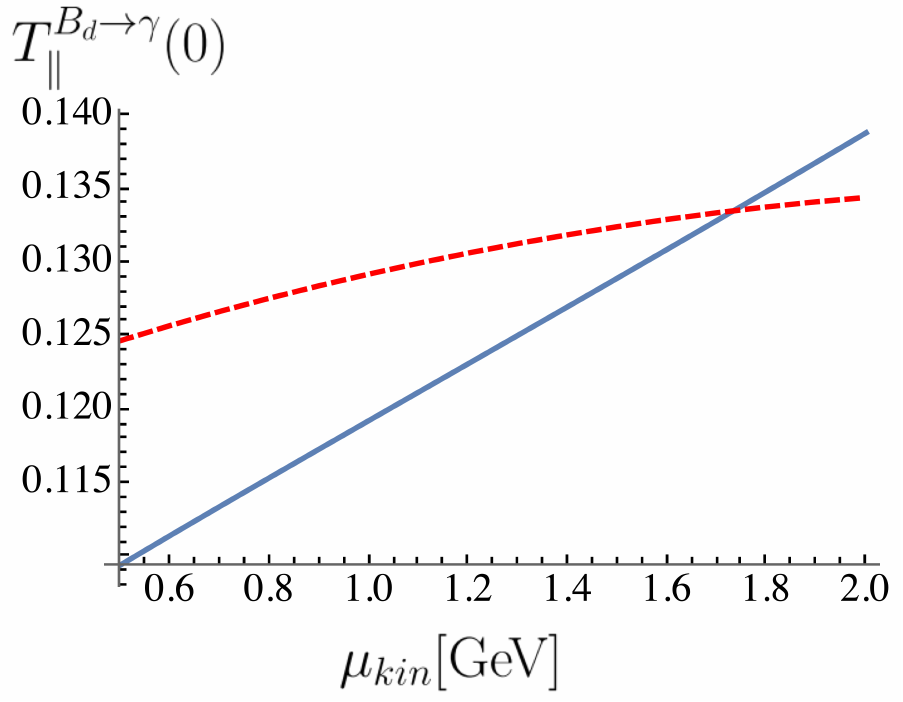}
    \end{minipage}\vspace{-1mm}
    \caption{\small Plots of a representative selection of form factors, evaluated at $q^2=0$. The plots serve to highlight the reduction in the kinetic scale dependence when radiative corrections to twist-1 and -2 are included (dashed red line) as compared to the LO result (solid blue line).}
    \label{fig:FFscale}
\end{figure}
The  merit of the NLO versus a LO  analysis shows in the 
mass scale dependence plots  in \FIG\ref{fig:FFscale}. The advantages of the 
kinetic-scheme over the pole- and $\MSbar$- schemes become apparent in \FIG\ref{fig:polekinetic} and \TAB\ref{tab:kin_vs_msbar}, respectively. 
\begin{table}[H]
  \centering
  \setlength\extrarowheight{-2pt}
  \begin{tabular}{c | l |r r|r r|r r}
{\small twist} & DA & \multicolumn{2}{c|}{$V_{\|}^{\bar B_u\to\gamma}(0)$} & \multicolumn{2}{c|}{$V_{\perp}^{\bar B_d\to\gamma}(0)$} & \multicolumn{2}{c}{$T_{\perp,\|}^{\bar B_u\to\gamma}(0)$}  \\\hline
  &                                  & \multicolumn{1}{c}{Kin} & \multicolumn{1}{c|}{$\MSbar$} & \multicolumn{1}{c}{Kin} & \multicolumn{1}{c|}{$\MSbar$} & \multicolumn{1}{c}{Kin} & \multicolumn{1}{c}{$\MSbar$}  \\\hline
1 & $\textrm{PT}\,\ORD{(\al_s^0)}$   & $45.7\%$                & $70.0\%$                      & $51.3\%$                & $86.2\%$                      & $51.4\%$                & $79.5\%$      \\
1 & $\textrm{PT}\,\ORD{(\al_s)}$     & $8.4\%$                 & $-6.3\%$                      & $14.9\%$                & $-18.0\%$                     & $16.5\%$                & $-10.6\%$     \\\hline
2 & $\phi_{\ga}(u)\,\ORD{(\al_s^0)}$ & $62.3\%$                & $119.1\%$                     & $28.5\%$                & $47.2\%$                      & $24.4\%$                & $35.8\%$     \\
2 & $\phi_{\ga}(u)\,\ORD{(\al_s)}$   & $9.5\%$                 & $-38.7\%$                     & $4.1\%$                 & $-16.8\%$                     & $4.3\%$                 & $-8.7\%$      \\\hline
3 & $\Psi_{(a)}(u)$                  & --                      & --                            & $0.6\%$                 & $-0.1\%$                      & $0.5\%$                 & $0.2\%$  \\
3 & $\Psi_{(v)}^{(1)}(u)$            & $2.5\%$                 & $3.7\%$                       & --                      & --                            & $0.8\%$                 & $0.9\%$  \\\hline
4 & $h_{\ga}^{(2)}(u)$               & $-9.2\%$                & $-15.3\%$                     & --                      & --                            & $-1.6\%$                & $-1.9\%$ \\
4 & $\mathbb{A}(u)$                  & $-12.1\%$               & $-20.2\%$                     & $-5.9\%$                & $-8.5\%$                      & $-4.2\%$                & $-5.1\%$  \\
4 & $Q_q\VEV{\bar q q}$              & $-8.8\%$                & $14.5\%$                      & --                      & --                            & --                      & --  \\
4 & $Q_b\VEV{\bar q q}$              & $-2.9\%$                & $-4.9\%$                      & $2.8\%$                 & $4.2\%$                       & $2.4\%$                 & $3.1\%$  \\\hline
  & Total$^*$                        & $-0.118$                & $-0.097$                      & $0.127$                 & $0.125$                       & $0.127$                 & $0.130$
  \end{tabular}
  \caption{\small A comparison of the different twist contributions in the kinetic and $\MSbar$ schemes (with $\mukin=1\GeV$ and $\mu_{\MSbar}=m_b(m_b)=4.18\GeV$) for a selection of illustrative form factors. Columns represent the percentage contribution of each DA to ``Total$^*$''. We remind the reader that the asterisk in total indicates it includes twist-$4$ contributions that do not close under the EOM (cf.~\SEC\ref{sec:EOMtest}). Contributions from 3-particle DAs have been dropped for brevity. Formally one should redetermine the sum rule parameters (cf. \TAB\ref{tab:FFSRparams}) for the $\MSbar$ evaluation, however we choose to use the same parameters as for the kinetic evaluation as this is sufficient to illustrate our point. We note the kinetic scheme shows greater suppression of the higher twist contributions. Additionally, in contrast to the kinetic scheme, radiative corrections in the $\MSbar$-scheme act to cancel with the LO results.}
  \label{tab:kin_vs_msbar}
\end{table}

\begin{figure}[h]
\centering
    \begin{minipage}{0.5\textwidth}
        \centering
        \includegraphics[width=1.\textwidth]{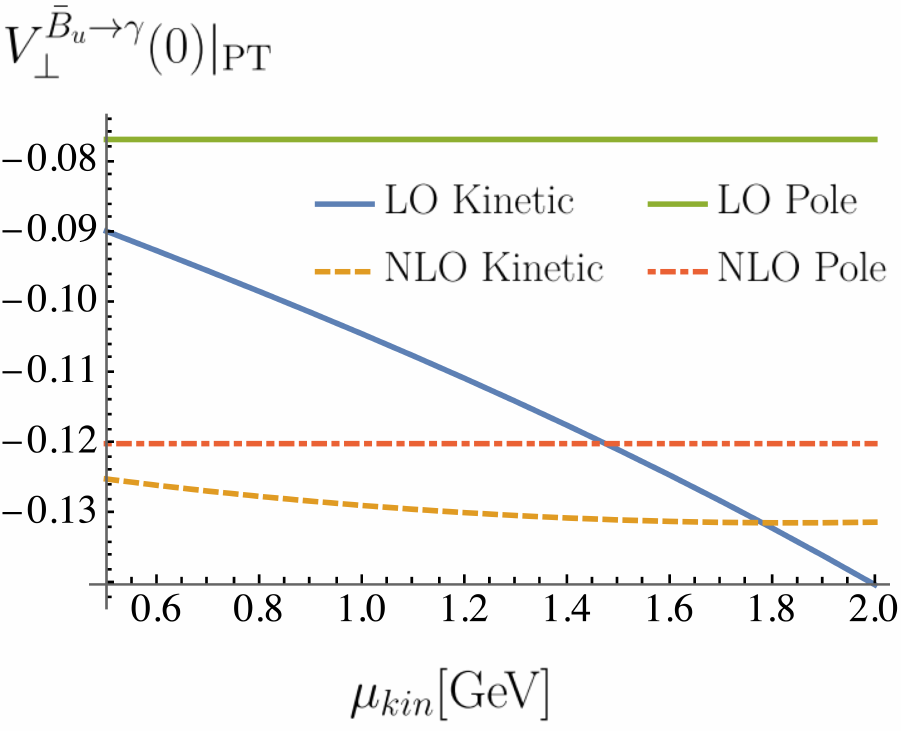}
    \end{minipage}\hfill
    \begin{minipage}{0.49\textwidth}
        \centering
        \includegraphics[width=1.\textwidth]{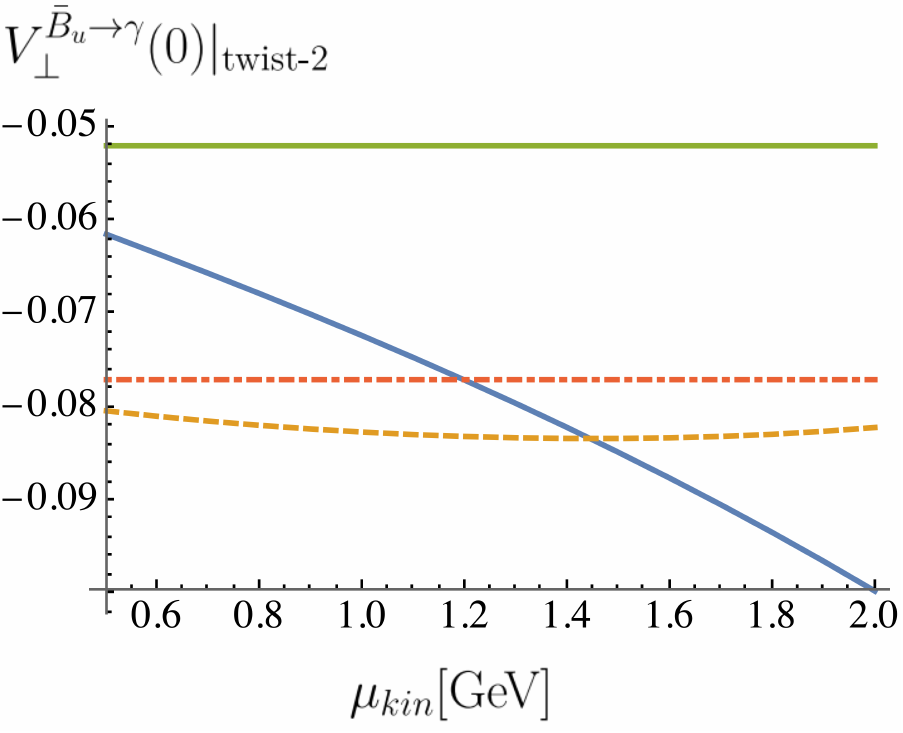}
    \end{minipage}\vspace{1mm}
        \begin{minipage}{0.49\textwidth}
        \centering
        \includegraphics[width=1.\textwidth]{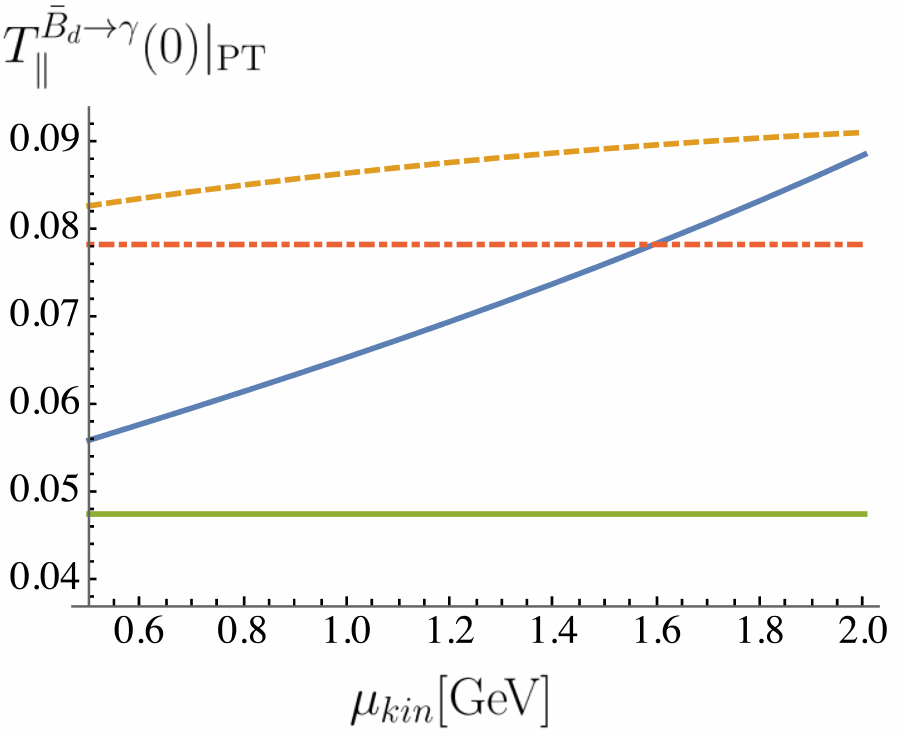}
    \end{minipage}\hfill
    \begin{minipage}{0.49\textwidth}
        \centering
        \includegraphics[width=1.\textwidth]{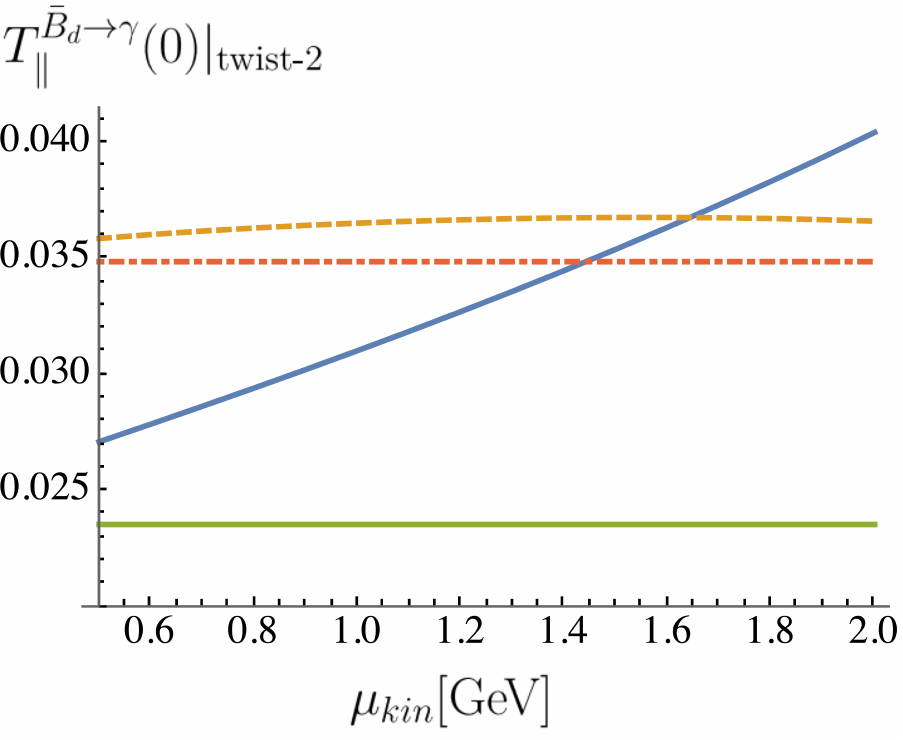}
    \end{minipage}
    \caption{\small Plots comparing the relative size of the radiative corrections to twist-$1$ and -$2$ in the pole- and kinetic-scheme, for two representative form factors. We observe large corrections in the pole scheme, with the $\ORD(\al_s)$ contributions being in the region of $50$-$60\%$ and $40$-$50\%$ of the LO result at twist-1 and -2, respectively. In contrast, using the kinetic scheme at $\mukin=1\GeV$, we find the NLO corrections are approximately $\sim20\%$ and $\sim10\%$ of the LO result at twist-1 and -2, respectively.}
    \label{fig:polekinetic}
\end{figure}

\section{Gauge (In)variance and the \texorpdfstring{$m_\ell =0$}{} Amplitude}
\label{app:GI}

In this appendix we intend to clarify a few issues around gauge invariance. 

\subsection{Gauge-variant part of the Charged Form Factor}
\label{app:GV}

In \SEC\ref{sec:QEDWI} it was shown how the  $\DelQ$-term in  \eqref{eq:ffs} appears
in the LCSR computation and below  its consistency is illustrated on general grounds. 
The $\DelQ$-term is present since 
 $O_\mu$ itself is not gauge invariant (charged) in the case where  the weak current is  charged. 
Let us write the matrix element on the LHS  of \eqref{eq:ffs} as follows
\begin{equation}
 \matel{ \gamma(k,\epsilon) }{O^V_\mu }{ \bar{B}_q (p_B) }  =  
 -i \eps^*_\rho \int_y e^{i k\cdot y} \matel{0}{T j^\rho(y) O^V_\mu(0) }{ \bar{B}_q (p_B) }   + \textrm{gauge-invariant} \;,
 \end{equation}
and then perform
a gauge tranformation  $\eps \to \eps + k$,
\begin{alignat}{2}
& -i k_\rho \int_y e^{i k\cdot y} \matel{0}{T j^\rho(y) O^V_\mu(0) }{ \bar{B}_q (p_B) }  &\;=\;& 
e \matel{0}{[Q,O^V_\mu(0)]}{\bar B_q}  \\[0.1cm]
& &\;=\;& \DelQ m_B \matel{0}{\bar q \ga_\mu (1-\ga_5) b} {\bar{B}_q}  = 
- i \DelQ m_B \pB{\mu} f_B \;,  \nonumber
\end{alignat}
where we have used $ \{ q_\al^\dagger(\vec{x},0) , q_\be(0) \} = \de^{(3)}(\vec{x}) \de_{\al \be}$. It is readily seen
that this matches the gauge variation of the Low-term on the RHS in \eqref{eq:ffs}.

\subsection{Gauge Invariance Restored, in the \texorpdfstring{$\bar B_u \to \ga \ell^- \bar  \nu$}{}-Amplitude}
\label{app:GIrestored}

Here we  illustrate that the total amplitude, 
emission from the $B$-meson (i.e. FF) and the lepton leg, 
 is gauge invariant.  The total amplitude ${\cal A}$ is proportional to 
\begin{equation*}
{\cal A}[\bar B_u \to \ell^- \bar \nu   \ga(\eps) ]  \propto e \eps^*_\rho \int_y e^{i ky} 
 \left( \matel{0}{T j^\rho(y) O_\mu^V }{\bar B_u}\matel{ \ell^- \bar \nu}{ L^\mu }{0}   + 
   \matel{0}{ O_\mu^V }{\bar B_u}\matel{ \ell^- \bar \nu}{T j^\rho(y) L^\mu}{0} 
   \right) \;,
\end{equation*}
where $H^{\textrm{eff}} \sim L^\mu O^V_\mu$ and
$L^\mu = \bar \ell \Gamma^\mu \nu$ is the charged lepton current with some unspecified 
Dirac matrix $\Gamma^\mu$. The total amplitude must be invariant under the residual gauge transformation 
$\eps \to \eps + k$, 
\begin{alignat}{2}
& \de_{\textrm{gauge}} {\cal A} &\;\equiv\;&  {\cal A}(\bar B_u \to \ell^- \bar \nu   \ga(\eps+k) ) - {\cal A}(\bar B_u \to \ell^- \bar \nu   \ga(\eps) ) 
\nonumber \\[0.1cm]
& &\;\propto\;&  i e\underbrace{(Q_b - Q_u - Q_{\ell})}_{=0}  \matel{0}{ O_\mu^V }{\bar B_u} \matel{ \ell^- \bar \nu}{ L^\mu }{0} = 0 \;,
\end{alignat}
which is the case if charge conservation is imposed ($Q_b = -1/3$, $Q_u = 2/3$ and $Q_{\ell} = -1$). 
 In conclusion, the total amplitude is indeed invariant under $\eps \to \eps + k$.

\subsection{The \texorpdfstring{$\bar B_u \to \ga \ell^- \bar  \nu$}{}-Amplitude at \texorpdfstring{$m_\ell \to 0$}{}}
\label{app:GIml0}

The $m_\ell$-case is special in that the emission term is local and allows a redefinition of the FF
 to absorb all effects \cite{Beneke:2011nf}. We follow this procedure in our notation as we consider
 it worthwhile to make contact with the existing literature. 
\begin{itemize}
\item In the $m_\ell = 0$ case the emission from the lepton is proportional
to the tree level matrix element.  Essentially 
\begin{equation}
\eps^*_\rho  
\matel{\ell(l_1) \nu(l_2)}{ \bar \ell \slashed{p}_B\frac{\slashed{k} + \slashed{l_1}}{(k+l_1)^2} \ga^\rho \nu }{0}  
= \eps^*_\rho   \matel{\ell(l_1) \nu(l_2)}{ \bar \ell \ga^\rho \nu }{0}  \;,
\end{equation}
 where $p_B = k+ l_1 +l_2$ and the  EOM was used for the $\nu$.  
 This allows to redefine the FF in such a way that the amplitude is proportional 
 to the FF only.  
 \item Focusing on the $\ga_5$-part of the FFs \eqref{eq:ffs}
 \begin{alignat}{3}
 \label{eq:axml=0}
& \matel{ \gamma(k,\epsilon) }{O^V_\mu }{ \bar{B}_q (p_B) }|_{\ga_5-\textrm{part}}
 &\;=\;&   s_e \left( -   \ppara{\mu}  
 \left(\FA(q^2)  + \DelQ \frac{m_B f_B}{k \cdot p_B}\right)   - \pB{\mu} i m_B \DelQ \frac{ \eps^* \cdot p_B}{k \cdot p_B}  f_{B_q}\right)   \nonumber \\[0.1cm]
&  &\;=\;& s_e (-  \ppara{\mu} \, \FA(q^2)   + i \eps^*_\mu m_B f_B \DelQ +
\ORD(  q_\mu) ) \;,
\end{alignat} 
 where the $\ORD( q_\mu)$ terms are irrelevant as they vanish in the massless case 
 when contracted with the vector or axial 
 lepton matrix element. 
\item By gauge invariance of the full amplitude, as argued above, it is then clear 
that the extra  $\eps_\mu^*$-term must cancel the emission term and thus the full amplitude
can be written as 
\begin{equation}
{\cal A}[\bar B_u \to \ell^-  \bar\nu   \ga(\eps) ] \big|_{m_\ell =0}  \propto s_e \big( \pperp{\mu} \, \FV (q^2)  -
  \ppara{\mu} \, \FA(q^2) \big)
\,  \matel{\ell(l_1) \bar \nu(l_2) }{\bar{\ell} \Gamma^\mu \nu}{0}  \;.
\end{equation}
It seems worthwhile to comment on why in the $m_\ell =0$ approximation the point-like structure (terms proportional to $f_B$  \eqref{eq:pt-like}) disappears. 
This is the case since the point-like term is given by the LO matrix element times $\eps^* \cdot p_B$ and other factors. This is a consequence of Low's theorem \cite{Low:1958sn}and means that as far as angular momentum conservation is concerned it behaves like the LO matrix element. The additional photon forms a Lorentz scalar with $p_B$ and is thus in a relative $p$-wave to the $B$-meson. Finally, since 
the LO matrix element is helicity suppressed (proportional to $m_\ell$) this contribution vanishes entirely in the $m_\ell \to 0$ limit.
\end{itemize}

\section{Correlation Functions and Equations of Motion}
\label{app:eom}

In this appendix we wish to clarify a few technical aspects regarding the EOM.

\subsection{Derivation of the Equations of Motion at Correlation Function Level}
\label{app:eomCorr}

To what extent the naive EOM\footnote{The QED-covariant derivative on $\sigma_{\mu\nu}$ is
  needed in case $Q_b \neq Q_q$. 
  With due apologies for the notation,  here $D_\nu$ (without arrows) is to be understood as a total derivative. }
\begin{equation}
\label{eq:eom}
  \bar q i (  D^\nu  i  \sigma_{\mu\nu} +  \overset{\leftrightarrow}{D}_\mu) [ \ga_5] b  + (m_q \pm   m_b) 
 \bar  q \ga_\mu [\ga_5] b  = 0      \;,
\end{equation}
($A \stackrel{\leftrightarrow}{\partial}_{\mu} B \equiv ({\partial}_{\mu} A) B - A ({\partial}_{\mu} B)$) 
are altered  in correlation functions by \CTs is a typical  field theory problem.
 Below we give the correlation function 
version of \eqref{eq:eom} with insertion of the photon and $B$-meson currents. 
The main result is that with the definition of $\Corr_\parallel$, compatible with the QED WI, 
the EOM are satisfied in the most straightforward way
\begin{alignat}{2}
\label{eq:EOMCF}
&   \Corr^T_\perp(p_B^2,q^2) +\sD  \Corr^{\cal D}_\perp(p_B^2,q^2)  &\;=\;& \frac{(m_b + m_q  )}{m_B} \Corr^{\cal V}_\perp(p_B^2,q^2)  \;,  \nonumber \\[0.1cm]
&  \Corr^T_\parallel(p_B^2,q^2)  + \sD \Corr^{\cal D}_\parallel(p_B^2,q^2)   &\;=\;&
 \frac{(m_b - m_q  )}{m_{B_q}} \Corr^{\cal V}_\parallel(p_B^2,q^2)     \;.
\end{alignat}
This equation is the basis for the simple relation \eqref{eq:EOMFF} and the usual statement
in the literature  
that the EOM hold on physics states. 
Below we show how \eqref{eq:EOMCF} emerges, despite a few \CTs on the way, 
 by starting with correlation functions in coordinate space.

\subsubsection{Current Insertion (hard Photon)}

By variation from the path integral under $q(x) \to \eps_q(x)$ and $\bar b(x) \to \bar \eps_b(x)$
and requiring independence in $\eps_q(x)$ and $ \eps_b(x)$ one gets two equations 
which can be combined to 
\begin{eqnarray}
\label{eq:EOMxspace}
\matel{0}{T \bar q  i  (   \overset{\leftarrow}{\slashed{D}}  \ga_\mu [\ga_5]  \mp   \ga_\mu [\ga_5] \overset{\rightarrow}{\slashed{D}} )b(x) 
j_B (y) j_\rho(0)}{0} &=& 
- (m_q  \pm m_b) 
\matel{0}{T \bar q \ga_\mu [\ga_5] b (x) 
j_B (y) j_\rho(0)}{0} \nonumber    \\[0.1cm] \nonumber
 & & -i \de(x-0)  \,  g_{\mu \rho} \SumQ 
\matel{0}{T j_B (y)  
j_{\tilde{B}[B]} (0) }{0} 
   \\[0.1cm] 
&  & + i  \de(x-y) \, \matel{0}{T  A[V]_\mu  (y) j_\rho (0)}{0} \;,
  \end{eqnarray} 
 where $\SumQ \equiv (Q_b+Q_q)$, 
  $j_{\tilde B[ B]} 
\equiv \bar b [\ga_5]  q$ and  $V[A]_\mu = \sum_{f=q,b} \bar f \ga_\mu [\ga_5] f$.
By using the identity 
\begin{equation}
 i D^\nu \bar q  i \sigma_{\mu \nu} [\ga_5]  b = -  i \bar q \overset{\leftrightarrow}{D}_\mu [\ga_5]  b 
+   \bar q (  i \overset{\leftarrow}{\slashed{D}} \ga_\mu [\ga_5] \mp  \ga_\mu [\ga_5]  i \overset{\rightarrow}{\slashed{D}}    )  b \;,  
\end{equation}
 one gets the useful forms for the $\perp$-direction
\begin{equation}
\label{eq:EOMperp}
 \matel{0}{T \bar q i (  D^\nu  i  \sigma_{\mu\nu} +   \overset{\leftrightarrow}{D}_\mu)  b (x) 
j_B(y) j_\rho(0)}{0} + (m_q +  m_b) 
 \matel{0}{T \bar q \ga_\mu  b (x) 
j_B(y) j_\rho(0)}{0} = 0   \;,
\end{equation}
and the $\parallel$-direction 
\begin{eqnarray}
\label{eq:EOMpara}
 & & \matel{0}{T \bar q i (  D^\nu  i  \sigma_{\mu\nu} +   \overset{\leftrightarrow}{D}_\mu)  \ga_5 b (x) 
j_B(y) j_\rho(0)}{0} + (m_q -   m_b) 
 \matel{0}{T \bar q \ga_\mu \ga_5 b (x) 
j_B(y) j_\rho(0)}{0}   =  \nonumber \\[0.1cm] 
& & -i \de(x-0)  \, g_{\mu \rho} \SumQ 
\matel{0}{Tj_B(y)  
j_B(0)}{0}  + i  \de(x-y) \,\matel{0}{T V_\mu (y) j_\rho(0)}{0} \;.
\end{eqnarray}
Above all derivatives are understood as being 
outside the timer-ordered products. This leads to particularly simple rules in momentum space. 
In order to make contact with our previous 
discussion we take the Fourier transformation $\int_x e^{i qx} \int_y e^{-ip_By}$. 
All \CTs are absent for   the $\perp$-direction,
while for the $\parallel$-direction 
the $\de(x-0)$- and the $\de(x-y)$-terms lead to correlation functions in the variables 
$p_B^2$ and $k^2$ respectively. Crucially, when contracting with the photon polarisation 
tensor $\eps^{*\rho}$ these terms are proportional to $\eps^{*\mu}$ and thus do not contribute to
our projection prescription \eqref{eq:recipe}.
Hence for the current insertion the EOM do indeed assume the simplest possible form \eqref{eq:eom}.

\subsubsection{Distribution Amplitude Part (soft Photon)}

If the photon is not created by the perturbative current then the analogues of 
\eqref{eq:EOMperp} and \eqref{eq:EOMpara} are given by
\begin{eqnarray}
 & & \matel{\ga}{ T \bar q i (  \partial^\nu  i  \sigma_{\mu\nu} +   \overset{\leftrightarrow}{D}_\mu) [ \ga_5] b (x) 
j_B(y)}{0} + (m_q \pm  m_b) 
 \matel{\ga}{T \bar q \ga_\mu [ \ga_5] b (x) 
j_B(y)}{0}   =   \\[0.1cm] \nonumber
& &   + i  \de(x-y) \, 
\underbrace{\matel{\ga}{  A[V]_\mu (y)}{0}}_{=0} \;,
\end{eqnarray}
where ${\matel{\ga}{  A _\mu (y)}{0}} =0$ by parity and 
${\matel{\ga}{  V_\mu (y)}{0}} = 0$ when the photon is not point-like, as is the case here.
Not surprisingly the EOM  assume the same simple form \eqref{eq:eom}
as the current insertion contributions. 

\subsection{Renormalisation of Correlation Functions and Equation of Motion}\label{app:renorm}

Since the (non)renormalisation of the EOM has only been briefly discussed in \cite{BSZ15}
and involves some mixing of operators we consider it worthwhile to revisit this matter, thereby 
showing the consistency of our results with the computation of the renormalisation matrix \cite{BSZ15}.

In order to clarify the notation we remind the reader of the renormalisation of 
a Wilson coefficient in the OPE. The Wilson coefficients correspond, after Fourier transformation, 
to the hard scattering kernels $T_H$ in the notation in \eqref{eq:collfac}. We may discuss the renormalisation 
in the language of the local OPE.  

We use the notation 
$O_i^{(0)}   = \bar Z_{ij} O_j  =  Z_q Z_{ij} O_j$ to denote the relation between bare and renormalised operators. 
The bare renormalisation constants include the wavefunction renormalisation $Z_q$. 
The local  OPE is parameterised by 
$ \vev{ O_A(x) O_B(0) }_\varphi =  C^C_{AB}(x) \vev{O_C(0)}_\varphi $
 where $\vev{\dots}_\varphi$  denotes a matrix element over physical states and summation over the index $C$ is understood.
 The  OPE equation may be written as 
   \begin{equation}
 \label{eq:master}
  (\bar Z^{-1})_{AA'} (\bar Z^{-1})_{BB'}   \vev{ O^{(0)}_{A'}(x) O^{(0)}_{B'}(0)}_\varphi  = C^{C}_{AB}(x) \vev{O_{C}(0)}_\varphi  \;,
\end{equation}
where it was used that the renormalisation matrix of the Wilson coefficient and the operator are 
each other's  inverse.  This is valid in mass-independent schemes such as DR with
$\MSbar$.

In our case: 
(i)  $O_B \to J_{B_q}$ which does not renormalise and thus $\bar Z_{BB'} \to 1$ in the formula 
above, (ii) $O_C  \to O_T 
\sim \bar q \sigma_{\mu \nu} q$ which renormalises multiplicatively, and we use
$Z_T = \bar Z_{TT} = Z_q  Z_{TT}$, (iii) $O_B \to O^D$ for which the renormalisation 
is more involved \cite{BSZ15}. 
We may simplify the task by absorbing the operator $O_4$ in   \cite{BSZ15},
  $(O_1)|_{\textrm{here}}  = (O_1 -O_4)|_{\!\!\mbox{\small{\cite{BSZ15}}}} $,\footnote{Here we focus on the basis relevant to 
the $\perp$-direction. The one for the $\para$-direction is obtained by $ b \to \ga_5 b$ and $m_q \to - m_q$.}
\begin{alignat}{1}
\label{eq:Os}
& O_1 = O_D   =  2 \bar q i \!\stackrel{\leftarrow}{D}_{\mu}  - i \partial_\mu (\bar q b) = 
 \bar q i \!\stackrel{\leftrightarrow}{D}_{\mu} b
\;, \quad    \;,  \nonumber  \\[0.1cm]
& O_2 = O_{\partial T} =   i \partial^\nu (\bar q i \sigma_{\mu \nu} b)     \;, \quad   
O_3 =  O_{mV} =   (m_q+m_b) \bar q \gamma_\mu b    \;,
\end{alignat}
for which the renormalisation matrix, computed in \cite{BSZ15}  cf. \APP A, reads
\begin{equation}
\label{eq:Z}
Z_{AA'} = \delta_{AA'} + \Delta  \left(  \begin{matrix} 2 & 2 & 6   \\ 0 & 0 & 0   \\ 0 & 0 & (2-6)      \end{matrix} \right)   \;, \quad \bar{Z}_{AA'} = \delta_{AA'} + \Delta \left(  \begin{matrix} 0 & 2 & 6   \\ 0 & -2 & 0   \\ 0 & 0 & -6      \end{matrix} \right) \;.
\end{equation}
Above and hereafter\footnote{This equation corrects an obvious factor of $2$ typo in v3 of \cite{BSZ15}.} 
\begin{equation}
 \Delta = C_F \frac{\al_s}{4 \pi} \frac{1}{4-d} = C_F \frac{\al_s}{4 \pi} \frac{1}{2 \eps_{\textrm{UV}}} \;,
 \end{equation}
 and $Z_q = 1 - 2 \Delta$ (Feynman gauge) was used to translate from $Z_{AA'}$  to $\bar Z_{AA'}$.
 In this basis the non-renormalisation of the EOM, 
\begin{equation}
\vev{(O^{(0)}_{\partial T}  + O^{(0)}_D + O^{(0)}_{mV}) \dots }_\varphi  = \vev{(O_{\partial T}  + O_D + O_{mV}) \dots }_\varphi  \;,
\end{equation}
is expressed by the columns of $\bar Z_{AA'}$ adding up to zero. Above the dots stand for single insertion of fields 
leading to \CTs on the LHS. If the field insertions are absent the EOM  simply  assume the form
$\vev{(O^{(0)}_{\partial T}  + O^{(0)}_D + O^{(0)}_{mV}) }_\varphi = \vev{(O_{\partial T}  + O_D + O_{mV})  }_\varphi  = 0$.

In summary, dropping the index $B$,  \eqref{eq:master} simplifies to 
\begin{equation}
C^{T}_{A}(x)   \vev{O_{T}(0)}_\varphi =     (\bar Z^{-1})_{AA'}   \vev{ O^{(0)}_{A'}(x) J_{B_q}^{(0)}(0)}_\varphi   \;.
\end{equation}

After this general discussion it remains to adapt to the operator basis used in this paper. 
The relation is as follows
\begin{equation}
\label{eq:basis}
 O^V = - \frac{1}{e}\frac{m_B}{(m_b+m_q)}  O_{mV}    \;, \quad O^{T,D} = \frac{1}{e}O_{\partial T, D}  \;.
\end{equation}
In the basis  \eqref{eq:basis}, in which our correlation function are presented,  
the  UV-counterterms assume the form 
\begin{equation}
\de \Pi^D =   \Delta \left(   - 2 \Pi^{T(0)}   + 6    \frac{m_b+m_q}{m_B} \Pi^{V(0)} \right)  \;,
\quad \de \Pi^T =    2 \Delta \Pi^{T(0)} \;, \quad \de \Pi^V =    0  \;,
\end{equation}
taking into account that
the  $O_V$-$O_{mV}$ relation  \eqref{eq:basis} implies a sign flip $\bar Z_{13'}$  and dropping  $+6$  in 
$\bar Z_{33}$
 (it corresponds to the mass renormalisation $Z_m = 1 - 6 \Delta$).
Concluding, we have explicitly checked that the renormalised $\Pi^{\type}$ quantity 
\begin{equation}
\label{eq:EOMPI}
\Pi^\type|_{\textrm{NLO}} \equiv \Pi^{\type(0)} + \Pi^{\type(1)} + \de \Pi^\type|_{m_b \to Z_m m_b } \;,
\end{equation}
 is free from UV-divergencies when expanded to order $\ORD(\al_s)$.
Above the superscripts $(0)$ and $(1)$ stand for LO and NLO respectively.  
We have explicitly verified \eqref{eq:EOMPI} in our NLO computations, that is twist-$1$ and twist-$2$.
As usual this shows consistency of the explicit computation and that  $Z_{AA'}$ is 
implicitly contained in the computation of this paper.

\section{Compendium for  the Photon Distribution Amplitude}
\label{app:DAs}

In this appendix we collect the information on the photon DAs. 
The main paper on the photon DA is \cite{Ball:2002ps} where the   powerful background field technique is 
used.  
The photon DA was originally introduced in \cite{Balitsky:1997wi}.
The reasons for writing such an appendix are at least threefold. 
First we collect the DAs used in this paper in more a compact way, second 
we include quark mass corrections and thirdly the DAs are presented 
in a format which is more useful for  computations (e.g. 
elimination of $1/kz$-factors through integration by parts and replacing $\eps^\perp$ etc by 
standard vectors). The most sizeable quark mass corrections are due to twist-$2$, 
the correction to the magnetic vacuum susceptibility \SEC\ref{sec:mqDA} ,and the mixing 
into the twist-$3$ DA via the EOM.

In reference \cite{Ball:2002ps} the photon DA are given for  an external 
photon field (i.e. $\matel{0}{\dots}{0}_{F_{\mu\nu}}$) as well as for a single 
photon state (i.e. $\matel{\ga(q,\eps)}{\dots}{0}$). The two are related 
by assuming a plane wave for the external field. In this appendix we
use the latter.  The photon DAs 
and their hadronic parameters are defined in  \SECs   \ref{app:photonDAdef} and  \ref{app:photonDA} 
respectively. A small dictionary between DAs of the photon and vector mesons is given 
in \TAB\ref{tab:dict}.
The light-cone propagator in the background gauge field is given in the \SEC \ref{app:FS} 
and which tangentially fits into this appendix.

\subsection{Matrix Elements of the Photon Distribution Amplitude}
\label{app:photonDAdef}

The photon DAs are largely analogous to the vector meson ones, cf. \TAB\ref{tab:dict}.   
We identify two main differences. 
Since the photon is massless, the pure $m_V$-terms are zero\footnote{Some $3$-particle 
DAs source the $\mathbb{A}$-DA via the EOM. For the vector meson DA such terms usually 
carry an artificial $m_V^2$ term in their matrix element definition which might led one to think that they vanish 
altogether for the photon.}
and $\matel{\ga}{\bar q F_{\mu\nu} q}{0}$ matrix elements play a distinguished role 
as they do not introduce new hadronic parameters.

It is convenient to first introduce $\eps^\perp$ and $g^\perp_{\mu \nu}$ \cite{BBKT98} 
\begin{equation}
g_\perp^{\mu \nu} = g^{\mu \nu} - \frac{1}{k  z} (k^\mu z^\nu + k^\nu z^\mu) \;, \quad 
\eps^\perp =
 \eps - \frac{\eps  z}{k  z}  k  - \frac{\eps  k}{k  z}  z  = 
 \eps - \frac{\eps  z}{k  z}  k \;,
\end{equation}
which are orthogonal to the light-like directions $z$ and $k$ (photon momentum):
$\eps^\perp k = \eps^\perp z =0 $ and $(g^\perp q)_{\mu}= (g^\perp z)_{\mu} =0$.\footnote{Above and hereafter we (often) suppress the contractions for brevity, e.g.
$k \cdot  z \to kz$.}  
It is also worth noting that $\eps^\perp$ 
is invariant under $\eps \to \eps + k$ which is the residual gauge transformation of the Lorenz 
gauge $\eps(k) \cdot k = 0$. 
In the formulae we use  $g_{\mu \nu}$ and $\eps_\mu$ rather than the $\perp$ version. 
Factors $1/kz$ that arise from these quantities are treated, 
as usual, via  integration  by parts
\begin{equation}
\label{eq:1/kz}
\frac{1}{kz}\int_0^1\!\! du\,e^{i\bar u[u] kz} f(u) \to i [-i] \!\int_0^1\!\! du
\,e^{i\bar u kz} I_1[ f](u),\quad  I_1[ f](u) \ =  \ \int_0^u\!\! dv\, f(v),
\end{equation}
where  $\int_0^1 dvf(v)=0$ is assumed and has to hold in order to produce a smooth $kz \to 0$ limit. 
More concretely, we use the notation 
\begin{alignat}{2}
& F^{(1)}(u) &\;\equiv\;&  I_1[F](u)=\int_0^u dw\, F(w) \;, \nonumber \\
& F^{(2)}(u)  &\;\equiv\;&  I_2[F](u)= I_1 \circ I_1[F](u) 
=\int_0^u dw\int_0^w dv\, F(v) \;.
\end{alignat}
 When comparing with \cite{Ball:2002ps} it should be taken into account that these authors 
 use $s_e|_{\textrm{BBK}}=-1$ and $s_g|_{\textrm{BBK}}=-1$ which is more of a standard in 
 the DA-literature.

\paragraph{Twist-$2$:}  The only twist-$2$ DA $\phi_\ga$ is  defined by
\begin{alignat}{2}
& \vev{\gamma(k,\epsilon)|\bar q (0) \sig_{\al \be} q (z)|0} &\; =\;&     {+}  i s_e e Q_q  \vev{\bar q q } \eps^*_{[\al} k_{\be]}
\int_0^1 du e^{i  \bar uk z}  \left( \chi_q \phi_\gamma(u)   + \frac{x^2}{4} {\mathbb A}(u)
 \right)
\nonumber \\[0.1cm]
&   &\;  \;&{-}\frac{i}{2} s_e e Q_q \vev{\bar qq}\left(k z\,\epsilon^*_{[\alpha}z_{\beta]}-\epsilon^* z\, k_{[\alpha}z_{\beta]}\right)\intu \, e^{i\bar uk z}\doublehat{h}_{\gamma}(u)
\label{eq:t2} \;,
\end{alignat} 
where   $k_{[\alpha}z_{\beta]} = k_\al z_\be -  k_\be z_\al $ hereafter. 
and Wilson lines  are not shown  for brevity. 
Here $\chi_q$ refers to the magnetic vacuum susceptibility for  a quark $q$.\footnote{Unlike in \cite{Lyon:2013gba} we do not adapt the sign of $\chi$  to the convention but  keep  $\chi > 0$.}   
The incorporation of quark mass effects is not-trivial as it requires 
a subtraction prescription and our  proposal is presented in the main text in  \SEC\ref{sec:mqDA}. 
The DAs $h_\ga$ and $\mathbb A$ are of twist-$4$ 
but are already mentioned at this point. 
The hadronic parameters of the other matrix elements  are hidden in the DAs, 
and revealed in the next section.
We consider it worthwhile to illustrate the procedure \eqref{eq:1/kz} for $h_\ga$
\begin{equation}
\frac{\vev{\bar q q}}{kz}  \eps^{\perp\,*}_{[\al} z_{\be]} 
\int_0^1 du e^{i   \bar u \,k z} h_\ga(u) =  -  \vev{\bar qq}\left(k z\,\epsilon^*_{[\alpha}z_{\beta]}-
\epsilon^* z\, k_{[\alpha}z_{\beta]}\right)\intu \, e^{i\bar u \,k z}\doublehat{h}_{\gamma}(u) \;.
\end{equation}
It is noted that RHS is still  explicitly gauge invariant.

\paragraph{Twist-$3$:} There are two $2$-particle DAs\footnote{The notation $\varepsilon_\mu( \eps^*,k,z)$
is a Levi-Civita tensor contracted with three vectors $\eps^*$, $k$ and $z$.}
\begin{alignat}{2}
\label{eq:t32}
& \matel{ \ga(k,\eps)}{ \bar q(0)   \ga_\mu q(z)    }{0}    &\;=&\; - i s_e e Q_q\fg\left(k z\, \epsilon^*_{\mu}-\epsilon^* z\, k_{\mu}\right)\intu \, e^{i \bar uk z}\singlehat{\Psi}_{(v)}(u)\\[0.1cm] \nonumber 
& \matel{ \ga(k,\eps)}{ \bar q(0)   \ga_\mu \ga_5 q(z)    }{0}  &\;=&\; {-}\frac{1}{4} s_e e Q_q f_{3 \ga}  \varepsilon_\mu( \eps^*,k,z)  
 \int_0^1 du e^{i \bar uk z}  \tilde{\Psi}_{(a)}(u)  
\end{alignat}
which are the analogue of $g^{(v,a)}_\perp$ for vector mesons however since the photon mass is zero 
all its contributions enter through the EOM. In addition there are mass corrections, for which we follow the notation in \cite{BSZ15}
\begin{equation}
\label{eq:tdepl}
\tilde{\Psi}_{(a)} = (1 - \tilde{\de}_+) {\Psi}_{(a)}  \;, \quad \tilde{\de}_+ = \frac{  f_{\ga q }  }{f_{3 \ga}} 2 m_q  \;,
\end{equation}
which generalises $\tilde{\de}_+(V) = f_V^\perp  2 m_q /(f_V m_V)$ from the vector meson case 
upon inspection of the matrix elements
\begin{equation}
( f_V^T,m_V f_V)  \leftrightarrow -s_e e Q_q (f_{\ga q}, f_{3\ga})  \;.
\end{equation}
Finally there are also the  $3$-particle DAs\footnote{The identification is $\al_{1} = \al_q$, $\al_2= \al_{\bar q}$ and $\al_3 = \al_g$. Their notation is not optimal 
for us since it is adapted to an incoming, rather than an outgoing, photon.}  
\begin{alignat}{2}
\label{eq:t33}
& \matel{ \ga(k,\eps)}{ \bar q(0) g G_{\rho\sigma}(vz)  \ga_\mu \ga_5  q(z)    }{0}  &\;=\;&
s_g s_e e Q_q f_{3 \ga} \, k_\mu \,  \varepsilon_{\rho\sigma}(k,\eps^*)
 \int {\cal D}\underline{\al}  e^{i  k z (\al_2 + v \al_3)}  {\cal A}(\al)    \;,\nonumber \\[0.1cm]
& \matel{ \ga(k,\eps)}{ \bar q(0) g G_{\rho\sigma}(vz)  \ga_\mu   q(z)    }{0}  &\;=\;& 
{-} i s_g s_e e Q_q f_{3 \ga} \, k_\mu  k_{[\rho} \eps^*_{\perp \sigma]} 
\int {\cal D}\underline{\al}  e^{i  k z (\al_2 + v \al_3)}  {\cal V}(\al)    \;,
 \end{alignat}
which also contribute via the EOM.
Above $\tilde G_{\al \be} \equiv \frac{1}{2} \eps_{\al \be \ga \de} G^{\ga \de}$ with 
$G_{\al \be} = \partial_{[\al}  A_{\be]} + ..$ and  ${\cal D}\underline{\al}  = 
d \al_1 d \al_2 d \al_3 \de(1- \sum_i \al_i)$ is the measure of the momentum fractions. 
And finally, the hadronic parameter  $f_{3\ga}$ is defined by
\begin{equation}\label{eq:f3gaDef}
 \matel{\ga(k, \eps) } { \bar q g \tilde G_{\mu\nu} \gamma_\alpha \gamma_5  q} { 0} =  
 - s_e s_g Q_q f_{3\gamma}  
 k_\al k_{[\mu} \eps_{\nu]} \;.
\end{equation}

\paragraph{Twist-4} 
A word of caution before we start out. The twist-$4$ sector requires the 
introduction of a yet unknown $4$-particle DA in order to be complete in all hadronic parameters cf. \SEC\ref{sec:EOMtest}. 
We therefore drop the twist-$4$ terms (other than the pure $Q_{b} \vev{\bar qq}$ ones) in 
the numerics but keep them in the analytic results. 
The twist-$4$ $2$-particle DAs $h_\ga$ and $\mathbb A$
are defined in \eqref{eq:t2} already and the quark condensate constitutes 
a third one. The $Q_q \vev{\bar qq}$ condensate terms are not well-defined in 
the OPE for on-shell photon since the come with a $1/k^2$-factor and hence these terms
  have  to be absorbed into
a DA.\footnote{This is only one of the technical possibilities cf. the discussion 
at the beginning of section $3$ in \cite{Ball:2002ps}.}. We arrive at the same expression 
as  \cite{Lyon:2013gba}, taking into account that $s_e|_{\textrm{LZ}}=-1$, with some more
detail   
 \begin{alignat}{2}
 \label{eq:qqfun}
 & \vev{\gamma(k,\epsilon)|\bar q (x)  q (z)|0}  &\;\to\;&   
 - i s_e e \int d^4 y e^{i kz} \eps^*_\rho  \matel{0}{T j^\rho(y)\bar q (x)  q (z)  }{0}
  \nonumber \\[0.1cm]
 & &\;=\;& +  s_e e Q_q  \frac{ \eps^*  (x-z)}{ k(x-z)} \vev{\bar q q }  \int_0^1 du e^{i  (\bar u kz + u  kx) }  (\de(u) - \de(\bar u)) \nonumber \\[0.1cm]
 & &\;=\;&  - i  s_e e Q_q   \vev{\bar q q } \eps^* (x-z) \int_0^1 du e^{i  (\bar u kz + u  kx) } \one|_{\one =1}  \;,
 \end{alignat}
 where $\one = 1$ is introduced as a flag in order to trace this term in the computation. 
 It is needed in order to satisfy the QED WI for example.
In addition to this exotic term there are the $3$-particle twist-$4$ DAs 
 given by 
\begin{alignat}{2}
\label{eq:S2}
& \langle  \ga(k,\eps) | \bar q(0)g   {G}_{\rho\sigma}(vz) q(z) | 0  \rangle
&\;=\;&
{+} i s_g  s_e e Q_{q} \vev{\bar q q} 
 \, k_{[\rho} \eps^*_{\perp \sigma]} 
\int {\cal D}\underline{\alpha} {S}(\underline{\alpha})
e^{i kz (\al_2 + v \al_3)} \,,
\nonumber \\  
& \langle  \ga(k,\eps) | \bar q(0) eQ_q   {F}_{\rho\sigma}(vz) q(z) | 0  \rangle
&\;=\;&
{+} i  e Q_{q} \vev{\bar q q} 
 \, k_{[\rho} \eps^*_{\perp \sigma]} 
\int {\cal D}\underline{\alpha} {S}_\ga(\underline{\alpha})
e^{i kz (\al_2 + v \al_3)} \,,
\nonumber \\
& \langle  \ga(k,\eps) | 
\bar q(0)gG_{\rho\sigma}(vz)i\gamma_5 q(z) |0 \rangle
&\;=\;&
{-}i s_gs_e e Q _{q}\vev{\bar qq}  \,  \varepsilon_{\rho\sigma}(k,\eps^*) \int {\cal D}\underline{\alpha} \tilde{{\cal S}}(\underline{\alpha})
e^{i kz (\al_2 + v \al_3)} \;,
\end{alignat}
and 
\begin{eqnarray}
\label{eq:Ti}
\langle \ga(k,\eps) | 
\bar q(0)g{G}_{\rho\sigma}(vz) \sigma_{\mu \nu} q(z) |0   \rangle  =
{+}s_g s_e  \frac{e Q_q\vev{\bar q q}}{k z} \sum_{n=1}^4 (t_n)_{[\rho\sig][\mu\nu]}
T_n(v,kz) \;, \nonumber \\[0.1cm] 
\langle \ga(k,\eps) | 
\bar q(0) e Q_q {F}_{\rho\sigma}(vz) \sigma_{\mu \nu} q(z) |0   \rangle  =
{+}   \frac{e Q_q\vev{\bar q q}}{k z} (t_4)_{[\rho\sig][\mu\nu]}
T_{4\ga} (v,kz) \;, 
\end{eqnarray}
with Lorentz structures $t_i$ 
\begin{alignat}{4}
& (t_1)_{\rho\sig\mu\nu}&\;=\;&
 \left(k z\,k_{\mu}\epsilon^*_{\rho}g_{\nu\sigma}-k_{\mu}\epsilon^*_{\rho}z_{\nu}k_{\sigma}-\epsilon z \,k_{\mu} k_{\rho}g_{\nu\sigma}\right)  \;, \quad  (t_3)_{\rho\sig\mu\nu}  &\;=\;& k_{\mu}k_{\rho}\epsilon^*_{\nu}z_{\sigma}
 \;, \nonumber \\[0.1cm]
 &  (t_2)_{\rho\sig\mu\nu}  &\;=\;& \left(k z\,k_{\rho}\epsilon^*_{\mu}g_{\nu\sigma}-k_{\rho}\epsilon^*_{\mu}z_{\sigma}k_{\nu}-\epsilon z\, k_{\mu}k_{\rho}g_{\nu\sigma}\right) \;, \quad  (t_4)_{\rho\sig\mu\nu} &\;=\;&
 k_{\mu}k_{\rho}\epsilon^*_{\sigma}z_{\nu}   \;,
\end{alignat}
and
\begin{eqnarray}
\label{Ti}
   T_i (v,kz) =\int {\cal D}\underline{\alpha} 
\,e^{i kz (\al_2 + v \al_3)}T_i(\underline{\alpha}) \;.
\end{eqnarray}
We note that $\{ {\cal S}_\ga ,T_{4 \ga} \}$ have the same form as
$\{ {\cal S} ,T_{4 } \}$ and thus it is convenient to define.
\begin{equation}
\label{eq:convenient}
 Q_q \overline{{\cal S}} =  Q_q {\cal S} +  Q_b {\cal S}_\ga \;, \quad    Q_q  \overline{T}_{4 }  = Q_q T_{4 }  + Q_b T_{4\ga } \;, 
\end{equation}
as the results will necessarily depend on this combination. 
The quantities $\{ {\cal S}_\ga ,T_{4 \ga} \}$ are special in that they do not introduce a further 
hadronic quantity other than $\vev{\bar q q }$ cf. subsection \SEC\ref{app:photonDAdef}.

The remaining $1/kz$-factor in \eqref{eq:Ti} requires special treatment because of the additional integral 
over $v$ originating from  the background gauge field propagator \eqref{eq:BGFprop}. 
Moreover, for the computation it proves efficient to write the integrals over the $3$-particle 
parameters $\al_i$ as an integral over a single parameters $u$ in complete analogy 
to the $2$-particle case
\begin{align}\nonumber
\left( 1, \frac{i}{kz} \right)\int_0^1 dv&\,G(v)\inta\, F(\au)e^{\frac{ikz}{2}\big(\alpha_2-\alpha_1+(2\bar v-1)\alpha_3+1\big )} \\[0.3em]\nonumber
=& \int_0^1 dv\,G(v)\int_0^1 d\alpha_3\int_0^{1-\alpha_3} d\alpha_2\, 
\left(F(\au),  \singlehat{F}(\au) \right)\,e^{ikz(\alpha_2+\bar v\alpha_3 )}\\[0.3em] 
=& \int_0^1 du\,e^{i\bar u kz} \left(F_\textrm{eff}(k,u),  \singlehat{F}_\textrm{eff}(k,u) \right) \;,
\end{align}
where 
\begin{equation}
\left(F_\textrm{eff}(k,u),  \singlehat{F}_\textrm{eff}(k,u) \right) \equiv
\int_0^1 d\alpha_3\int_{\text{Max}[0,\bar u -\alpha_3]}^{\text{Min}[1-\alpha_3,\bar u]} d\alpha_2\,\frac{1}{\alpha_3}G(v(\au,u)) \left(F(\au),  \singlehat{F}(\au) \right) \;,
\end{equation}
with  $v(\au,u)=(\alpha_2+\alpha_3-\bar u)/\alpha_3$ and
\begin{equation}
\singlehat{F}(\au)=\int_0^{\alpha_2} d\omega\,F(1-\omega-\alpha_3,\omega,\alpha_3) \;.
\end{equation}
For the $1/kz$ relation,   $\int_0^1 d\alpha_3\,e^{ikz(1-v \alpha_3)}\int_0^{1-\alpha_3} d\alpha_2 F(\au)=0$, is required and has to hold, as before, for a smooth $kz \to 0$   limit.
The vanishing of this boundary terms has been explicitly verified for all $T_i(\au)$. 
In practice the function $G(v) \to \{1,v \}$.

\begin{table}
\centering
  \resizebox{\columnwidth}{!}{
\begin{tabular}{C | C |  L |  L  | L |  L  | L    }
   \textrm{twist}  &   j_{\textrm{conf}}  &  \ga\textrm{-DA}  &  \textrm{vector meson-DA}     &   \ORD(m_V^n) [\ga] & \textrm{partons}  & \textrm{sym}   \\ \hline
2 & 2^{\phantom{*}}   &  \phi_{\ga} \; \mbox{\eqref{eq:t2}}  & \phi_{\perp} = \phi_{2,V}^\perp      &  0 &  \bar q \sigma_{\mu\nu}  q  & (u,\bar u)_S \\
2 & 2^{\phantom{*}}   &  - &  \phi_{\parallel} = \phi_{2,V}^\para& 1 & \bar q \ga_\mu q  & (u,\bar u)_S  \\ 
\hline
3 & \frac{3}{2}^*  &  \Psi^{(v)} \; \mbox{\eqref{eq:t32}}   & g_\perp^v  = \mathbb{B}  = \phi_{3,V}^\perp & 1 \;[\textrm{eom } {\cal V}]  &  \bar q \ga_\mu  q  & (u,\bar u)_S  \\
3 & \frac{3}{2}^*  & \Psi^{(a)} \; \mbox{\eqref{eq:t32}}   & g_\perp^a = \mathbb{D} =  \psi_{3,V}^\perp & 1\; [\textrm{eom } {\cal A}] & \bar q  \ga_\mu\ga_5  q  & (u,\bar u)_S \\
3 & \frac{3}{2}^*  & - & h^{(s)}_\para= e = \psi_{3,V}^\para& 2 \;  & \bar q q   & (u,\bar u)_S \\
3 & \frac{3}{2}^*  & -  &  h^{(t)}_\para = h_L = \phi_{3,V}^\para  & 2  & \bar q \sigma_{\mu\nu}q  
& (u,\bar u)_S \\
3 & \frac{9}{2}^{\phantom{*}}  & {\cal V} \; \mbox{\eqref{eq:t33}}  & {\cal V} = \Phi^\parallel_{3,V} & 1 [\bar q G q] & \bar q \ga_\mu G_{\al\be} q  & (\al_1,\al_2)_A \\
3 & \frac{7}{2}^{\phantom{*}}  & {\cal A}\; \mbox{\eqref{eq:t33}}   & {\cal A} = \tilde{\Phi}^\parallel_{3,V}& 1 [\bar q \tilde{G} q] & \bar q \ga_\mu \ga_5 \tilde{G}_{\al\be} q &  (\al_1,\al_2)_S \\
3 & \frac{9}{2}^{\phantom{*}}  & - & {\cal T} = \tilde{\Phi}^\perp_{3,V}& 1 & \bar q \sigma_{\mu\nu} {G}_{\al\be} q  
& (\al_1,\al_2)_A  \\  \hline
    4 &  3^{\phantom{*}}   & \mathbb{A} \; \mbox{\eqref{eq:t2}} & \mathbb{A}_{T,\perp} = \chi_\perp   = \phi_{4,V}^\perp   & 2 \;[\chi^{-1}] & \bar q \sigma_{\mu\nu}  q  & (u,\bar u)_S  \\ 
4 & 3^{\phantom{*}}   & -  & \mathbb{A}_\para  = \chi_\para  = \mathbb{A} = \phi_{4,V}^\para & 3 & \bar q  \ga_{\mu}  q  & (u,\bar u)_S  \\ 
4 &1^{\phantom{*}}  &  - &  g_{\parallel,3}=  \psi_{4,V}^\para   & 3 & \bar q \ga_\mu q   & (u,\bar u)_S \\
4 & 1^{\phantom{*}}  & h_\ga  \; \mbox{\eqref{eq:t2}} &  h_{\perp,3} =  \psi_{4,V}^\perp  & 2 \; [\chi^{-1}] & \bar q \sigma_{\mu\nu}  q  & (u,\bar u)_S \\
4 & 1^{\phantom{*}}   & \vev{\bar q q}  \; \mbox{\eqref{eq:qqfun}} &  - & - [\vev{\bar q   q}]  & \bar qq   & (u,\bar u)_S\\
4 &\frac{7}{2}^{\phantom{*}}   & {\cal S},{\cal \tilde{S}}\; \mbox{\eqref{eq:S2}}   & {\cal S},{\cal \tilde{S}} = \Psi^\perp_{4,V},\tilde{\Psi}^\perp_{4,V}& 2 [\bar q \tilde{G} q] & \bar q  ({G},\ga_5 \tilde{G})_{\al\be} q & (\al_1,\al_2)_A  \\
4 & 4,\frac{7}{2},4,\frac{7}{2}  & T_{1..4} \; \mbox{\eqref{eq:Ti}}   & {\cal T}_{1..4} = \tilde{\Phi}^{(1..4)\perp}_{4,V}& 2 [\bar q \tilde{G} q] & \bar q  \sigma_{\mu\nu} {G}_{\al\be}  q  & (\al_1,\al_2)_A
 \end{tabular}
 }
 \caption{Brief overview of correspondence of photon and vector meson DAs and their properties. 
 Columns from left to right signify the twist, conformal spin (where applicable),  photon DA,  
 vector meson DA, 
 order in the vector meson mass [reason for being non-vanishing for photon DA
 when vector meson mass present], parton structure and symmetry properties of the DA. 
 The symmetry properties refer to the DA-variables as used in this work and for the vector mesons $G$-parity 
 odd states are assumed in accordance with the photon quantum numbers. $S$ and $A$ stand for symmetric and antisymmetric respectively as can be inferred  from the explicit DAs given in \SEC\ref{app:photonDA}.
 The twist and conformal spin are given by $t = l-s$ and $j  = (l+s)/2$, where $l$ is the dimension of the operator 
 and $s$ its spin whereby the $k$, $\perp$ and $z$ directions count as $+1$, $0$ and $-1$ with regard to spin.
 The conformal spin takes into account the lowest non-vanishing hadronic parameter which is different 
 for particles of even or non-definite $G$-parity. E.g. for the $j_{{\cal V}_{K^*}} = 7/2 $ and not
$j_{{\cal V}_{\rho}} = 9/2 $.  The asterisk on the conformal spin indices states that 
 the DA is an admixture of the conformal states e.g. $(j_{\bar q} , j_{q}) = (1,0) \pm (0,1)$. 
 We do not show the twist-$5$ DA $\mathbb{G}^{v,a}_\perp$ introduced in  \cite{BSZ15}.
 For the vector meson we give some of the various names given to the DAs in the literature are shown
 e.g. \cite{BBKT98},  \cite{Ball:1998fj} \cite{BBL07}. For the photon DA we use the \cite{Ball:2002ps}-notation which is standard.}
 \label{tab:dict}
\end{table}

\subsection{Explicit Distribution Amplitudes}
\label{app:photonDA}

In this section we give the form of the DA used in this paper 
based on the results in \cite{Ball:2002ps} with added quark mass corrections. 
As before we present in increasing twist.
The hadronic parameters with respect to their twist and values are summarised in 
\TABs \ref{tab:twist} and \ref{tab:input_params} respectively.

\begin{table}[ht]
\centering
\begin{tabular}{ c|c|c|c }
	Twist & 2 & 3 & 4 \\  \hline
	2-Particle & $f_{\ga q } \equiv \chi_q \vev{\bar qq}, a_2(\ga), a_4(\ga)$ & $\fg$, $\wg V$, $\wg A$ &  $\qbarq$, $\qbarq \kappa^+$\\ \hline
	3-Particle & - &$\fg$, $\wg V$, $\wg A$ & $\qbarq$$\big($$\kappa$, $\kappa^+$, $\zeta_1$, $\zeta_1^+$, $\zeta_2$, $\zeta_2^+$$\big)$ 
\end{tabular}
\caption{\small Hadronic parameters and their association with the twist. The conformal spin of the $2$-particle 
and $3$-particle DA-parameters is 
$j_{\chi \vev{\bar qq} a_n} = 2 +n$ (with $a_0\equiv 1$) and $j_{( \fg, \wg V, \wg A)} = (7/2,9/2,9/2)$ respectively. The bottom right corner does not close under the EOM, it requires the inclusion 
of $4$-particle DAs cf. the discussion in \SEC\ref{sec:EOMtest}.
}
\label{tab:twist}
\end{table}

\paragraph{Twist-2:} The  $\phi_\gamma$ DA 
takes  the same form as the   $\rho_0$ meson DA 
\begin{equation}
\label{eq:phigamma}
\phi_\gamma(u) = 6\bar uu \left( 1 + \sum_{n \geq 1} a_{2n}(\ga) C_{2n}^{3/2}(\xi) \right) \;,
\end{equation}
where $C_{n}^{3/2}$  is a Gegenbauer polynomial, $\xi \equiv 2u-1$,
 formally  $a_0(\ga) =1$ and all  odd Gegenbauer moments vanish ($a_{2n+1}(\ga) =0$)
as the photon is $G$-parity odd. 
The corresponding hadronic parameter $\chi_q$ is given in \eqref{eq:t2} is determined from a sum rule 
or lattice QCD for $m_q =0$ and the mass correction to it is new to this work and presented in \SEC\ref{sec:mqDA}.
It can be understood as the linear response 
to an external electromagnetic field 
$\vev{\bar q \sigma_{xy} q}_{B_z} =  e \chi_q \vev{ \bar q q} B_z + \textrm{non-linear}$. Hence the name 
susceptibility.  In \cite{Ball:2002ps} the parameter  $a_2(\ga)$ is investigated but 
it is concluded that  no reliable information on  $a_2(\ga)$ or any higher order Gegenbauer moment 
can be extracted from the sum rules considered.  We take 
$a_2(\ga)|_{2\GeV}  = 0(0.1)$ as a reference value for its uncertainty given that one of the determinations
in  \cite{Ball:2002ps} yields $a_2(\ga)|_{1\GeV} = 0.07$.\footnote{The instanton model \cite{Petrov:1998kg} 
photon DA, quoted in  \cite{Ball:2002ps}, is normalised at $\mu_0 = 0.6 \GeV$ and gives 
$a_0(\ga,\mu_0) = 0.96$ and $a_2(\ga,\mu_0) = 0.03$ with all others Gegenbauer moments vanishing.}

\paragraph{Twist-3:} The twist-$3$ DAs are characterised by $3$-particle DA parameters of which the leading 
term is defined in \eqref{eq:f3gaDef}. 
The DA are given by ($\xi \equiv 2u -1$)
\begin{alignat}{2}
& {\cal V} (\underline{\alpha}) &\;=\;& 
540\,  \omega^V_\gamma
 (\alpha_1-\alpha_2)\alpha_1 \alpha_2\alpha_3^2\;,
\nonumber\\
& {\cal A} (\underline{\alpha})&\;=\;& 
360\, \alpha_1 \alpha_2 \alpha_3^2 
\left[ 1+ \omega^A_\gamma\,\frac{1}{2}\,(7\alpha_3-3)\right]\ \;,
\nonumber\\
& \psi^{(v)}(u)&\;=\;& 10 C^{1/2}_2(\xi) + \frac3{8} C^{1/2}_4(\xi)
(15\omega^V_\gamma-5\omega^A_\gamma )  + \frac{3}{2} \tilde{\de}_+ ( 2+ \ln u + \ln \bar u) \;,
\nonumber\\
& \psi^{(a)}(u) &\;=\;&  \frac53(1-\xi^2) C^{3/2}_2(\xi)\left(1+\frac9{16}\omega^V_\gamma-
\frac3{16}\omega^A_\gamma \right) + 
6 \tilde{\de}_+ (3 u \bar u + u \ln u + \bar u \ln \bar u)
\;,
\end{alignat}
where $\tilde{\de}_+$ is defined in \eqref{eq:tdepl}.
These matrix elements match those in eq.~5.4 in \cite{BBKT98} upon 
application of the following  map ($\zeta_{3V}^{V,A} = f_{3V}^{V,A}/(m_V f_V)$ as defined in eq 4.5 \cite{BBKT98})
\begin{equation}
(\omega^A_{1,0} , f_{3V}^A, f_{3V}^V ) \leftrightarrow ( \omega_\ga^A,  f_{3 \ga}, \frac{3}{28}  f_{3 \ga} \omega_\ga^V) 
\end{equation}
and dropping all other hadronic parameters as they vanish for the photon.
Moreover, they agree with  \cite{Ball:2002ps} (where the definition of 
the hadronic parameters $\omega^{V,A}_\gamma$ can be found)  
in the limit  of no quark mass corrections $\tilde{\de}_+ \to 0$ as expected.

\paragraph{Twist-4:} 

 The expression for twist-$4$ are given in eq.~4.27 and eq.~4.42 in  \cite{Ball:2002ps}
\begin{alignat}{2}
\label{eq:Tampl}\nonumber
& h_{\gamma}(u) &\; =\;& -10(1+2\kappa^+)C_2^{1/2}(\xi) = -10(1+2\kappa^+)(6u^2-6u+1)\\ &\nonumber
T_1(\underline{\alpha}) &\; =\;& {-}120(3\zeta_2+\zeta_2^+) (\alpha_2-\alpha_1)
\alpha_2\alpha_1\alpha_3,\nonumber \\ &
T_2(\underline{\alpha}) &\; =\;& 30\alpha_3^2 (\alpha_2-\alpha_1)
\left[
 (\kappa-\kappa^+) +\,(\zeta_1-\zeta_1^+)\, (1-2\alpha_3) +
  \zeta_2(3-4\alpha_3)
\right],\nonumber \\ &
T_3(\underline{\alpha}) &\; =\;& -120(3\zeta_2-\zeta_2^+) (\alpha_2-\alpha_1)
\alpha_2\alpha_1\alpha_3,\nonumber \\ &
T_4(\underline{\alpha}) &\; =\;&
30\alpha_3^2 (\alpha_2-\alpha_1) \left[
 (\kappa+\kappa^+) + \,(\zeta_1+\zeta_1^+)\, (1-2\alpha_3) +
  \zeta_2(3-4\alpha_3)
\right],\nonumber \\ &
T_{4\ga}(\underline{\alpha}) &\; =\;&
- 60\alpha_3^2 (\alpha_2-\alpha_1)  (4-7(\al_1+\al_2)) \;
,\nonumber \\ &
{ S}(\underline{\alpha}) &\; =\;& 30\alpha_3^2\! \left\{
(\kappa+\kappa^+)\,(1-\alpha_3) +
  (\zeta_1+\zeta_1^+) (1-\alpha_3)(1-2\alpha_3)\right.\nonumber\\ &
       & & {}\left.+
   \zeta_2[3(\alpha_2-\alpha_1)^2-\alpha_3 (1-\alpha_3)]\right\} ,\nonumber \\ &
   { S}_{\ga}(\underline{\alpha}) &\; =\;& 60 \al_3^2 (\al_1+\al_2)(4-7(\al_1+\al_2)) \;,\nonumber \\ &
\widetilde{ S}(\underline{\alpha}) &\; =\;& -30\alpha_3^2\! \left\{
 (\kappa-\kappa^+) \,(1-\alpha_3)  +
(\zeta_1-\zeta_1^+) (1-\alpha_3)(1-2\alpha_3)\right.\nonumber \\ &
& & {}\left.+
   \zeta_2[3(\alpha_2-\alpha_1)^2-\alpha_3 (1-\alpha_3)]
 \right\},\nonumber \\ &
\mathbb{A}(u) &\; =\;& 40u^2\bar{u}^2(3\kappa-\kappa^++1) +8\left(\zeta_2^+-3\zeta_2\right)\Big[u\bar u(2+13u\bar u)\nonumber \\ &
& & +2u^3(10-15u+6u^2)\ln u+2\bar{u}^3(10-15\bar u+6\bar u^2)\ln\bar u\Big],
\end{alignat}
where the 
$\zeta_i$'s are defined   eq.~4.30 in  \cite{Ball:2002ps} and 
the $\kappa$'s are originate from 
 \begin{alignat}{2}
& \langle  \ga | \bar q g{G}_{\mu\nu}  q |0 \rangle &\;=\;&
s_g s_e Q_q  \vev{\bar qq} (\kappa^+ + \kappa)\,
 i k_{[\mu} \eps^*_{\nu]} \;, \quad  \nonumber  \\[0.1cm]
& \langle  \ga | \bar q g\widetilde{G}_{\mu\nu}\gamma_5  q |0 \rangle &\;=\;&
s_g s_e Q_q \vev{\bar qq} \, (\kappa^+ - \kappa)\,  i k_{[\mu} \eps^*_{\nu]}\;.
\end{alignat}
Note that we have the opposite sign in $T_{4\ga}$ which is indeed what one finds upon using 
the expressions in  \cite{Ball:2002ps} eq.~4.35-4.36 and we have checked that it is this form that satisfies
the EOM in eq~4.45.  This is however not relevant at a numerical level since we have to neglect this contribution as it requires the inclusion of $4$-particle DAs cf. the discussion in \SEC\ref{sec:EOMtest}.
The terms with no new hadronic parameter in $h_\ga$ and $\mathbb A$ come from
 \begin{equation}
 \label{eq:F3}
 \matel{\ga}{\bar q(x) F_{\mu \nu} (vx) q (0) }{0} = \vev{\bar q q}  i k_{[\mu} \eps^*_{\nu]} e^{-i v kx}  + \ORD(x^2) 
 \end{equation}
 through the EOM with $S_\ga,T_{4\ga}$ (cf. eq 4.16 \cite{Ball:2002ps}).
 The lack of hadronic parameter is somewhat unfamiliar and maybe ought to be compared best to the vector meson mass
 corrections for the vector meson DAs.

\subsection{Quark Propagator on the Light-cone in  Fock-Schwinger Gauge}
\label{app:FS}

The Fock-Schwinger gauge is defined by $(x - x_0) \cdot A^a(x) = 0$.
The power of the technique is that one can replace usual by covariant derivatives 
$(x- x_0) \cdot \partial \equiv (x - x_0) \cdot D $.
For $x_0 = 0$  a formal solution is given by 
\begin{equation}
A^a_\mu(x) = \int_0^1 d \alpha \, \alpha x^\omega G^a_{\omega \mu} (\alpha x) 
 = \frac{1}{2}   x^\omega G^a_{\omega \mu} (0)  + \dots   \, ,
\end{equation}
which is used to compute diagram $A_{D_1})$ in \FIG\ref{fig:DA-deriv-graphs}.

For the computation of the 3-particle DA contribution the propagator in the 
gluon background field is of use and its light-cone version has been derived 
in \cite{Balitsky:1987bk} which we write for a massive propagator in the form
$\matel{0}{T q(0) \bar q(x)}{0}_A$ rather than $\matel{0}{T q(x) \bar q(0)}{0}_A$
which amounts to a replacement of $v \to \bar v$ in the gluon field.  One expands 
\begin{eqnarray}
 \matel{0}{T q(0) \bar q(x)}{0}_A = i\int \frac{d^4k}{(2\pi)^4} e^{ i k\cdot x} S_q(k) \;, \quad S_c(k) = S_c^{(0)}(k) + S_c^{(2)}(k,x) + \dots\;,
\end{eqnarray}
where we only need the first correction $S_q^{(2)}$ which can be written in a symmetric form,
\begin{eqnarray}
\label{eq:BGFprop}
S^{(0)}_q &=& \frac{\slashed{k}+m_q}{k^2-m_q^2} \;, \nonumber \\
S^{(2)}_q &=&  s_g g_s \frac{1}{2}\int _0^1 dv \Big( v\sigma\!\cdot\! G(\bar v x)  
\frac{\slashed{k}+m_q}{(k^2-m_q^2)^2} + 
\bar v   \frac{\slashed{k}+m_q}{(k^2-m_q^2)^2} \sigma\!\cdot\! G( \bar v x) \big)
\,.
\end{eqnarray}
The variable $s_g$ is the sign of the covariant derivative \eqref{eq:cD} which we prefer to keep 
explicit.  Note further terms in the expansion can be read off from eq.~A.16 in  \cite{Balitsky:1987bk}, 
and even though given for a massless propagator the inclusion of quark mass would seem rather 
straightforward.

\section{Results of Correlation Functions}
\label{app:results}
In this appendix we collect the result of our computation in terms of correlations functions.

\subsection{Leading order Correlation Functions}

We introduce the following shorthands
\begin{alignat}{2}
& \den  &\;\equiv\;&  m_b^2-u p_B^2-\bar u q^2 \;, \nonumber \\
& \denbar &\;\equiv\;&  m_b^2+(\al_2+\bar v \al_3-1)p_B^2-(\al_2+\bar v \al_3)q^2\equiv m_b^2-u_{\text{eff}} \,p_B^2-\bar u_{\text{eff}} \,q^2 \;,
\end{alignat}
and remind the reader that $\one$ is a flag that has to be set to one cf. \eqref{eq:qqfun}. We denote $C_0(0,p_B^2,q^2,m_q^2,m_q^2,m_b^2)$ in the limit that $m_q$ is small, as $\Czz$. 
In terms of PV functions the LO correlators read
\begin{align}
s_e\,\Pi_{\perp}^V(p_B^2,q^2)=&\frac{m_b^2 m_B N_c}{4 \pi ^2}\Bigg[
   \DelQ\left(1-\frac{m_q}{m_b}\right)
   \frac{B_0\left(p_B^2,0,m_b^2\right)-B_0\left(q^2,0,m_b^2\right)}{p_B^2-q^2}\nonumber \\[0.2cm]
   &+Q_b\,\Cz+\frac{m_q}{m_b}Q_q\,\Czz\Bigg]\nonumber \\[0.2cm]
&{+}m_b m_B Q_q \qbarq\intu\frac{\chi\phi_{\gamma}(u)}{\den}{+}\frac{1}{2} m_b^2m_B Q_q f_{3\gamma}\intu\frac{\Psi_{(a)}(u)}{\den^2}\nonumber \\[0.2cm]
&{-}\frac{1}{4}m_b m_B Q_q \qbarq\intu\frac{2m_b^2+\den}{\den^3}\mathbb{A}(u)\nonumber \\
&+\,s_g^2m_b\,m_B\,Q_q\,\qbarq\intv\inta\Bigg[ \frac{{+}\overline{{\cal S}}(\au){+}\tilde{{\cal S}}(\au)(2v-1)}{\denbar^2}\nonumber \\ 
&{-}\frac{2(p_B^2-q^2)}{\denbar^3}\Big(-\singlehat{T}_1(\au)+\singlehat{T}_2(\au)+(2v-1)\singlehat{T}_3(\au)-(2v-1)\singlehat{\overline{T}}_4(\au)\Big)\Bigg]\nonumber \\
&+\frac{m_bm_B Q_b\qbarq}{(m_b^2-p_B^2)(m_b^2-q^2)}\;,
\end{align}

\begin{align}
s_e\,\Pi_{\parallel}^V(p_B^2,q^2)=&\frac{m_b^2 m_B N_c}{4 \pi ^2
   q^2 (p_B^2-q^2)^2}
   \Bigg[\left(1-\frac{m_q}{m_b}\right)\Bigg\{B_0\left(q^2,0,m_b^2\right) \left(m_b^2(p_B^2-2 q^2) \DelQ-q^4 \SumQ\right)\nonumber \\[0.1cm]
   &+B_0\left(p_B^2,0,m_b^2\right)q^2 \left(m_b^2 \DelQ+q^2 \SumQ\right)-(p_B^2-q^2)\left(A_0\left(m_b^2\right)\DelQ-q^2 \SumQ\right)\nonumber\Bigg\} \\[0.3cm]
   &+C_0\left(0,p_B^2,q^2,m_b^2,m_b^2,0\right) q^2 (p_B^2-q^2) \left(2m_b^2\left(1-\frac{m_q}{m_b}\right)-(p_B^2-q^2)\right)Q_b\nonumber \\[0.1cm]
   &+\Czz \frac{m_q}{m_b}q^2(p_B^2-q^2)^2Q_q\Bigg]\nonumber \\[0.1cm]
  &{+}m_b m_B Q_q \qbarq\intu\frac{\chi\phi_{\gamma}(u)}{\den}-2m_b^2m_BQ_q \fg\intu\frac{\Psi^{(1)}_{(v)}(u)}{\den^2}\nonumber \\
&+2 m_b m_B Q_q \qbarq\intu\frac{\bar u\,\one}{\den^2}{+}2m_b\,m_B\,Q_q\qbarq\intu\frac{2m_b^2+\den}{\den^3}\,h^{(2)}_{\gamma}(u)\nonumber \\
&{-}\frac{1}{4}m_b m_B Q_q \qbarq\intu\frac{2m_b^2+\den}{\den^3}\mathbb{A}(u)\nonumber \\
&\,{-}s_g^2m_b\,m_B\,Q_q\,\qbarq\intv\inta\Bigg[\frac{\overline{{\cal S}}(\au)(2v-1)+\tilde{{\cal S}}(\au)}{\denbar^2} \nonumber \\
&{-}\frac{2(p_B^2-q^2)}{\denbar^3}\Big(-\singlehat{T}_1(\au)-(2v-1)\singlehat{T}_2(\au)+(2v-1)\singlehat{T}_3(\au)+\singlehat{\overline{T}}_4(\au)\Big)\Bigg]\nonumber \\
&-\frac{m_bm_B Q_b\qbarq}{(m_b^2-p_B^2)(m_b^2-q^2)}\;,
\end{align}

\begin{align}
s_eiX^V(p_B^2,q^2)=&\frac{m_b^2m_B N_c\DelQ}{4 \pi ^2 {p_B^2} {q^2}(p_B^2-q^2)} \bigg[{q^2} \left({p_B^2}-m_b^2\right) {B_0}\left({p_B^2},0,m_b^2\right)-{p_B^2} \left({q^2}-m_b^2\right)
   {B_0}\left({q^2},0,m_b^2\right)\nonumber\\&+{A_0}\left(m_b^2\right) ({q^2}-{p_B^2})\bigg]+2 m_b m_B Q_q \qbarq\intu\frac{\one}{\den^2}-\frac{2m_bm_B Q_b\qbarq}{(m_b^2-p_B^2)(m_b^2-q^2)}\;,
\end{align}

\begin{align}
s_e\,\Pi_{\perp}^T(p_B^2,q^2)=&\frac{m_b N_c}{8 \pi ^2
   (p_B^2-q^2)}
   \Bigg[B_0\left(p_B^2,0,m_b^2\right) \left(m_b^2 \DelQ-p_B^2 \SumQ\right)\nonumber \\
  & +B_0\left(q^2,0,m_b^2\right) \left(m_b^2 \DelQ+q^2\SumQ\right)-(p_B^2-q^2)\SumQ\nonumber \\
  &+2m_q m_b(p_B^2-q^2)\left(Q_b\Cz+Q_q\Czz\right)\Bigg]\nonumber \\
&{+}m_b^2\,Q_q\,\qbarq\intu\frac{\chi\phi_{\gamma}(u)}{\den}\nonumber \\
& -m_b\,Q_q\,\fg \intu\left({-}\frac{1}{4}\,\frac{\den+q^2+m_b^2}{\den^2}\,\Psi_{(a)}(u) +\frac{\den-q^2+m_b^2}{\den^2}\,\Psi^{(1)}_{(v)}(u)\right)\nonumber \\
&{-}2m_b^2\,Q_q\,\qbarq\intu\frac{q^2-m_b^2}{\den^3}\,h^{(2)}_{\gamma}(u){-}\frac{1}{2}m_b^4 Q_q \qbarq\intu\frac{\mathbb{A}(u)}{\den^3}
\nonumber \\
&+s_g^2m_b\,Q_q\,\fg\intv\inta\frac{v(p_B^2-q^2)}{\denbar^2}\left(\mathcal{A}(\au){+}\mathcal{V}(\au)\right)\nonumber \\
&+\,s_g^2m_b^2\,Q_q\,\qbarq\intv\inta\Bigg[\frac{{+}\overline{{\cal S}}(\au){-}\tilde{{\cal S}}(\au)}{\denbar^2}\nonumber \\
&{-}\frac{2(p_B^2-q^2)}{\denbar^3}\Big(-\singlehat{T}_1(\au)+\singlehat{T}_2(\au)-\singlehat{T}_3(\au)+\singlehat{\overline{T}}_4(\au)\Big)\Bigg]\nonumber \\
&+\frac{m_b^2 Q_b\qbarq}{(m_b^2-p_B^2)(m_b^2-q^2)}\;,
\end{align}
    
\begin{align}
s_e\,\Pi_{\parallel}^T(p_B^2,q^2)=&\frac{m_bN_c}{8 \pi ^2(p_B^2-q^2)^2}
   \Bigg[B_0\left(p_B^2,0,m_b^2\right) \left(m_b^2(p_B^2+q^2)\DelQ-p_B^2 (p_B^2-3 q^2)\SumQ\right)\nonumber \\[0.1cm]
   &+B_0\left(q^2,0,m_b^2\right) \left(m_b^2 (p_B^2-3 q^2) \DelQ-q^2(p_B^2+q^2)\SumQ\right)\nonumber \\[0.1cm]
   &+4 \,C_0\left(0,p_B^2,q^2,m_b^2,m_b^2,0\right) \,m_b^2\, q^2 (p_B^2-q^2)Q_b- 2A_0\left(m_b^2\right) (p_B^2-q^2)\DelQ\nonumber \\[0.1cm]
  & +\left(p_B^2-q^2\right)\left(p_B^2+q^2\right)\SumQ\nonumber \\[0.1cm]
    &+2m_q m_b(p_B^2-q^2)^2\left(Q_b\Cz+Q_q\Czz\right)\Bigg]\nonumber \\
&{+}m_b^2\,Q_q\,\qbarq\intu\frac{\chi\phi_{\gamma}(u)}{\den}\nonumber \\
& -m_b\,Q_q\,\fg \intu\left({-}\frac{1}{4}\frac{\den-q^2+m_b^2}{\den^2}\,\Psi_{(a)}(u)+\frac{\den+q^2+m_b^2}{\den^2}\,\Psi^{(1)}_{(v)}(u) \right)\nonumber \\
&{+}2m_b^2\,Q_q\,\qbarq\intu\frac{q^2+m_b^2}{\den^3}\,h^{(2)}_{\gamma}(u){-}\frac{1}{2}m_b^4 Q_q \qbarq\intu\frac{\mathbb{A}(u)}{\den^3}\nonumber \\
&+s_g^2m_b\,Q_q\,\fg\intv\inta\frac{v(p_B^2-q^2)}{\denbar^2}\left(\mathcal{A}(\au){+}\mathcal{V}(\au)\right)\nonumber \\
&{-}\,s_g^2m_b^2\,Q_q\,\qbarq\intv\inta\Bigg[\frac{-\overline{{\cal S}}(\au)+\tilde{{\cal S}}(\au)}{\denbar^2}\nonumber \\
&{-}\frac{2(p_B^2-q^2)}{\denbar^3}\Big(-\singlehat{T}_1(\au)+\singlehat{T}_2(\au)-\singlehat{T}_3(\au)+\singlehat{\overline{T}}_4(\au)\Big)\Bigg]\nonumber \\
&+\frac{m_b^2 Q_b\qbarq}{(m_b^2-p_B^2)(m_b^2-q^2)}\;,
\end{align}

\begin{align}
s_e\,\Pi_{\perp}^D(p_B^2,q^2)=&\frac{m_b N_c}{8 \pi ^2 (p_B^2-q^2)}
   \Bigg[B_0\left(p_B^2,0,m_b^2\right)\left(m_b^2 \DelQ+p_B^2\SumQ\right)\nonumber \\
   &-B_0\left(q^2,0,m_b^2\right) \left(m_b^2 \DelQ+q^2 \SumQ\right)\nonumber \\
  & +2C_0\left(0,p_B^2,q^2,m_b^2,m_b^2,0\right) m_b^2(p_B^2-q^2) Q_b+(p_B^2-q^2)\left(Q_b+Q_q\right)\Bigg]\nonumber \\
   &{-} \frac{m_b Q_q \fg}{2}\intu\frac{\Psi_{(a)}}{\den}-2s_g^2m_bQ_q\fg\intv\inta\frac{v(p_B^2-q^2)\mathcal{A}(\au)}{\denbar^2} \nonumber\\[0.1cm]
   &{+}4s_g^2m_b^2Q_q\qbarq\intv\inta\,\frac{v\tilde{{\cal S}}(\au)}{\denbar^2} ,
\end{align}

\begin{align}
s_e\,\Pi_{\parallel}^D(p_B^2,q^2)=&\frac{m_b N_c}{8 \pi ^2 q^2
   (p_B^2-q^2)^2}
   \Bigg[B_0\left(q^2,0,m_b^2\right) \Big[2m_b^4\left(1-2\frac{m_q}{m_b}\right) (p_B^2-2 q^2) \DelQ\nonumber \\[0.1cm]
   &+m_b^2q^2 (q^2(Q_b-5Q_q)- p_B^2\DelQ))+q^4(p_B^2+q^2+4m_q m_b) \SumQ\Big]\nonumber \\[0.2cm]
& + B_0\left(p_B^2,0,m_b^2\right)q^2 \Big[2 m_b^3(m_b-2m_q)\DelQ+m_b^2 (q^2(Q_b+3Q_q)- p_B^2\DelQ)\nonumber \\[0.2cm]
&+(p_B^2(p_B^2-3 q^2)-4m_q m_b q^2) \SumQ\Big]\nonumber \\[0.2cm]
 &+2C_0\left(0,p_B^2,q^2,m_b^2,m_b^2,0\right) m_b^2q^2(p_B^2-q^2) \left(2 m_b^2-4m_qm_b-p_B^2-q^2\right) Q_b \nonumber \\[0.2cm]
   &-2 A_0\left(m_b^2\right)\left(m_b^2-2m_q m_b-q^2\right) (p_B^2-q^2)\DelQ \nonumber \\[0.1cm]
   &+q^2 (p_B^2-q^2) \left(2m_b^2-4m_qm_b-p_B^2-q^2\right)\SumQ\Bigg]\nonumber \\
   &-2m_b^2 Q_q \qbarq \intu \frac{\xi\,\one}{\den^2} +2m_b Q_q \fg\intu\frac{\singlehat{\Psi_{(v)}}}{\den}\nonumber\\[0.2cm]&{-}2s_g^2m_bQ_q\fg\intv\inta\frac{v(p_B^2-q^2)\mathcal{V}(\au)}{\denbar^2}
   {-}4s_g^2m_b^2Q_q\qbarq\intv\inta\,\frac{v\overline{{\cal S}}(\au)}{\denbar^2} \nonumber\\[0.2cm]
   & -\frac{2m_b^2Q_b\qbarq}{(m_b^2-p_B^2)(m_b^2-q^2)}.
\end{align}

\begin{align}
s_eiX^D(p_B^2,q^2)=&\frac{m_b^3 N_c\DelQ}{4 \pi ^2 {p_B^2} {q^2}(p_B^2-q^2)} \bigg[{q^2} \left({p_B^2}-m_b^2\right) {B_0}\left({p_B^2},0,m_b^2\right)-{p_B^2} \left({q^2}-m_b^2\right)
   {B_0}\left({q^2},0,m_b^2\right)\nonumber\\&+{A_0}\left(m_b^2\right) ({q^2}-{p_B^2})\bigg]+2 m_b^2 Q_q \qbarq\intu\frac{\one}{\den^2}-\frac{2m_b^2 Q_b\qbarq}{(m_b^2-p_B^2)(m_b^2-q^2)}\;,
\end{align}

\subsection{Next-to-leading Order at twist-\texorpdfstring{$1$}{} (Perturbative Diagrams)}

The computation is too lengthy to present in a paper, instead we include the results in the ancillary file \texttt{corr\_PT\_NLO.m}. The results are split according to both FF and charge i.e. \texttt{PiVperpQq} corresponds to the contribution from $\Pi_{\perp}^V(p_B^2,q^2)|_{\text{PT}}$ proportional to $Q_q$. The full perturbative correlation function can be obtained by summing the contribution from each charge, e.g. $\Pi_{\para}^T(p_B^2,q^2)|_{\text{PT}}=$\texttt{PiTparaQq}$+$\texttt{PiTparaQb}.

\subsection{Next-to-leading Order at twist-\texorpdfstring{$2$}{}}
\label{app:resultsNLOt2} 

The $\ORD(\alpha_s)$ corrections to the twist-$2$ amplitude can be decomposed into the following form,
\begin{equation}
\label{eq:t2NLO}
s_e \Pi^{\type}(q^2,p_B^2)={-}  K_{\type} m_B Q_q \chi\qbarq  \frac{C_F \alpha_s}{4\pi}\intu\,\phi_{\gamma}(u)\left(\calpha^{\type } A_0+\cbeta_i^{\type}B_0^i+\cgamma_i^{\type} C_0^i+R^{\type}\right),
\end{equation}
with $(K_{V},K_{T,D})   = (m_b ,m_B)$. For brevity we drop the reference to the projection as both directions yield the same result. The PV functions that appear, along with their corresponding coefficients, are given below with $R^{\type}$ denoting rational terms.

\begin{flalign}
A_0&=A_0(m_b^2)\;,	&	B_0^0&=B_0(0,0,0)\;,			&	C_0^0&=C_0(0,p_B^2,u p_B^2+\bar u q^2,0,0,m_b^2)\;,\nonumber \\
	&			&	B_0^1&=B_0(p_B^2,0,m_b^2)\;, 	&	C_0^1&=C_0(0,q^2,u p_B^2+\bar u q^2,0,0,m_b^2)\;,\nonumber \\
	&  			&	B_0^2&=B_0(q^2,0,m_b^2)\;,	&	&\nonumber \\
	&			& 	B_0^3&=B_0(u p_B^2+\bar u q^2,0,m_b^2)\;,	& 
\end{flalign}

\begin{flalign*}
&\calpha^V=\frac{2m_b^2-\den}{\den^2(m_b^2-\den)} \;,	\qquad\quad	 \cbeta^V_0=-\frac{3}{\den}\;,	\;\qquad\quad	\cbeta^V_1=\frac{2m_b}{\bar u(p_B^2-q^2)\den }\;,		\qquad\quad	\cbeta^V_2=\frac{m_b^2-3q^2}{u(p_B^2-q^2)\den }\;,&
\end{flalign*}
\vspace{-0.3cm}
\begin{flalign*}
&\cbeta^V_3=\frac{4 m_b^6 \bar{u}- m_b^4 (2 \den  (5-4 u)+4 q^2 \bar{u})+2 \den m_b^2 (4 q^2 \bar{u}+\den  (3-2 u))-3 \den ^2 q^2 \bar{u}}{\bar{u} u\left(p_B^2-q^2\right) \left(m_b^2-\den \right)\den^2}\;, & 
\end{flalign*}
\vspace{-0.3cm}
\begin{flalign}
&\cgamma^V_0=\frac{2(p_B^2-m_b^2)}{\den}\;,	\;\;\qquad\quad	\cgamma^V_1=\frac{2(q^2-m_b^2)}{\den}\;,	\qquad\quad R^V=\frac{3\den-2\,m_b^2}{\den^2}\;,&\nonumber 
\end{flalign}
\begin{flalign}
\calpha^T&=\frac{2}{\den^2}\;,	&	\calpha^D&=\frac{1}{\den(m_b^2-\den)}\;,\nonumber \\
\cbeta^T_0&=-\frac{3}{\den}\;,	&	\cbeta^D_0&=0\;,\nonumber\nonumber \\
\cbeta^T_1&=\frac{2(m_b^2-\den)}{\bar u(p_B^2-q^2)\den}\;,	&	\cbeta^D_1&=\frac{2}{\bar u(p_B^2-q^2)}\;,\nonumber \\
\cbeta^T_2&=-\frac{2q^2}{u(p_B^2-q^2)\den }\;,	&	\cbeta^D_2&=\frac{m_b^2-q^2}{u(p_B^2-q^2)\den }\;,\nonumber \\
\cbeta^T_3&=-\frac{2\left(m_b^2-\den\right)\left(\den-2\bar u(m_b^2-q^2)\right)}{\bar u u(p_B^2-q^2)\den^2}\;,		&	\cbeta^D_3&=\frac{2(\den-m_b^2)+\bar u q^2}{u \bar u (p_B^2-q^2)(m_b^2-\den)}\;,\nonumber \\
\cgamma^T_0&=\frac{2\left(p_B^2-m_b^2\right)}{\den}\;,	&	\cgamma^D_0&=0\;,\nonumber\\
\cgamma^T_1&=\frac{2\left(q^2-m_b^2\right)}{\den}\;,		&	\cgamma^D_1&=0\;,\nonumber\\
R^T&=\frac{2(\den-m_b^2)}{\den^2}\;, 	&	R^D&=\frac{1}{\den}\;.&
\end{flalign}

\

\section{Multiple-Polylogarithms}
\label{app:mpl}

MPLs, or Goncharov functions, can be viewed as generalisations of the classical polylogarithms. MPLs arise naturally when using the iterative method of differential equations to compute Feynman integrals. The development of the method of differential equations and advances in the understanding of MPLs represent a major breakthrough for multi-loop calculations \cite{Duhr:2014woa}.
The MPLs of weight $n$ are defined iteratively by the integral,
\begin{equation}
    G(w_1,\dots w_n;u)=\int_0^u\frac{ dt}{t-w_1}G(w_2,\dots,w_n;t).
\end{equation}

\noindent Some useful equivalences with the classical polylogarithms  are given below.
\begin{align}
    G(\mathbf{0}_n;u)&=\frac{1}{n!}\log^n(u)  \;,&G(\mathbf{a}_n;u)&=\frac{1}{n!}\log^n\left(1-\frac{u}{a}\right) \;,\nonumber \\
    G(\mathbf{0}_n,a;u)&=-\mathrm{Li}_{n+1}\left(\frac{u}{a}\right)  \;,&
\end{align}
for $a\neq0$. For the case where $w_n\neq 0$ the following scaling property holds,
\begin{equation}\label{eq:Gdiv}
    G(w_1,\dots,w_i,\dots,w_n;u)=G\left(\frac{w_1}{w_i},\dots,1,\dots,\frac{w_n}{w_i};\frac{u}{w_i}\right) \;,
\end{equation}
which proves useful in performing the photon on-shell limit $k^2\to 0$.

\subsection{Discontinuities}
\label{app:discG}

To identify the discontinuity of an MPL across some branch cut we require two key ingredients. The first is that since the discontinuity operator preserves weight and the discontinuity itself must be proportional to $2\pi i$ the discontinuity of a weight-$n$ MPL can only be comprised of objects of weight-$(n\!-\!1)$. \footnote{$\pi$ is of weight one  since $ \log(x\pm i \eps)=\log|x|\pm i \pi \Theta(-x)$ } The second key ingredient is a piece of machinery from the mathematics of algebras and coalgebras, whose utility arises from the fact that MPLs form a Hopf algebra. That is, an MPL $f$, satisfies the relation 
\begin{equation}\label{eq:coaction}
    \Delta(\text{Disc}\, f)=(\text{Disc} \otimes \text{id})\Delta(f).
\end{equation}
The coaction on MPLs $\Delta$ has been used without definition, however we direct the interested reader to \cite{Duhr:2014woa,Duhr:2012fh} for a detailed discussion of the coaction and the MPL Hopf algebra. The following calculation is facilitated by the \texttt{Mathematica} package \texttt{PolyLogTools} \cite{polylogtools}.
Proceeding by means of  example, we consider a weight-4 MPL which is the highest weight appearing in the calculation. To keep expressions manageable we consider only the kinematic region of interest, $p_B^2>m_b^2>q^2$. From the first ingredient we know that the discontinuity must take the following form,
\begin{equation}\label{eq:disc_gen}
    \text{Disc}\;G\!\left(0,1,\frac{p_B^2}{q^2},1,\frac{p_B^2}{m_b^2}\right)=2\pi i\left(g_3+\pi g_2+\pi^2g_1+\alpha\pi^3\right),
\end{equation} 
where the $g_i$ are some combination of weight-$i$ objects and $\alpha$ is a constant. Information on the $g_i$ can then be gained from looking at different components of the coaction on \eqref{eq:disc_gen}, once the functional dependence is known the remaining constant can be determined numerically. To determine $g_3$ we inspect the $1,3$ component of the coaction on \eqref{eq:disc_gen},
\begin{equation}\label{eq:D13}
    \Delta_{13}\left(\text{Disc}\;G\!\left(0,1,\frac{p_B^2}{q^2},1,\frac{p_B^2}{m_b^2}\right)\right)=2 i (\pi\otimes g_3).
\end{equation}
Using \eqref{eq:coaction}, the LHS of \eqref{eq:D13} can be rewritten as follows,
\begin{align}\label{eq:disc_D13}\nonumber
    \Delta_{13}\left(\text{Disc}\;G\!\left(0,1,\frac{p_B^2}{q^2},1,\frac{p_B^2}{m_b^2}\right)\right)=\,&(\text{Disc} \otimes \text{id})\Delta_{13}\left(G\!\left(0,1,\frac{p_B^2}{q^2},1,\frac{p_B^2}{m_b^2}\right)\right)\\\nonumber
    =&\,2i\,\Theta\!\left(\frac{p_B^2}{m_b^2}-1\right)\pi\otimes\Bigg[G\!\left(0,1,\frac{p_B^2}{m_b^2}\right)G\!\left(1,\frac{q^2}{p_B^2}\right)\\ \nonumber
    & -2\,G\!\left(0,\frac{p_B^2}{m_b^2}\right)G\!\left(1,\frac{q^2}{p_B^2},\frac{q^2}{p_B^2}\right) -G\!\left(0,1,\frac{p_B^2}{q^2},\frac{p_B^2}{m_b^2}\right) \\
    & +G\!\left(0,\frac{q^2}{p_B^2},1,\frac{q^2}{p_B^2}\right)\Bigg],
\end{align}
where we have used the fact that $\text{Disc} \,G(1,p_B^2/m_b^2)=-2\pi i\,\Theta(p_B^2/m_b^2-1)$. Comparing \eqref{eq:D13} and \eqref{eq:disc_D13} the function $g_3$ can be easily read off,
\begin{align}
g_3=\Theta\!\left(\frac{p_B^2}{m_b^2}-1\right)\Bigg[&G\!\left(0,1,\frac{p_B^2}{m_b^2}\right)G\!\left(1,\frac{q^2}{p_B^2}\right)
     -2\,G\!\left(0,\frac{p_B^2}{m_b^2}\right)G\!\left(1,\frac{q^2}{p_B^2},\frac{q^2}{p_B^2}\right)\nonumber\\
     &-G\!\left(0,1,\frac{p_B^2}{q^2},\frac{p_B^2}{m_b^2}\right)
      +G\!\left(0,\frac{q^2}{p_B^2},1,\frac{q^2}{p_B^2}\right)\Bigg].
\end{align}
To obtain $g_2$ we repeat the process but this time consider the $2,2$ component,
\begin{equation}\label{eq:D22}
    \Delta_{22}\left(\text{Disc}\;G\!\left(0,1,\frac{p_B^2}{q^2},1,\frac{p_B^2}{m_b^2}\right)\right)=\Delta_{22}(2\pi i \,g_3)+2 i (\pi^2\otimes g_2).
\end{equation}
The LHS can again be evaluated using \eqref{eq:coaction}, however this time the discontinuity operator will act on weight-2 MPLs. These discontinuities can be computed in an identical manner to the one outlined here, however only the $1,1$ component and a numerical fit of a constant will be required. In this manner, computing a discontinuity is an iterative process building on the knowledge of discontinuities of lower weight MPLs. Evaluating \eqref{eq:D22} we find $g_2=0$. Proceeding we identify $g_1$ through the $3,1$ component,
\begin{equation}
    \Delta_{31}\left(\text{Disc}\;G\!\left(0,1,\frac{p_B^2}{q^2},1,\frac{p_B^2}{m_b^2}\right)\right)=\Delta_{31}(2\pi i \,g_3)+2 i (\pi^3\otimes g_1).
\end{equation}
This time the evaluation of the LHS will require discontinuities of weight-3 MPLs which can be determined by considering the $1,2$ and $2,1$ components of their respective coactions and numerically fitting a constant. Inserting all the relevant quantities we find,
\begin{equation}
    g_1=\frac{1}{6}\Theta\!\left(\frac{p_B^2}{m_b^2}-1\right)G\!\left(1,\frac{q^2}{p_B^2}\right).
\end{equation}
Finally we numerically determine $\alpha=0$, such that the full discontinuity in the region of interest is given by,
\begin{eqnarray}
 &   & \text{Disc}\;G\!\left(0,1,\frac{p_B^2}{q^2},1,\frac{p_B^2}{m_b^2}\right)=2\pi i\, \Theta\!\left(\frac{p_B^2}{m_b^2}-1\right)\Bigg[ \left(G\!\left(0,1,\frac{p_B^2}{m_b^2}\right)+\frac{\pi^2}{6}\right)G\!\left(1,\frac{q^2}{p_B^2}\right)\nonumber \\
 &   &  -2\,G\!\left(0,\frac{p_B^2}{m_b^2}\right)G\!\left(1,\frac{q^2}{p_B^2},\frac{q^2}{p_B^2}\right)
    -G\!\left(0,1,\frac{p_B^2}{q^2},\frac{p_B^2}{m_b^2}\right)+G\!\left(0,\frac{q^2}{p_B^2},1,\frac{q^2}{p_B^2}\right)\Bigg].
\end{eqnarray}

\subsection{The MPL Basis}
As discussed in section \ref{sec:PT} it was necessary to compute additional master integrals to those of 
\cite{DiVita:2017xlr} in order for large cancellations to become manifest. These additional integrals were related 
to the ones of \cite{DiVita:2017xlr} by the swapping of external legs ($p_B^2\leftrightarrow q^2$). Whilst both 
the $Q_q$ (A type) and $Q_b$ (B type) computations required these additional integrals it was only necessary 
to explicitly compute them for the B-type diagrams. For the A-type MI the relative simplicity of the alphabet 
translates to there being just five unique MPL arguments,
\begin{align}
w_1&=-\frac{k^2}{p_B^2} \;,    &w_2&=\frac{2 k^2}{k^2-q^2-p_B^2+\Lambda}\;, \nonumber \\
w_3&=\frac{2 k^2}{k^2-q^2-p_B^2-\Lambda } \;,   &w_4&=-\frac{k^2}{q^2} \;,\qquad\qquad \qquad\quad w_5=-
\frac{k^2}{m_b^2} \;,
\end{align}
where $\Lambda=\sqrt{(k^2-p_B^2-q^2)^2-4p_B^2q^2}$. Under $p_B^2\leftrightarrow q^2$ the arguments 
$w_{2,3,5}$ map onto themselves and $w_1\leftrightarrow w_4$. Consequently the basis of MPLs for the A-
type MIs is closed under swapping of external legs. To generate the additional A-type MIs we can therefore 
simply take $p_B^2\leftrightarrow q^2$ in the appropriate MIs of \cite{DiVita:2017xlr} and large cancellations of 
MPLs will still be manifest. For the B-type MIs there are 19 possible arguments of the MPLs. These arguments 
do not close under leg swapping so a similar procedure cannot be applied.

\bibliographystyle{utphys}
\bibliography{../../Refs-dropbox/References_FF-Bgamma.bib}

\end{document}